\documentclass[aps,reprint,twocolumn,superscriptaddress,longbibliography,floatfix]{revtex4-2}

\usepackage{graphicx}
\usepackage{amsmath,amssymb,bm}
\usepackage{booktabs}
\usepackage[colorlinks=true,linkcolor=blue,citecolor=blue,urlcolor=blue]{hyperref}

\newtheorem{theorem}{Theorem}
\newtheorem{lemma}{Lemma}
\newtheorem{corollary}{Corollary}
\newtheorem{proposition}{Proposition}

\newtheorem{result}{Result}  

\makeatletter
\def\@opargbegintheorem#1#2#3{\trivlist
  \item[\hskip\labelsep{\bfseries #1\ #2}]{\bfseries(#3)}\hspace{0.45em}\itshape\ignorespaces}
\makeatother

\newcommand{\chitwo}{\chi^{(2)}}
\newcommand{\avg}[1]{\langle #1 \rangle}

\begin{document}

\title{A charge selection rule fixes what a squeezed-light reservoir\\ computer can compute and afford}

\author{Daniel Soh}
\affiliation{Wyant College of Optical Sciences, University of Arizona, Tucson, AZ, USA}

\date{\today}

\begin{abstract}
Reading an optical quantum reservoir costs repetitions growing super-exponentially with feature order. For reservoirs encoding data in a parametric pump's phase, one conservation law fixes what is readable and its cost. Pairwise photon exchange conserves an integer phase charge: across an ensemble of input masks and squeeze settings, order-$D$ readout reaches exactly the assemblies of at most $D$ unit-charge kernels; degree-one homodyne readout is universal for fading-memory functionals along weak-squeezing families, at shot cost polynomial in accuracy; and at fixed squeezing no finite degree reaches every sector, at a computable distance. Nonlinearity sits in the optics; detector variance is fixed at every order. In a hardware-faithful twin the entangled-register machine outperforms a task-tuned classical reservoir at matched readout dimension and noise by 1.7--6.1$\times$ on nonlinear benchmarks and by 7.1 points on open RF data, and a displacement-encoded control at identical photon number loses 17.7 points, as the charge algebra predicts.
\end{abstract}

\maketitle

\section{Introduction}
\label{sec:intro}

Reservoir computing fixes a dynamical system, drives it with the
input, and trains only a linear readout on the
trajectory~\cite{Jaeger2004,Appeltant2011,Larger2012,Brunner2013};
quantum reservoir computing carries the idea to quantum
dynamics~\cite{Fujii2017,Govia2021}, with continuous-variable optics a
natural host. A cornerstone is the theorem of Nokkala \emph{et al.}:
Gaussian states under linear symplectic dynamics already provide
universal reservoir computing~\cite{Nokkala2021,GarciaBeni2023}. The
theorem is exact---and silent on the question that dominates any
laboratory realization: \emph{what does each feature cost to
measure?}
In a Gaussian scheme all nonlinearity must come from the readout,
through higher moments of the quadratures ($a=X+iP$); the single-shot
variance of an estimator of $\avg{X^m}$ is
\begin{equation}
\mathrm{Var}(X^m)\sim(2m-1)!!\,\sigma_X^{2m},
\label{eq:factorial}
\end{equation}
to leading order (exact coefficients in Methods): averaging $X^m$
is dominated by rare large excursions, so the budget to resolve a
high-order feature grows catastrophically, while any scheme avoiding
the ladder through a weak transducer pays a fixed referral of its own
[Eq.~\eqref{eq:crossover}]. Universality theorems say what a machine can \emph{express},
not what it can \emph{afford}~\cite{Ehlers2025,Hua2025,Ahmed2025}.

Two recent developments sharpen the question. Paparelle \emph{et al.}
have demonstrated an optical continuous-variable reservoir in which
data are written into the pump phase of a parametric process and
retrieved by mode-selective homodyne detection, with controlled fading
memory and a high-fidelity digital twin~\cite{Paparelle2026}: the
architecture class treated here now exists in the laboratory. Hu
\emph{et al.} have characterized what a physical learner can resolve
under a finite sampling budget~\cite{Hu2023}, establishing measurement
cost rather than expressivity as the operative constraint, and further
work improves what a fixed measurement ensemble
buys~\cite{Hahto2025}. What none of these supplies is an exact
statement of \emph{which} functions the dynamics makes readable, and at
what shot cost; that is what one conservation law delivers here.

\begin{figure*}[!t]
\centering
\includegraphics[width=0.9\textwidth]{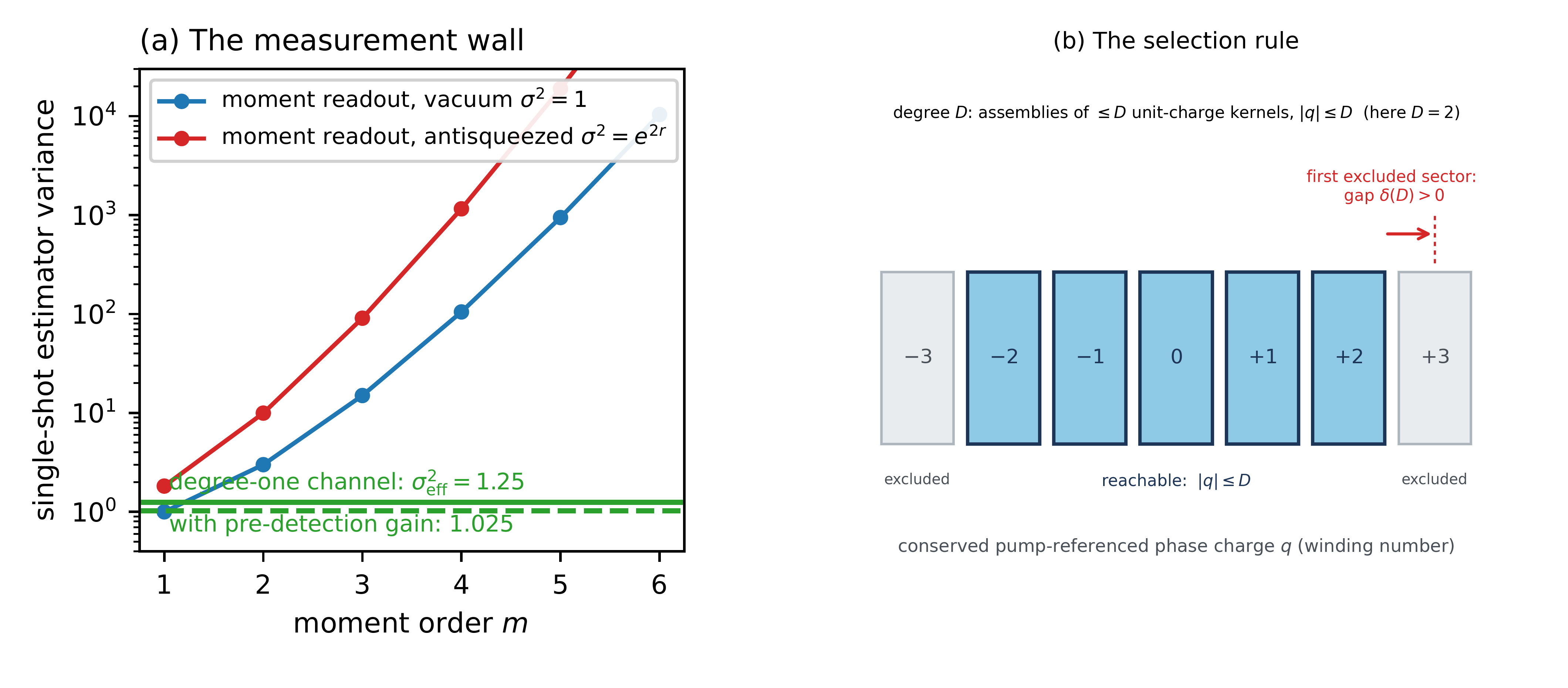}
\caption{\textbf{One conservation law prices a quantum machine.}
(a)~\emph{The measurement wall.} Single-shot variance of an order-$m$
moment estimator, $(2m{-}1)!!\,\sigma^{2m}$ [Eq.~\eqref{eq:factorial}],
for vacuum ($\sigma^2{=}1$) and antisqueezed ($\sigma^2{=}e^{2r}$,
$r{=}0.3$) readout, against the machine's order-one channel, whose
estimator variance is pinned at order one ($\sigma_{\rm eff}^2=1.25$
at the campaign point, $1.025$ with pre-detection phase-sensitive gain
suppressing the detection-loss penalty; Methods). (b)~\emph{The selection rule.} The conserved
pump-referenced phase charge organizes the reachable function space:
order-$D$ polynomial post-processing reaches exactly the $D$-block
assemblies of
unit-charge kernels, confined to the sectors $|q|\le D$
(Theorem~\ref{thm:main}), with the first excluded sector held at the
explicit gap $\delta(D)>0$ [Eq.~\eqref{eq:gap}]. }
\label{fig:concept}
\end{figure*}

This paper's central result is that for pump-phase-encoded parametric
reservoirs the affordability question has an exact answer, supplied by
one conservation law. A squeezer exchanges photons with the pump only in pairs, so every
circulating quantity owes an integer number of pump quanta---its
\emph{charge}---which loss, delays, and passive optics conserve. We prove (Sec.~\ref{sec:reachability})
that the functionals reachable at polynomial order $D$ are exactly the
assemblies of at most $D$ unit-charge
kernels, confined to the charge sectors $|q|\le D$; universality, a gapped no-go for linear readout, falsifying kernels,
and a polynomial cost bound follow (Fig.~\ref{fig:concept}).
This sharpens the Fourier-spectrum program of quantum machine
learning~\cite{Schuld2021,PerezSalinas2020,Casas2023,Meyer2023}: here
the spectrum is organized by one conserved charge of the optics. The characterization is two-sided and
priced: an exact reach with an explicit inapproximability gap [Eq.~\eqref{eq:gap}], a
per-character measurement cost [Eq.~\eqref{eq:budget}], and a
classical-light machine proven to compute the same functions. Tested on a hardware-faithful twin against a task-tuned classical digital reservoir at matched readout dimension and measurement noise, the machine leads on every nonlinear benchmark under budget (Table~\ref{tab:results}, Figs.~\ref{fig:t2}--\ref{fig:t3}); the classical-light theorem is what lets us say the lead is not an expressivity effect.

In a nutshell: encoding in the squeeze angle makes histories enter as
non-commuting products, so the state carries
\emph{characters}---products of input oscillations---at every depth and
degree (Sec.~\ref{sec:reachability}). The $\chitwo$ transducer converts the covariance holding them into a
first moment of the harmonic field, so the detector only ever measures
order-one statistics---yet each linear reading contains polynomials of
the encoded input at every degree, formed by the dynamics before
detection. Combining features extends the reach across charge sectors,
one unit per feature, up to $|q|\le D$ (Theorem~\ref{thm:main});
deconvolving the known one-pole filter and multiplying isolates each
character, dense by Stone--Weierstrass---universality at polynomial
shot cost, the excluded sectors held off by an explicit gap.

The design principle is \emph{put the polynomial where it is cheap}:
nonlinearity lives in the dynamics, compounded by recursion for free,
while the measurement stays at order one, where estimator variance is
bounded (Methods). We price the asymmetry with the
information processing capacity~\cite{Dambre2012} at fixed shot budget
(Methods), every claim tested against matched baselines; the machine is
an illustrative room-temperature chip design---no device was built---at
a conservative operating point that lower-bounds the hardware.

\section{Results}

\subsection{The minimal machine}
\label{sec:machine}
\label{sec:architecture}
\label{sec:model}

\begin{figure*}[t]
\centering
\includegraphics[width=\textwidth]{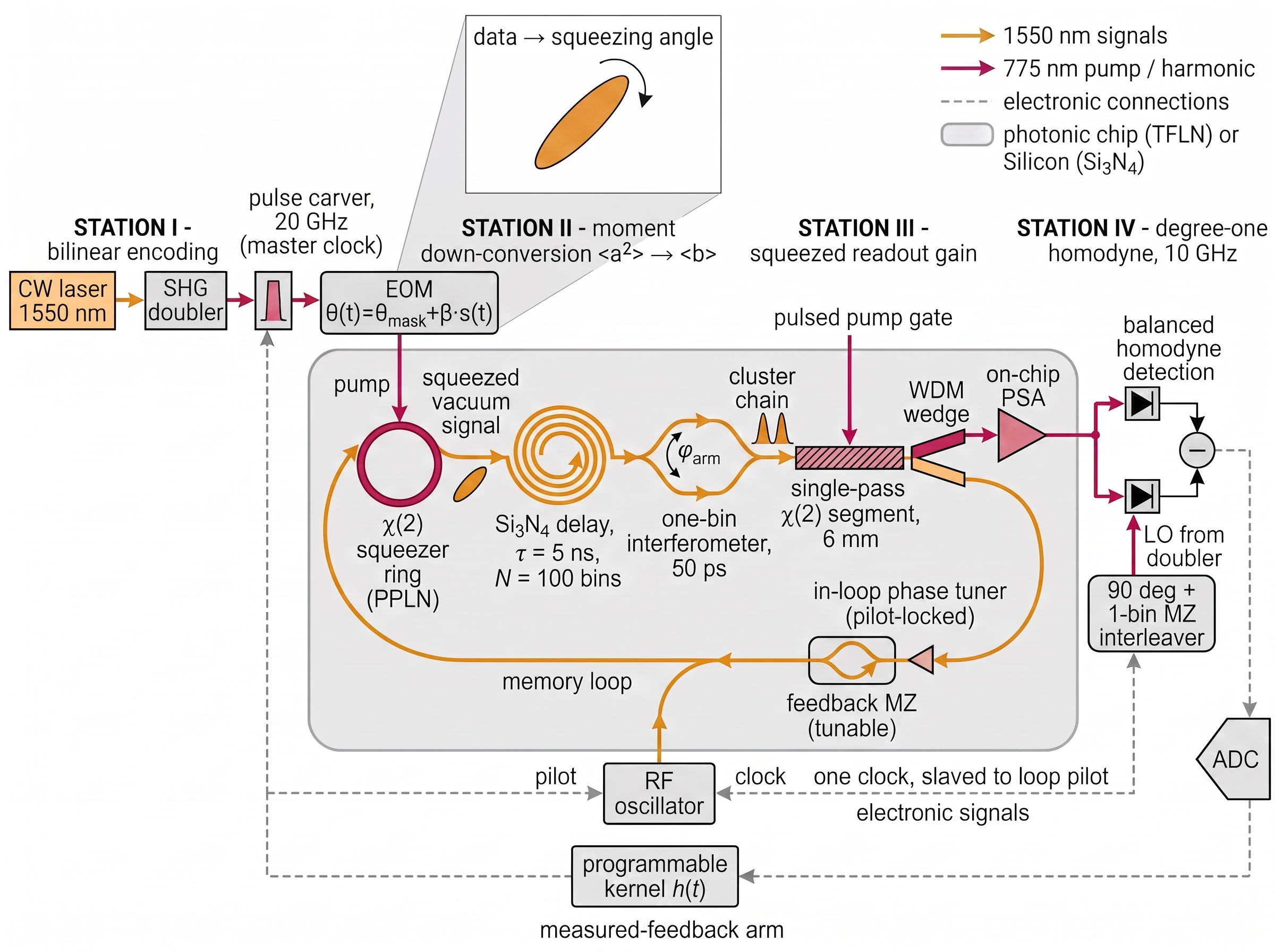}
\caption{\textbf{The measurement-economical squeezed-light reservoir.}
Signal flow runs left to right. False color throughout: amber is the
1550-nm fundamental band, crimson the 775-nm harmonic band, dashed gray
the electronics.
\emph{Encoding (station~I).} A continuous-wave 1550-nm laser is frequency
doubled. The 775-nm pump is carved into a 20-GHz pulse train that serves
as the master clock, and an electro-optic modulator writes the data into
the pump phase, and hence into the squeezing angle,
$\theta(t)=\theta_{\rm mask}+\beta s(t)$.
\emph{Squeezing.} On a hybrid TFLN/Si$_3$N$_4$ chip the pulsed pump
drives a periodically poled squeezer ring ($\chitwo$ OPO). It generates
squeezed vacuum at 1550~nm whose ellipse orientation carries the encoded
history (overlay).
\emph{Memory.} The fundamental circulates through a buried Si$_3$N$_4$
spiral delay ($\tau=5$~ns; spiral loss $\approx0.1$~dB of the
$\approx0.5$~dB round trip), which hosts $N=100$ time-bin nodes per round
trip (glowing packets). A one-bin unbalanced interferometer
($\Delta=50$~ps) chains neighboring bins into an entangled cluster chain.
\emph{Transduction (station~II).} A poled, group-velocity-matched
\emph{single-pass} $\chitwo$ segment, gated by the pulsed pump, converts
$\avg{a^2}$ into the 775-nm mean field. A WDM splitter then peels the
harmonic band into the readout chain, while the fundamental continues
through a pilot-locked in-loop phase tuner to a tunable Mach--Zehnder
coupler that re-injects it \emph{into the squeezer ring}, closing the
memory loop.
\emph{Readout (stations~III and~IV).} An on-chip phase-sensitive
amplifier boosts the extracted field, and balanced homodyne detectors
read it at order one. The 775-nm local oscillator is picked off after
the doubler and passively interleaved between the $X$ and $P$ bases.
\emph{Timing.} All timing derives from one RF oscillator slaved to the
loop's pilot tone. The dashed electronic path implements the slower
measured-feedback arm,
$\varepsilon_2(t)=\varepsilon_0 p(t)\big[1+G_{\rm fb}\!\int\!
h(t')\,Y(t-t')\,dt'\big]$.}
\label{fig:architecture}
\end{figure*}

Figure~\ref{fig:architecture} shows the machine: the polynomial is
computed in the dynamics by a single-pass $\chitwo$ segment in the
delay loop, and the measurement is pinned at polynomial order one
(walkthrough in SI Sec.~S8; Table~\ref{tab:params} separates hardware
targets from the conservative campaign point behind every result).

\emph{Bilinear encoding.} The phase-modulated pump meets the squeezer
at angle
\begin{equation}
\theta_k = \theta_{\mathrm{mask}}(k \bmod N') + \beta\, s_n ,
\label{eq:encoding}
\end{equation}
with $\theta_{\mathrm{mask}}$ a fixed pseudorandom mask of period
$N'=N\pm1$, braiding the $N$ virtual nodes into one
cycle~\cite{Appeltant2011}. With $(\bm m,\bm\Sigma)$ the register mean
and covariance, the per-bin symplectic update is
\begin{equation}
\bm\Sigma \to S_k\,\bm\Sigma\,S_k^{\top} + \mathcal N_k,\qquad
S_k\big|_{\rm sq} = R(\tfrac{\theta_k}{2})\,Z(r)\,R(\tfrac{\theta_k}{2})^{\top},
\label{eq:update}
\end{equation}
with $R$ the rotation, $Z(r)$ the squeeze, and $\mathcal N_k$ the
loss noise (SI Sec.~S1). The data sit \emph{inside} $S_k$, never as a displacement:
displacement \emph{adds}, making a filter, while steering the map
\emph{multiplies} non-commuting operations, so the ordered product
$S_M\cdots S_1$ forms input products of its own
accord~\cite{LloydBraunstein1999}.

\emph{Coherent, cluster-chain memory.} The fundamental circulates
through a delay spiral~\cite{Blumenthal2018} and is re-injected
\emph{into the squeezer ring}---the cluster-state generator of
Refs.~\cite{Yokoyama2013,Asavanant2019} serving as reservoir
memory---and a one-bin interferometer chains neighbors, charge-free.
One dial, the round-trip amplitude $G$, sets echo state and depth
$M\approx(1-G)^{-1}$. A slower electronic \emph{measured-feedback} arm modulates the pump
with a delayed kernel of the homodyne record; comparing them at
matched budget isolates the value of entangled memory
(Sec.~\ref{sec:narma}).

\emph{Coherent moment down-conversion.} Each circulation passes a
poled single-pass TFLN waveguide~\cite{Wang2018,Lu2019,Nehra2022};
with $b$ the harmonic mode and $\mu\ll1$ the conversion amplitude, the
perturbative action per pass is
\begin{equation}
\Delta\avg{b} = -i\mu\,\avg{a^2}
= -i\mu\big[(\Sigma_{XX}-\Sigma_{PP}) + 2i\,\Sigma_{XP}\big] :
\label{eq:downconversion}
\end{equation}
a \emph{linear read-off of three covariance entries}, doing optically
in one pass what moment readout does statistically across shots
(Methods; SI Sec.~S1). The entries are
nonlocal---the field is a superposition of epochs---so
\begin{equation}
\avg{a^2(t)} = \sum_{m,n} c_m c_n\, \avg{\tilde a(t-m\tau)\,\tilde a(t-n\tau)} .
\label{eq:kernel}
\end{equation}
The cross terms are the machine's delayed nonlinear multiplications;
past the bound of Refs.~\cite{Duan2000,Simon2000}, they are
entanglement correlators. Delay assembles, $\chitwo$ multiplies,
recursion compounds: the quantum limit of the Ikeda
system~\cite{Ikeda1979}.

\emph{Degree-one readout.} A phase-sensitive amplifier applies
noiseless gain \emph{before} detection loss---suppressing that penalty
to $(1-\eta_{\rm det})/(\eta_{\rm det}G)$---and balanced homodynes read
at order one. Parity keeps $\avg{a}=0$, so this is the \emph{only} first-moment
port: a first moment of $b$ whose value is a second moment of $a$, at
variance bounded by the arriving vacuum floor (Methods).

Every claim below is tested on a hardware-faithful digital twin---an
\emph{exact} Markovian embedding on $N$ bin
modes~\cite{Grimsmo2015,PichlerZoller2016} with the $\chitwo$
back-action tracked (Methods; SI Sec.~S1). Gaussian dynamics is
efficiently simulable~\cite{MariEisert2012}: entanglement here cannot
change the complexity class, only the economics, which $C(B)$
measures.

\subsection{Exact reachability: the charge selection rule}
\label{sec:reachability}

Every proof is bookkeeping around one idea: $\avg{a^2}$ destroys two
signal photons---the yield of one pump photon---so it owes one pump
quantum; that integer debt is its \emph{charge}. Only the oriented
squeezers change a debt; the charge is conserved, loss and all, and
this law generates the function space as a selection rule organizes a
spectrum.

We prove the structural result on the \emph{reduced single-loop
model}: one circulating mode, an oriented squeeze $r>0$ per symbol at
angle $\theta_t=\varphi_t+\beta s_t$, then transmission $\eta$ with
vacuum injection. Nothing is lost at this $N{=}1$ caricature:
Proposition~\ref{prop:register} transports every structural clause to
the register.

Two variables carry it: the read feature $m=\avg{a^2}$ (absolute
quadrature units throughout the theory; the twin's harvest is the same object
rescaled, Methods)
[Eq.~\eqref{eq:downconversion}] and the energy channel
$J=\avg{a^\dagger a}+\tfrac12$: a linear recurrence whose
\emph{weights}, not inputs, carry the data. Inputs are one-sided
sequences on the compact $\mathcal U$; closures are uniform on
$C(\mathcal U)$. Standing assumptions: stability guard
$\bar\rho:=\eta e^{2r}<1$, encoding injectivity
$0<\beta s_{\max}<\pi$; complex $m_t$ is recorded at every slot.

The natural basis is Fourier: for finitely supported integer
patterns $n$, write
\begin{equation*}
E_n(s)=\prod_te^{i\beta n_ts_t}
\end{equation*}
for the input \emph{characters}, with \emph{charge} $q(n)=\sum_tn_t$
(the net winding number), \emph{weight} $\|n\|_1$, and $\mathcal V_q$
the closed span at charge $q$. The reachable class $\mathcal A_D$ is
the closure of all order-$\le D$ polynomials in the read features over
finite readout sets and a finite ensemble of \emph{settings}: mask phases $\varphi$ and guard-compatible squeeze strengths $r$, at fixed $(\eta,\beta)$---standard input-mask multiplexing. The three clauses quantify over different families, and the
proofs fix which. Clause (I) is an equality for the class defined over a designed mask ray of $5^L$ epochs and guard-compatible squeeze settings, $L$ the number of lags in the resolved window (SI Sec.~S3): at any fixed setting the upper inclusion is exact and each assembleable character is reached to within the residual bound of Theorem~S1, which vanishes only along $r\to0$; whether one frozen mask suffices is an open hypothesis. Clause~(II) is a statement about the
\emph{family} $\{\mathcal A_D(r)\}_{r>0}$ as $r\to0$, not about
$\mathcal A_D$ at any one $r$: at fixed hardware the residual is bounded but
nonzero, and the bound is quoted with the clause. Clause~(III) holds at every
fixed setting individually, hardware included.

\subsubsection{The structure lemma}

The first step is to put the exact
dynamics in a form that exposes the charge. The exact covariance
dynamics of the reduced loop closes on the three channels $v_t=(m_t,\bar m_t,J_t)$:
\begin{widetext}
\begin{equation}
v_t=T(\theta_t)\,v_{t-1}+b_{\rm vac},\qquad b_{\rm vac}=\Big(0,\,0,\,\tfrac{1-\eta}{2}\Big)^{\!\top},
\qquad
T(\theta)=\eta\!
\begin{pmatrix}
\cosh^2\!r & e^{2i\theta}\sinh^2\!r & -e^{i\theta}\sinh2r\\[1pt]
e^{-2i\theta}\sinh^2\!r & \cosh^2\!r & -e^{-i\theta}\sinh2r\\[1pt]
-\tfrac12e^{-i\theta}\sinh2r & -\tfrac12e^{i\theta}\sinh2r & \cosh2r
\end{pmatrix}.
\label{eq:Tmap}
\end{equation}
\end{widetext}

\begin{lemma}[Structure of the one-step map]
\label{lem:structure}
The map \eqref{eq:Tmap} has the following three properties.

\smallskip\noindent\textbf{(a) Gauge grading.} With
$Q=\mathrm{diag}(1,-1,0)$ assigning charges $(+1,-1,0)$ to $(m,\bar m,J)$,
\begin{equation}
T(\theta+\chi)=e^{i\chi Q}\,T(\theta)\,e^{-i\chi Q},\qquad
e^{i\chi Q}b_{\rm vac}=b_{\rm vac},
\label{eq:gauge}
\end{equation}
for all $\chi$---the conservation law. In the Laurent decomposition
$T(\theta)=\sum_{|d|\le2}T_de^{id\theta}$ (SI Sec.~S3\,B),
$[Q,T_d]=d\,T_d$ and $T_d=0$ for $|d|\ge3$; the readable row is
holomorphic in the pump phase, $e_m^\dagger T_0=x\,e_m^\dagger$,
$x:=\eta\cosh^2\!r$.

\smallskip\noindent\textbf{(b) Contraction.} Every row of the
entrywise-absolute map sums to $\bar\rho<1$; the stationary solution is
unique, bounded by $v_\infty=\tfrac{1-\eta}{2(1-\bar\rho)}$, with
fading memory at rate $\bar\rho$---echo state at a hardware-fixed
rate.

\smallskip\noindent\textbf{(c) Dyson grading.} All input dependence
sits in $T_{\rm od}:=T-T_0$ with small parameter
$\sigma:=\eta(\sinh2r+\sinh^2\!r)$; the Dyson expansion grades the
stationary state by the number $k$ of input engagements,
$\|v^{(k)}\|=O(\sigma^k)$, and the grade-one readable content is the
one-pole filter
\begin{equation}
m^{(1)}_t=\sum_{j\le t}\varkappa\,x^{t-j}e^{i\theta_j},\qquad
\varkappa=-\eta\sinh(2r)\,J_{\rm ss},
\label{eq:onepole}
\end{equation}
with $J_{\rm ss}=\tfrac{1-\eta}{2(1-\eta\cosh2r)}$; the grade-two
readable content is the single admissible family
\begin{equation}
m^{(2)}_t=\eta\sinh^2\!r\;\varkappa\sum_{i<j\le t}x^{t-j}\,x^{j-1-i}\,
e^{i(2\theta_j-\theta_i)} .
\label{eq:gradetwo}
\end{equation}
\end{lemma}

\noindent\emph{All proofs are in SI Sec.~S3, self-contained, with a
reader's guide and glossary, assuming no quantum-optics background.}

\begin{corollary}[The three kernel laws]
\label{cor:laws}
Every character in the expansion of $m_t$ obeys:
\emph{(selection)} $q(n)=+1$;
\emph{(ordering)} the latest engaged slot has $n_{t^*}\in\{+1,+2\}$, never
conjugated;
\emph{(per-slot)} $|n_t|\le2$.
\end{corollary}

\begin{figure*}[!t]
\centering
\includegraphics[width=\textwidth]{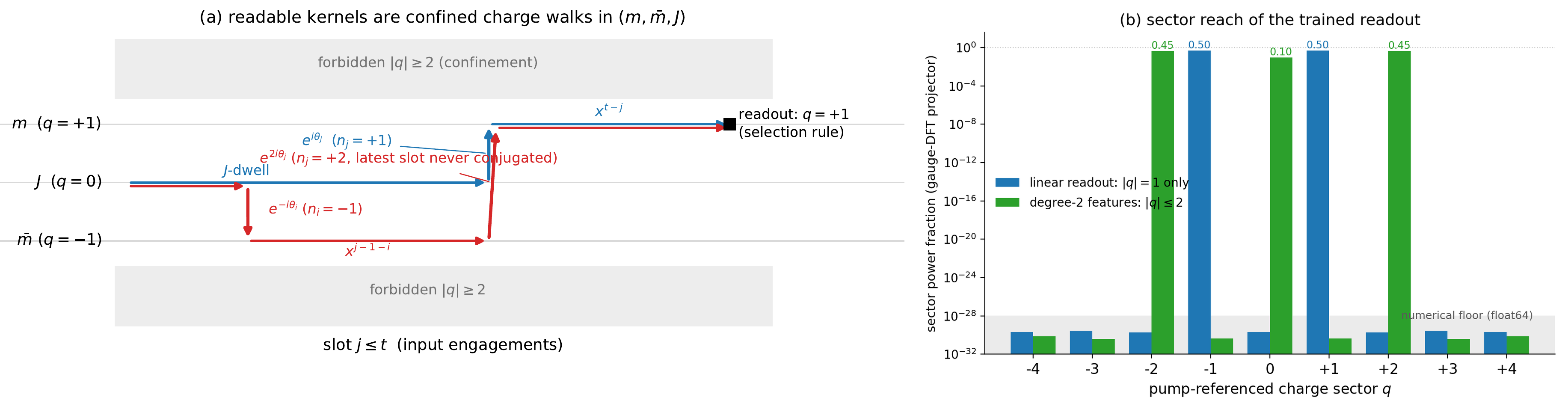}
\caption{\textbf{The charge-confinement law, pictorially.} (a) Every readable
kernel monomial of the exact dynamics is a walk of pump-referenced phase
charge on the three channels $(m,\bar m,J)$ of Eq.~\eqref{eq:Tmap}, confined
to $|q|\le1$ (shaded regions are unreachable in flight). Blue: the grade-one
walk of Eq.~\eqref{eq:onepole} --- a $J$-dwell, a single engagement
$e^{i\theta_j}$, and geometric decay $x^{t-j}$ to the readout. Red: the
grade-two family of Eq.~\eqref{eq:gradetwo} --- an early conjugated
engagement $e^{-i\theta_i}$ into $\bar m$, a dwell, and the double-charged
step $e^{2i\theta_j}$ at the latest engaged slot, which the ordering law
forbids to be conjugated. Every walk terminates in $m$ with net charge
$q=+1$: the selection rule of Corollary~\ref{cor:laws}. (b) The same algebra
measured: sector power of the trained readout from the gauge-DFT
projector at $K{=}12$ mask shifts (alias-free through $|q|\le5$).
Linear features load only $|q|=1$; order-2 features add exactly
$\{0,\pm2\}$; the sectors $|q|\ge3$ are empty to numerical precision
(shaded band)---the falsifier channels of Corollary~\ref{cor:falsify},
silent as the law requires.}
\label{fig:chargewalks}
\end{figure*}

\noindent Figure~\ref{fig:chargewalks} renders the three laws as
charge walks on the channels of Eq.~\eqref{eq:Tmap}.
\subsubsection{Main theorem}

Call a pattern $n$ \emph{admissible} if it obeys the three kernel laws
of Corollary~\ref{cor:laws}, \emph{co-admissible} if $-n$ is, and
\emph{$D$-assembleable} ($n\in\mathfrak N_D$) if it is a sum of at most
$D$ such patterns. Assembly confines---$|q(n)|\le D$, per-slot
$|n_t|\le2D$---but the converse fails (SI, \emph{What polynomial order
does not buy}).

\begin{theorem}[Charge-sector reachability]
\label{thm:main}
Under the standing assumptions, for all $D\ge0$, with each clause quantified
over the family named in it:
\begin{itemize}
\item[\textbf{(I)}] \emph{(Exact reach: block decomposition, over a designed mask ensemble and guard-compatible squeeze settings; at any fixed setting the inclusion $\subseteq$ is exact and the reverse holds to within the residual of Theorem~S1.)}
$\displaystyle
\mathcal A_D=\overline{\operatorname{span}}\{E_n:\ n\in\mathfrak N_D\}$,
the closed span of the $D$-assembleable characters; consequently
$\mathcal A_D\subseteq\overline{\operatorname{span}}
\bigcup_{|q|\le D}\mathcal V_q$ ($\mathcal V_q$ the closed span of
charge-$q$ characters), the sector containment strict
(SI, \emph{What polynomial order does not buy}).
\item[\textbf{(II)}] \emph{(Universality of the family
$\{\mathcal A_D(r)\}_{r>0}$ as $r\to0$; not a statement at any fixed $r$.)}
$\bigcup_D\mathcal A_D$ is dense in $C(\mathcal U)$: every
fading-memory target is realized to any accuracy by some finite polynomial order, settings, and
readout, the error controlled along
guard-compatible families with $r\to0$; at any fixed point the
reconstruction error is bounded by the analytic FIR residual
($10.8\%$ at the campaign point; Corollary~\ref{cor:exchange}).
\item[\textbf{(III)}] \emph{(No-go with explicit gap, at every fixed
setting including hardware.)} For every pattern $n$
with $q(n)=D{+}1$---so that $E_n$ is an eigenfunction of the
uniform-shift action, $E_n(s+\tau\mathbf 1)=e^{iq(n)\chi}E_n(s)$
with $\chi=\beta\tau$, oscillating at frequency $D{+}1$ outside
the band $|q|\le D$ that order-$D$ readout can reach---
\begin{align}
\operatorname{dist}_\infty\!\big(E_n,\mathcal A_D\big)&\;\ge\;
\delta(D,\beta s_{\max})>0,\nonumber\\
\delta(D,a)&:=\prod_{j=1}^{2D+1}\sin\!\Big(\frac{j\,a}{4(2D+1)}\Big).
\label{eq:gap}
\end{align}
\end{itemize}
\end{theorem}

\begin{figure}[!t]
\centering
\includegraphics[width=0.95\columnwidth]{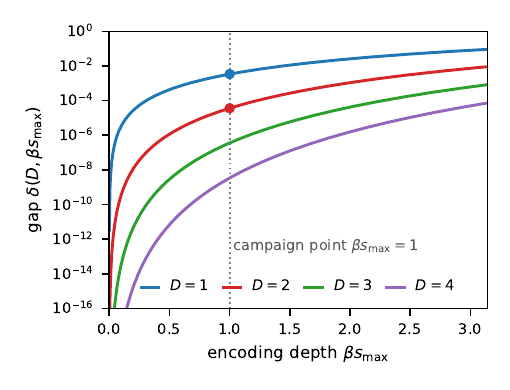}
\caption{\textbf{The no-go gap, quantitatively.} The explicit lower bound
$\delta(D,\beta s_{\max})$ of Theorem~\ref{thm:main}(III) [Eq.~\eqref{eq:gap}]
across polynomial order $D$ and encoding depth. The gap is strictly
positive throughout the injectivity window $0<\beta s_{\max}<\pi$ --- no
finite polynomial order closes the next charge sector --- but it decays steeply with
$D$, quantifying why polynomial post-processing (Theorem~\ref{thm:main}(II))
is cheap in practice: at the campaign operating point $\beta s_{\max}=1$,
$\delta(1,1)\approx3.4\times10^{-3}$ and
$\delta(2,1)\approx3.7\times10^{-5}$---each added polynomial order buys
roughly two
orders of magnitude of the remaining distance. The no-go is exact but
quantitatively modest at these depths; its role is structural, not
practical.}
\label{fig:gapsweep}
\end{figure}

In words: order-$D$ buys exactly the $D$-kernel assemblies, all
orders together exhaust every fading-memory functional, and the first
excluded sector sits at a computable distance (Fig.~\ref{fig:gapsweep}).

The proof architecture and route map are given in SI Sec.~S3.

\subsubsection{Register, falsifiers, and resource corollaries}

\begin{proposition}[Register neutrality]
\label{prop:register}
In the full register every added element---passive mixing, per-bin
loss, ancilla reset, static phases---commutes with the gauge action
\eqref{eq:gauge}; only the oriented squeezers carry pump charge. The
selection rule, ordering law, upper inclusion of (I), and clause (III)
hold verbatim (SI Sec.~S3).
\end{proposition}

\begin{proposition}[Classical-light equivalence]
\label{prop:classical-equivalence}
Let the classical-light twin be the identical architecture and
encoding with the squeeze map followed by the reclassicalizing
channel
$\bm\Sigma \to \bm\Sigma + \bm{\mathcal N}_{\rm cl}(\theta_k)$, where
\begin{equation*}
\bm{\mathcal N}_{\rm cl}(\theta)=\eta_{\rm esc}(1-e^{-2r})\,
R(\theta/2)\,\mathrm{diag}(1,0)\,R(\theta/2)^{\top}
\end{equation*}
restores the minor axis to vacuum---the $P$-representability
boundary (Methods). Then:

\emph{(i)--(ii)} The reclassicalizing source lies \emph{inside} the
gauge-graded algebra, so every readable twin monomial obeys the three
kernel laws and all twin features lie in $\mathcal A_D$: a proven
containment (SI Sec.~S3).

\emph{(iii)} Span \emph{equality} of the noiseless order-$D$ readouts
holds for every $D$ for almost every mask (Lemma~S2, SI Sec.~S3), the
operating masks assumed generic.

No noiseless separation between machine and twin exists at any
polynomial order.
\end{proposition}

\begin{corollary}[Detection attribution]
\label{cor:detection-attribution}
Under span equality the noiseless feature families carry identical
information, related by a linear map; every finite-budget difference is
that map's interaction with the shared detection noise. Conditioning,
not span, carries the margins.
\end{corollary}

\begin{corollary}[Three forbidden families]
\label{cor:falsify}
Measured harmonic-band cross-correlation kernels vanish identically
for every pattern violating the selection rule, the ordering law, or
the per-slot bound; a violation beyond its measurement uncertainty refutes the operating
model, not a benchmark.
\end{corollary}

\begin{corollary}[Echo state at hardware rate]
\label{cor:esp}
Under the guard, the stationary functional is unique with
input-uniform fading memory at rate $\bar\rho=\eta e^{2r}$, fixed by
hardware, independent of register size and input statistics.
\end{corollary}

\begin{corollary}[Affordable universality]
\label{cor:exchange}
For every character of weight $k$ and every $\epsilon\in(0,1)$ there
exist guard-compatible settings and an order-$k$ polynomial estimator,
built from the FIR-deconvolved clause-(II) features, with RMS error
$\le\epsilon$ uniformly, at per-feature shot budget
\begin{equation}
B \;\le\; c\,\Big(\frac{\sigma_{\rm eff}}{\mu_T}\Big)^{\!2}
k^{3}\,(1+v_\infty)^{3(k-1)}\;\epsilon^{-3}.
\label{eq:budget}
\end{equation}
with $c$ absolute, $\mu_T$ the transduction gain, $\sigma_{\rm eff}$
the per-shot readout deviation.
\end{corollary}

Every fading-memory functional is thus realized at shot cost
polynomial in $1/\epsilon$, order three, against the super-exponential
$(2m-1)!!$ of moment readout; constants at the audited operating point,
and the caveat that the campaign sits outside the small-$\sigma$ regime
of the proof, are in Methods and SI Sec.~S3.

\subsection{Property characterization: fading memory and convergence}
\begin{figure*}[!t]
\centering
\includegraphics[width=0.95\textwidth]{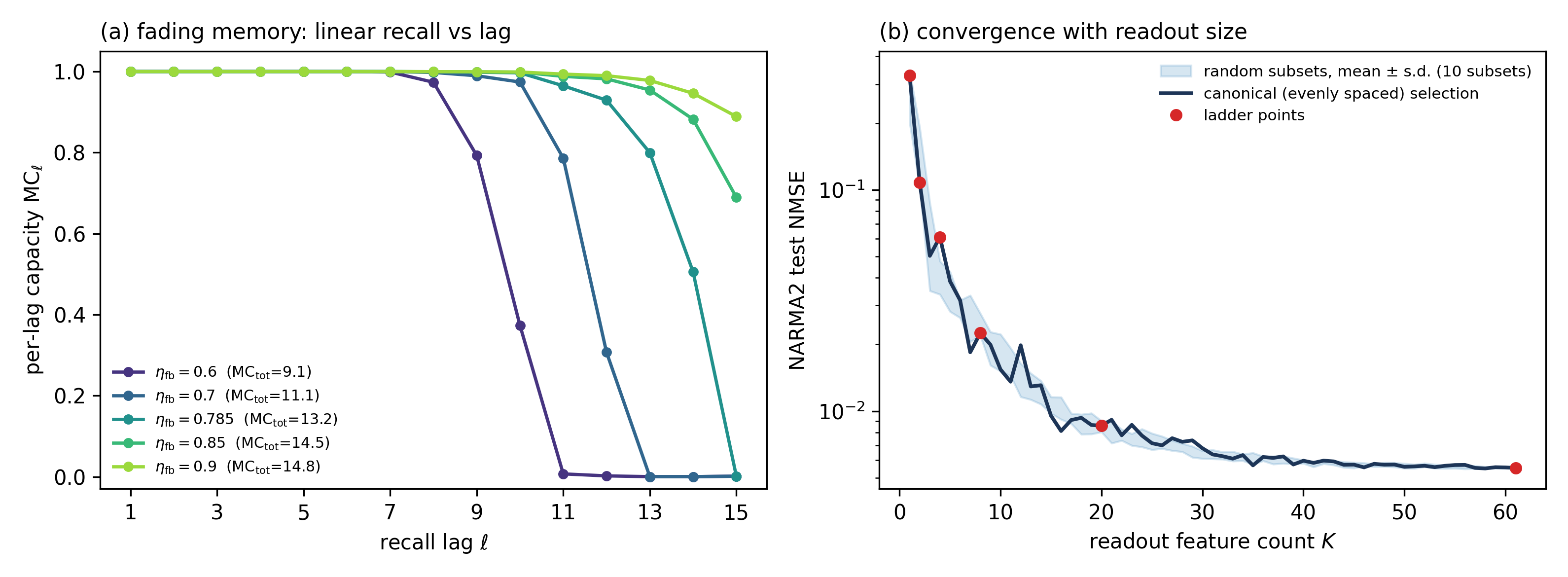}
\caption{\textbf{Property characterization.} (a)~Per-lag linear memory
capacity $\mathrm{MC}_\ell$ across the feedback ladder
$\eta_{\rm fb}=0.6$--$0.9$ ($\eta_{\rm fb}$ is the Mach--Zehnder \emph{power}
transmissivity retained by the head bin, SI Sec.~S2; loop loss is the twin of record's distributed form,
$\eta_\ell=\eta_L^{1/N}$ per bin step, and the dependence of capacity on
the loss-compounding convention is discussed in Subsec.~S1\,D).
Total capacity and the recall shoulder grow with feedback transmission.
(b)~NARMA2 test NMSE versus readout feature count $K$ (noiseless
protocol, single mask, 61 bins; canonical selection with random-subset
mean$\,\pm\,$s.d.). The plateau height at this session length is not a
floor: it contains a finite-sample component that recedes with training
length ($0.0055\!\to\!0.0030$ from $T_{\rm tr}=1500$ to $6000$).}
\label{fig:t1}
\end{figure*}

\label{sec:memory}

This subsection measures a property of the machine, not a competition:
the reachability theorem presumes convergence to a unique trajectory
with fading memory; both are verified directly [Fig.~\ref{fig:t1}(a)]:
per-lag recall is near-exact at short lags, collapses by
$\ell\gtrsim12$, and the settle test converges at every admissible
point. Linear recall is deliberately not
benchmarked against a classical reservoir---linear memory is the
currency this architecture allocates away from
(Sec.~\ref{sec:mechanism}). Two sensitivities are stated rather than
smoothed: total capacity depends on the loop-loss compounding
convention (qualitative properties hold under both), and on the mask at
the several-tens-of-percent level, so values are mask ensembles.

The saturation with readout size [Fig.~\ref{fig:t1}(b)] is feature
redundancy, not noise: the effective dimension is of order two against
$122$ components, so added bins re-express the span, while the plateau
height retains a finite-sample component that recedes with training
data (SI, \emph{Convergence with readout size}).

\subsection{Synthetic benchmark under matched resources}
\begin{figure*}[!t]
\centering
\includegraphics[width=0.95\textwidth]{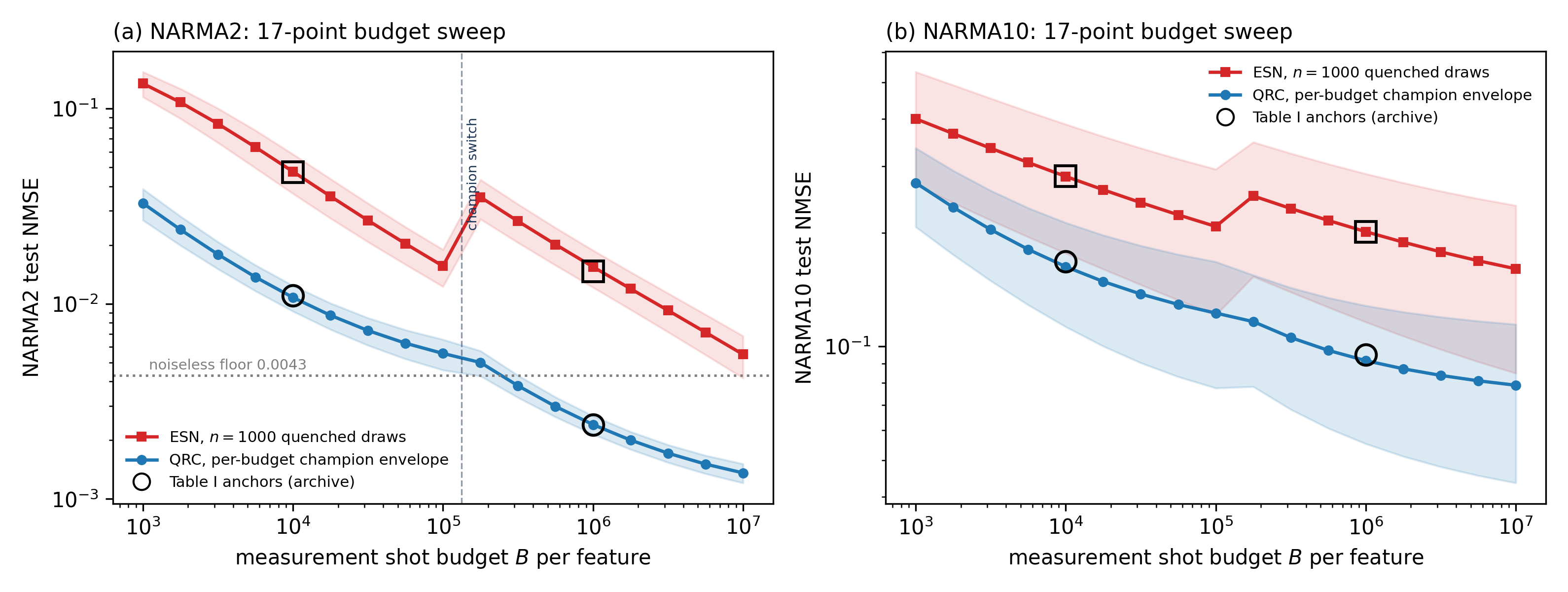}
\caption{\textbf{Measurement economics across four budget decades.}
Test NMSE versus per-feature shot budget $B$ (17 points per task) for
(a)~NARMA2 and (b)~NARMA10. Machine: per-budget frozen-champion
envelope (held-out seeds $\{12,13\}\times$ mask seeds $\{100..104\}$,
five shot realizations; intermediate budgets evaluate the frozen
champions without re-optimization). Baseline: task-tuned,
dimension-matched digital reservoir ensemble, $n{=}1000$ distinct
quenched draws (disjoint reservoir seed ranges), three shot
realizations per draw, hyperparameters selected noiselessly on a
single tuning draw and frozen per task, per-feature signal-to-noise
matched to the machine's selection-seed linear feature power. Bands
are quenched ensemble s.d.; open markers: the corresponding
Table~\ref{tab:results} values (agreement $\le5\%$). The
baseline discontinuity at the champion switch is the matched-noise
reference following the machine's operating point, drawn rather than
smoothed. On NARMA10 the ensemble means are separated by tens of
standard errors while $11$--$14\%$ of individual baseline draws beat
the machine; both statistics are reported.}
\label{fig:t2}
\end{figure*}

\label{sec:narma}

Benchmarks price the machine against a named classical baseline under a
stated matching. NARMA2,
is the validation instrument, inside the proven reach by design;
NARMA10 adds depth. The baseline is a task-tuned echo-state network at matched readout
dimension and matched per-feature measurement noise, an ensemble of
$1000$ quenched realizations against the machine's mask ensemble
(protocol in Methods and the Fig.~\ref{fig:t2} caption); both tuning
budgets are counted, the machine's exceeding the baseline's---an
asymmetry that favors the machine (SI Sec.~S6).

\emph{Parity at infinite budget, exactly as the theory demands}: at
the optimized operating points machine and classical-light twin agree
noiselessly to within $0.4\%$ (ratios $0.996$,
$1.001$)---Proposition~\ref{prop:classical-equivalence} made manifest, a registered prediction confirmed. What follows prices the machine against the classical digital baseline.

\emph{Priced at finite budget, the machine leads the digital baseline on both nonlinear tasks} (Table~\ref{tab:results}). The ordering is protocol-invariant but the
magnitude is not: on NARMA2 the machine leads the baseline by $4.3\times$ and $6.1\times$
when each machine is given its own per-task optimum, and by $1.8\times$ and
$5.4\times$ under the symmetric tuning protocol that hands both machines an
equal search budget. Both are tabulated; the equal-search figures are the protocol-matched
comparison and the conservative reading. The machine is the more
search-sensitive of the two: moving from its own optimum to the equal-search
grid costs it a factor of nine in NMSE at $B=10^{4}$ against the baseline's
factor of four, so the per-task-optimum row should be read as what a tuned
machine achieves, not as what a fixed one delivers. At the
tabulated budgets, on NARMA10 by $1.7\times$ and $2.1\times$. The NARMA10 statement is
distributional rather than uniform: $11$--$14\%$ of individual baseline draws
beat the machine, so the result is a shift in the mean of a broad comparator
ensemble, not a machine that wins every draw. Across four budget
decades the ordering is uniform, the machine ahead of the digital baseline by
$2.6$--$6.1\times$ [Fig.~\ref{fig:t2}].

\emph{At finite budget the sign of the machine--twin difference depends on task, operating point and search protocol}, so under pre-registered rule DR-7 every margin reported here is over the digital baseline and not over classical light; the twin comparisons are reported in full in SI Sec.~S7.

\emph{The boundaries land where a crossover law puts them} (SI
Sec.~S5): dynamical placement wins over moment readout only when the factorial penalty
exceeds the transduction referral, $(2m-1)!!>\mu_T^{-2}$, i.e.\ above
\begin{equation}
m^\ast(\mu_T)=\min\{m:(2m-1)!!>\mu_T^{-2}\},
\label{eq:crossover}
\end{equation}
which is $m^\ast=6$ at the campaign $\mu_T=10^{-2}$---above every
moment order a lossy Gaussian register computes at all. The transduction
amplitude is a \emph{design target} for the $6$-mm poled segment of
Table~\ref{tab:params}, not a measured device value, and the crossover is
sensitive to it in a way worth stating: $m^\ast=6$ holds on
$9.8\times10^{-3}<\mu_T\le3.3\times10^{-2}$, becoming $5$ above that window
and $7$ below it. The campaign value $\mu_T=10^{-2}$ sits only $2\%$ above the
lower edge, so a device delivering slightly less conversion than the target
moves the crossover to $m^\ast=7$. The crossover order is therefore a
target-dependent statement, and we flag it as the parameter to which
Eq.~\eqref{eq:crossover} is least robust. The referred-to-physical budget conversion
$B_{\rm phys}=B/\mu_T^{2}$ and the energy accounting scale as $\mu_T^{-2}$
and are quoted at the target value. In the
encoding-limited regime, moment orders $2$--$3$ cube affordably while
order $4$ fails at every budget even as the bilinear machine retains
genuine input-order-$4$ content: pre-detection products are an
\emph{input-order multiplier}. This prediction was staked before the
runs and its high-order component was falsified; it is reported as a
mechanism finding (SI Sec.~S5).

\emph{The economics claim has a boundary, stated here rather than
found later}: when one scalar is spread redundantly over many bins with
a known mask, a vacuum-limited measure-then-compute front end can
demodulate and average, gaining $\sqrt{\text{bins per symbol}}$ and
winning outright on the NARMA encoding. What it cannot do is exploit
information carried at high rate per bin, where the RF corpus lives.
Measurement economy is a claim about physically embodied, high-rate
inputs; the synthetic tasks above stand as validation instruments, not
economic evidence, and the real-data tests of
Sec.~\ref{sec:results} carry the structural claims.

\subsection{Why the pattern has this shape}
\label{sec:mechanism}

The pattern of the margins over the digital baseline is the theory's own prediction:
capacity is \emph{allocated} to nonlinear structure by placing
nonlinearity in the dynamics and holding the readout at degree
one, and the allocation is paid on linear recall. The measurements agree where they exist: the margin over the digital baseline is largest where the task is most nonlinear relative to its memory demand (NARMA2) and smaller where linear recall dominates (NARMA10) (Table~\ref{tab:results}); on pure linear memory the theory predicts the inversion, where a contracting $\tanh$ recurrence is the better instrument, and that benchmark is deliberately not run here (Sec.~\ref{sec:memory}). A theory that predicted only successes
would be weaker evidence than one that predicts, as this one does, the
single benchmark on which the architecture should lose.

\begin{table}[!htb]
\caption{Benchmark performance at per-task, per-budget optimized
operating points against the matched classical baseline. Machine
entries are mask-ensemble means over ten mask draws with quenched s.d.\
in parentheses (shot-noise dispersion reported separately in SI
Sec.~S5); baseline entries are ensemble means over $1000$ fresh
reservoir realizations with quenched s.d.\ in parentheses. Baseline
matching: $305$ features against the machine's $305$, identical ridge
and splits, per-feature measurement noise matched to the machine's shot
level; hyperparameters selected once per task and budget on a
validation block, frozen before the held-out seeds. All entries are
held-out. The \emph{equal-search} rows apply the symmetric tuning protocol
of Methods---machine on a $24$-point guard-respecting hardware grid, baseline
re-selected per reservoir draw over five draws---at matched per-feature SNR;
they are the protocol-matched comparison, the rows above them the per-task
optimum of each machine. Margins are baseline over machine, so values below unity favour the baseline:
noiselessly the equal-search ESN is marginally ahead ($0.9\times$), which is
the expected behaviour of a machine whose advantage is measurement economics
rather than expressivity. Noiseless machine--twin parity is reported
separately in SI Sec.~S7 and is not a column of this table. NMSE, lower is
better.}
\label{tab:results}
\begin{ruledtabular}
\footnotesize\setlength{\tabcolsep}{2pt}
\begin{tabular}{lccc}
 & noiseless & $B=10^4$ & $B=10^6$\\
\colrule
\multicolumn{4}{l}{\emph{NARMA2}}\\
QRC, optimized & 0.0043(1) & \textbf{0.0110(13)} & \textbf{0.0024(2)}\\
ESN ensemble, $n=1000$ & --- & 0.0473(128) & 0.0147(33)\\
margin & --- & $4.3\times$ & $6.1\times$\\
QRC, equal-search & 0.0013 & 0.0987(58) & 0.0076(7)\\
ESN, equal-search & 0.0012(3) & 0.1820(386) & 0.0412(443)\\
margin, equal-search & $0.9\times$ & $1.8\times$ & $5.4\times$\\
\colrule
\multicolumn{4}{l}{\emph{NARMA10}}\\
QRC, optimized & --- & \textbf{0.168(13)} & \textbf{0.095(6)}\\
ESN ensemble, $n=1000$ & --- & 0.282(105) & 0.201(85)\\
margin & --- & $1.7\times$ & $2.1\times$\\
\end{tabular}
\end{ruledtabular}
\end{table}

\subsection{Real-data structure tests}
\begin{figure*}[!t]
\centering
\includegraphics[width=0.95\textwidth]{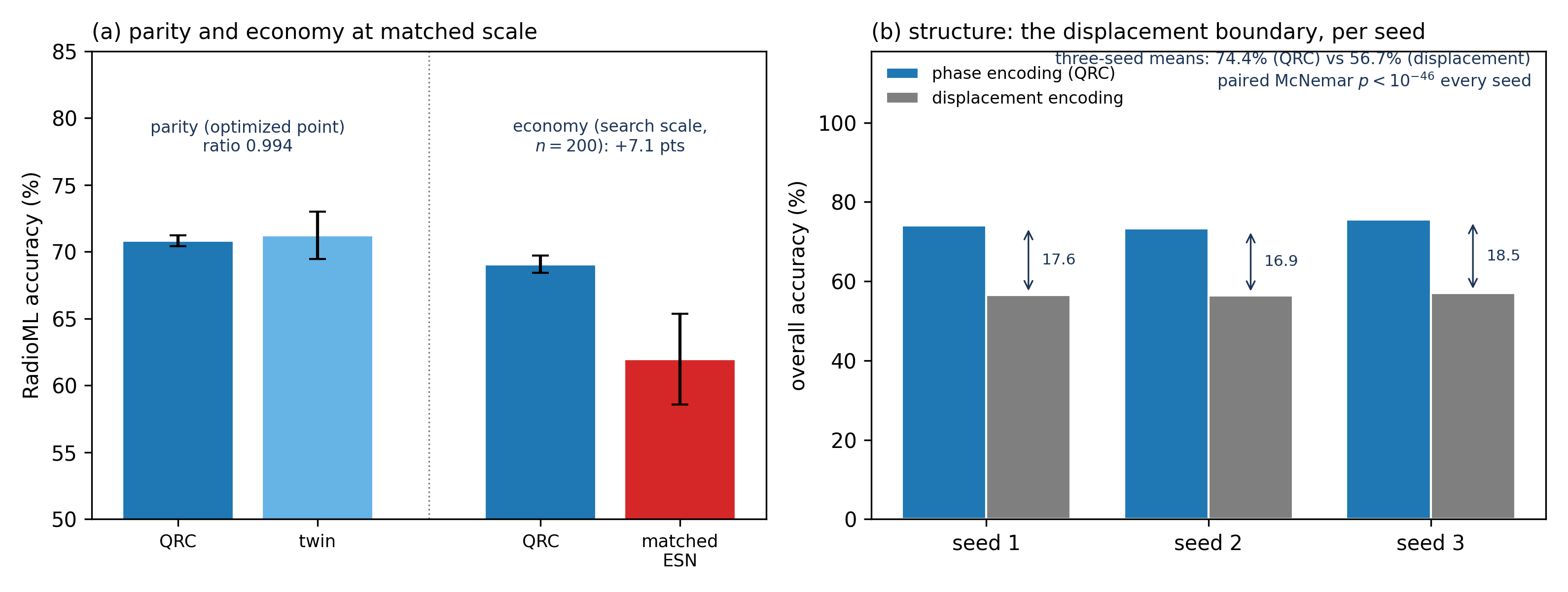}
\caption{\textbf{Real-data structure tests (RadioML).} (a)~Parity at
the optimized operating point ($70.8\pm0.4\%$ machine vs
$71.3\pm1.8\%$ classical-light twin, ratio $0.994$) and economy at
matched search scale ($69.1\pm0.6\%$ vs architecture-matched ESN bank
$62.0\pm3.4\%$, $n{=}200$ draws); both are search-scale statements,
scale-fenced from the full-corpus rows (Sec.~\ref{sec:results}).
Error bars are quenched dispersions (standard deviations across draws, not
standard errors): $n{=}10$ mask draws for the machine and the twin,
$n{=}200$ draws for the ESN bank, $n{=}3$ split seeds in~(b). (b)~The displacement boundary, per
split seed: phase encoding versus the displacement-encoded control on
identical hardware and photon number ($74.4\%$ vs $56.7\%$ three-seed
means; per-seed deficits $17.6$, $16.9$, $18.5$ points, with $1000$-draw
paired bootstrap intervals $[15.6,19.6]$, $[14.6,19.0]$, $[16.4,20.7]$
points, none containing zero). Every value in both panels is read at render
time from the archived result files; none is transcribed.}
\label{fig:t3}
\end{figure*}

\label{sec:results}

With the twin validated, we test the structural claims---parity,
charge algebra, displacement boundary---on an open RF corpus the
machine was not designed for. This section carries no budget-axis claim: the budget sweep exists only for the synthetic benchmarks (Table~\ref{tab:results}, Fig.~\ref{fig:t2}), which validate the mechanism but do not evidence the embodied-input claim; the corpus-scale budget axis awaits a re-propagation not performed in this work.

\subsubsection{Parity at no extra cost}

At the optimized operating point and infinite budget, machine and
classical-light twin agree to within mask dispersion [ratio $0.994$;
Fig.~\ref{fig:t3}(a)]---Proposition~\ref{prop:classical-equivalence}
measured rather than assumed.

\subsubsection{Economy at finite budget against the matched baseline}

The comparator is architecture matched---a bank of echo-state
instances mirroring the machine's mask bank, identical read channels,
noise injection, feature expansion and readout, $6142$ features against
$6142$ (Methods).

At matched measurement SNR the machine leads the matched bank by
$7.1$ points [Fig.~\ref{fig:t3}(a)], $1.8\%$ of individual baseline
realizations ahead; both figures are quoted at the scale at which the
two machines were run identically, not to be combined with the
corpus-scale accuracies above. \emph{Selection.} Three machine accuracies
exist on this corpus and are all reported here: search scale ($69.1\%$),
the displacement-comparison scale ($74.4\%$), and full corpus ($85.9\%$).
The comparison reported as the economy margin is the search-scale one,
because it is the only scale at which machine and baseline were run
identically. We did not pre-register which scale would carry the comparison, and we report all
three so that the choice is inspectable. Optimization makes the task tractable at low budget: the optimized
machine sits within two points of its own noiseless ceiling at
$B=10^{4}$.

\subsubsection{Structure: the charge algebra on real data}

The displacement-encoded control---identical hardware and photon
number, data entering additively---reaches $56.7\%$ against the
machine's $74.4\%$ [Fig.~\ref{fig:t3}(b)]: a $17.7$-point deficit,
consistent across all three split seeds ($17.6$, $16.9$, $18.5$ points;
$1000$-draw paired bootstrap intervals $[15.6,19.6]$, $[14.6,19.0]$,
$[16.4,20.7]$). On the reduced twin the control is
\emph{not} deficient; the gap opens only with interferometer and
braiding active: multi-epoch products, not the squeezer alone, are what
bilinear encoding buys.

\begin{figure*}[!t]
\centering
\includegraphics[width=\textwidth]{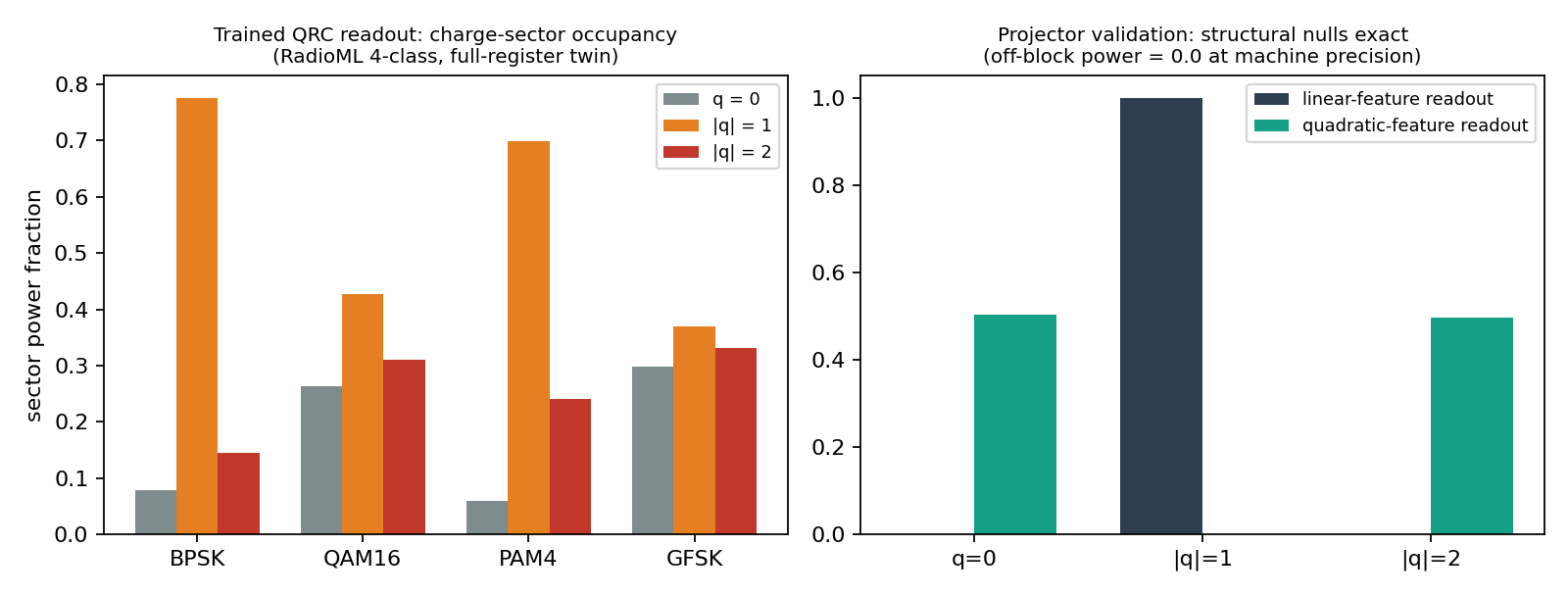}
\caption{\textbf{Charge-sector occupancy of trained readouts.} Left: real-world
readouts exercise the entire reachable spectrum $|q|\le2$. Right: restricting the readout to linear or to quadratic features places
the power exactly on $|q|=1$ or on $q\in\{0,\pm2\}$.}
\label{fig:occupancy}
\end{figure*}

Gauge-projected trained readouts occupy the entire reachable
spectrum $|q|\le2$ (Fig.~\ref{fig:occupancy}), and the deficit is not
localized by sector: the control fails on input-order multiplication
\emph{within} sectors---confined to $|q|=1$ the machine retains genuine
input-order-$4$ content on every seed while the displacement control
shows none in any sector (two pre-registered diagnostics, SI
Sec.~S5).

\subsubsection{What outperforms the machine, and the energy accounting}

A feature-matched classical
network reaches comparable accuracy noiselessly, and convolutional models
higher still---capacity, not noise, binds at this size; those machines buy
accuracy with full state access, unlimited SNR, and joules. The machine's own
full-corpus accuracy is $85.9\%$ at $B=10^{4}$, above the search-scale figures
of Fig.~\ref{fig:t3}(a): those are scale-fenced comparisons run identically for
both machines and are not to be read against full-corpus numbers.

The energy accounting is claimed narrowly: $1.0\times10^{-9}$~J of
optical signal energy per inference at referred $B=10^4$
($B_{\rm phys}=10^{8}$; the energy is computed under the physical
convention)---exact,
small, not a system metric; wall-plug is dominated by detection
electronics, so no comparative energy claim is made (Methods).

\subsection{The register is nonclassical and entangled; the twin is neither}

The classical-light twin spans the same functions
(Proposition~\ref{prop:classical-equivalence}) but is not the same physical
object, and the difference is measurable on the state rather than inferred
from performance. Witness definitions and the evaluation procedure are in
Methods; the twin is evaluated as a null control under the identical
procedure.

\begin{result}[Witnessed nonclassicality and entanglement]
\label{res:witness}
At both campaign operating points the stationary register is nonclassical
mode by mode---$47$--$52$ of $60$ bins carry a sub-vacuum quadrature---and
multipartite entangled: every single-bin-versus-remainder bipartition
violates the Gaussian PPT criterion, with logarithmic negativity up to
$E_N=0.59$, while no bin pair is entangled. The reclassicalized twin, run
through the identical procedure, passes every witness: no sub-vacuum mode,
no violated bipartition. This is an empirical statement about the
simulated stationary covariance, not a theorem, and not a measurement: no
device was built. It is also close to true by construction, since
$\bm{\mathcal N}_{\rm cl}$ is defined precisely to restore the minor axis to
vacuum. Its content is therefore that the twin construction is faithful---the
reclassicalization removes nonclassicality and nothing else---and that the
witness values are predictions for a platform of this class, computable now and
measurable on the architecture of Ref.~\cite{Paparelle2026}.
\end{result}

\noindent The entanglement is multipartite and not pairwise, which is what
the architecture predicts: the interferometer chains neighbours into a
cluster while the delay distributes correlation across epochs, so no
two-bin block carries it. Result~\ref{res:witness} establishes that the operating register is a quantum object and the control's is not, so the twin is a genuine classical control rather than a relabeling; whether the witnessed entanglement sets the finite-budget margin is the open question this architecture is built to test (SI Sec.~S7).

\section{Discussion}
\paragraph*{Registered predictions and their outcomes.}
Predictions frozen before compute were kept regardless of outcome
(SI, \emph{Registered predictions and their outcomes}): the
budget-migration and mask-knee predictions confirmed; an analytically
derived stability wall and a delay-independence prediction falsified
and retracted. Measure stability and memory, never infer them from
worst-case bounds, is part of the method.

\paragraph*{Scope of the quantum claims.} This paper claims three things about the quantum machine. Its operating register is nonclassical mode by mode and multipartite entangled, and the classical-light control is neither (Result~\ref{res:witness}). It outperforms a task-tuned classical digital reservoir under measurement budget on every nonlinear task tested, by 1.7--6.1$\times$ on the synthetic benchmarks and by 7.1 accuracy points on open RF data, at matched readout dimension and measurement noise (Table~\ref{tab:results}, Figs.~\ref{fig:t2}--\ref{fig:t3}). And its reachable function class is exactly characterized, with classical light reaching the same class (Proposition~\ref{prop:classical-equivalence}), which locates any finite-budget contribution of the squeezed state in feature conditioning (Corollary~\ref{cor:detection-attribution}) rather than expressivity---an attribution this architecture can test directly and which is left open here (SI Sec.~S7).

\label{sec:outlook}

The architecture assembles demonstrated
components~\cite{Lu2019,Nehra2022,Wang2018,Blumenthal2018,Yokoyama2013,Asavanant2019,Appeltant2011,Larger2012};
what is new is the principle arranging them: measurement cost as
primary design constraint, selecting nonlinearity, encoding, readout,
and simulation strategy alike.

We close with the claim as the data support it: \emph{minimality is
not a compromise}. A loop holding a sixth of a photon per bin, read
through one first-order channel, is a universal reservoir within
Theorem~\ref{thm:main}'s scope; under measurement budget it outperforms a task-tuned classical digital reservoir on every nonlinear task tested, on open real-world data matches its classical-light twin of identical hardware noiselessly to within 0.6\% as the theory predicts, and the finite-budget mechanism is measured
[Figs.~\ref{fig:t2}--\ref{fig:t3}]. What the encoding buys is order
multiplication inside the reachable sectors; the displacement control
cannot access it, and fails exactly where the charge algebra says it
must.

The boundary of the economics claim was stated where the benchmarks are
reported (Sec.~\ref{sec:narma}) rather than discovered afterwards:
measurement economy is a claim about physically embodied, high-rate
inputs, and the synthetic tasks are validation instruments.

The theory is written to be tested, and the platform for testing it
already exists~\cite{Paparelle2026}. What remains to be measured are
the diagnostics this theory turns on: the three forbidden kernel
families of Corollary~\ref{cor:falsify}, the parameter-free charge
covariance, the finite-budget margin over the digital baseline of Fig.~\ref{fig:t2}a, and the
sector-occupancy spectrum---every one defined operationally on
measured records (SI Sec.~S5). The non-Gaussian frontier remains open---correlating the tracked
$\chitwo$ back-action with capacity is the sharpest available probe of
whether non-Gaussianity buys processing power---and the natural
successor is the trilinear quantized-pump architecture.

\section{Methods}
\subsection{Entanglement and nonclassicality witnesses}
Witnesses are evaluated on the stationary covariance of the undriven
mask orbit at the two campaign operating points (campaign mask seed;
vacuum units $\sigma_{\rm vac}=\openone$). Nonclassicality:
a mode counts as sub-vacuum when the smaller eigenvalue of its
$2\times2$ covariance block lies below the vacuum value. Entanglement:
for every single-bin-versus-remainder bipartition the covariance is
partially transposed (momentum sign flip on the singled-out bin) and
the minimal symplectic eigenvalue $\tilde\nu$ computed; for Gaussian
states and $1\times N$ bipartitions this criterion is necessary and
sufficient, and the logarithmic negativity is
$E_N=\max(0,-\ln\tilde\nu)$. Bin pairs are tested identically on
$4\times4$ blocks. The reclassicalized twin is evaluated as a null
control under the identical procedure.
\subsection{RadioML task and protocol}
\label{sec:radioml}

RadioML~2016.10a~\cite{OShea2016} is an open corpus of $128$-sample
complex baseband snapshots at graded SNR. It was obtained from the public Zenodo mirror of the DeepSig release (RML2016.10a, CC BY-NC-SA 4.0). An eight-class variant falls
below the pre-registered acceptance threshold and is retained in the
record; we report the four-family
scope: BPSK, QAM16, PAM4, GFSK at $+18$, $+8$, $0$~dB---$12{,}000$
snapshots, split $60/20/20$. Data enter only through
Eq.~\eqref{eq:encoding} ($I$/$Q$ interleaved, $256$ bins); per bin the
machine reports $m_k=\tfrac12(\Sigma_{XX}-\Sigma_{PP})+i\Sigma_{XP}$.
Features, matching, ridge, the four baselines and the noise inventory are specified below.

\subsection{Optimal hyperparameters are task dependent}
\label{sec:tuning}

The machine has no single operating point: per-task, per-budget
optimization (Bayesian search, SI Sec.~S6; the objective never touches
a test partition or held-out seed) sends the NARMA family to strong
feedback near the measured stability boundary and the RF task to weak
squeezing at the injectivity limit. Relative to a frozen configuration, per-task optimization improves
NARMA10 five-fold, memory capacity by an order of magnitude, and RF
classification by $25$~points; the optimum migrates with budget in the
direction the cost model predicts, and benchmark comparisons at a fixed
configuration understate whichever architecture was frozen---the reason
every number here is a per-task optimum with its search budget
disclosed.

The search protocol, champion rule and shrinkage accounting are specified in the following subsection and in SI Sec.~S6.

\subsection{Hyperparameter search and champion rule}
Stability is measured, never inferred. The admissibility of a candidate
point is decided by a settle test on the covariance recursion, and the
search maps the boundary empirically---a substantial fraction of
evaluations return non-convergent and are recorded as such. An earlier
attempt to place that boundary analytically from an amplitude bound was
falsified by direct measurement and is retracted (SI Sec.~S6): under a
generic mask the growth rate is set by non-commuting rotated squeezes
and is strictly smaller than the worst-case aligned-axis bound, so
guard conditions of that form are conservative certificates rather than
boundaries.

\label{sec:methods}

\subsection*{Architecture parameters and timing}
Table~\ref{tab:params} lists the fast-clock parameter set, giving
hardware targets alongside the conservative campaign operating point at
which every reported result is computed. The campaign point is not a
tuned choice but a bound. Injected squeezing is held at $r=0.3$ so that
the stability guard (round-trip amplitude $G=0.785$, covariance guard
$\bar\rho=G^{2}=0.616$), the recirculating
antisqueezed-quadrature budget, and the Gaussian-closure gauge
$\varepsilon_G$ all retain wide margins. Quoted performance therefore
lower-bounds the hardware-target machine.

All timing derives from one RF oscillator slaved to the loop's pilot
tone, the synchronously pumped OPO discipline. We do not attempt
simultaneous conjugate readout: the Arthurs--Kelly
cost~\cite{ArthursKelly1965} is paid by
passive $X/P$ interleaving, at a halved per-quadrature rate, and carried
into the energy accounting as $2B$ circulations per symbol.

\begin{table}[t]
\caption{Fast-clock parameter set: component-level hardware targets versus
the conservative campaign operating point used for every reported result.
The campaign point is deliberately conservative, chosen where the twin's
guarantees bind simultaneously: the stability guard, stated either as the
round-trip amplitude $G=e^{r}\sqrt{\eta}<1$ or equivalently as the
covariance contraction $\bar\rho=\eta e^{2r}=G^{2}<1$; the antisqueezing
recirculation budget; and Gaussian-closure validity ($\varepsilon_G$); all quoted performance therefore lower-bounds
the target hardware. Squeezing figures are on-chip values. Readout is interleaved $X/P$ homodyne at 10~GHz per
quadrature with on-chip PSA in both columns.}
\label{tab:params}
\begin{ruledtabular}
\scriptsize\setlength{\tabcolsep}{2pt}
\begin{tabular}{@{}lll@{}}
 & Hardware target & Campaign point\\
\colrule
Bin period / clock & $\Delta=50$~ps / 20~GHz & same\\
Pump pulse (gate) & 10--15~ps & same\\
Poled segment & 6~mm; 2-ps walk-off & same\\
Loop delay / nodes & $\tau=5$~ns; $N=100$ & $N=60$ ($N'=61$)\\
Loop loss / circ. & $\approx0.5$~dB & 0.5~dB ($\eta_L=0.89$)\\
Escape efficiency & $>0.95$ (over-coupled) & $0.95$\\
Injected squeezing & 5--8~dB & 2.6~dB ($r=0.3$)\\
Feedback coupling & MZ-tunable & $\eta_{\rm fb}=0.40$\\
Round-trip amplitude $G$ & $<1$ & $0.785$ ($\bar\rho=G^2=0.62$)\\
Memory depth & $\sim(1-G)^{-1}$ & $\approx4$--$5$ circ.\\
Readout & intlv.\ $X/P$, 10~GHz & same\\
Timing jitter & $<50$~fs & gain noise $<1\%$\\
\end{tabular}
\end{ruledtabular}
\end{table}

\subsubsection*{Derivations of the two quoted identities}

\emph{Variance of moment estimators [Eq.~\eqref{eq:factorial}].} For a
zero-mean Gaussian variable $X$ with variance $\sigma_X^2$, the Isserlis
(Wick) theorem gives $\avg{X^{2m}}=(2m-1)!!\,\sigma_X^{2m}$, the double
factorial counting the pairings of $2m$ factors~\cite{Isserlis1918}.
Hence $\mathrm{Var}(X^m)=\avg{X^{2m}}-\avg{X^m}^2$ equals
$(2m-1)!!\,\sigma_X^{2m}$ exactly for odd $m$, where the odd moment
vanishes, and $[(2m-1)!!-((m-1)!!)^2]\,\sigma_X^{2m}$ for even $m$, whose
leading term is the same. This is the scaling quoted in
Eq.~\eqref{eq:factorial}; for the near-Gaussian output statistics of the
machine it holds to the accuracy of the Gaussian-closure gauge
$\varepsilon_G$.

\emph{Covariance identity in Eq.~\eqref{eq:downconversion}.} With
$a=X+iP$,
\begin{equation*}
a^2=X^2-P^2+i\,(XP+PX),
\end{equation*}
where the cross terms produce the symmetrized combination $XP+PX$
automatically, so no commutator term appears. Taking expectations in a
zero-mean state ($\bm m=0$, guaranteed by parity) and using
$\avg{X^2}=\Sigma_{XX}$, $\avg{P^2}=\Sigma_{PP}$,
$\tfrac12\avg{XP+PX}=\Sigma_{XP}$ gives
$\avg{a^2}=(\Sigma_{XX}-\Sigma_{PP})+2i\,\Sigma_{XP}$. In the
vacuum limit $\Sigma_{XX}=\Sigma_{PP}=\tfrac14$ and $\Sigma_{XP}=0$, so
$\avg{a^2}=0$, consistent with parity.
\emph{Normalization, stated once.} Two scalings of the same covariance appear
in this work. Absolute quadrature units, used in this derivation, put the
vacuum at $\bm\Sigma_{\rm vac}=\tfrac14\openone$. Vacuum units, used by the
twin and by every witness evaluation, rescale to $\bm\sigma=4\bm\Sigma$ so
that $\bm\sigma_{\rm vac}=\openone$; the symplectic blocks are invariant under
the rescaling and every reported feature and noise variance is a vacuum-unit
ratio. The per-bin harvest quoted in the RadioML protocol below,
$m_k=\tfrac12(\sigma_{XX}-\sigma_{PP})+i\,\sigma_{XP}$, is written in vacuum
units and equals $2\avg{a^2}$ in absolute units. No reported number depends on
the choice: the transduction amplitude $\mu$ is defined against $\avg{a^2}$ in
Eq.~\eqref{eq:downconversion} and never multiplies the harvested feature,
because the referred-budget convention sets $\mu_T$ to unity in the
simulation. The kick prefactor
$\Delta\avg{b}=-i\mu\avg{a^2}$ follows from perturbative integration of
the traveling-wave equations over the gated transit, at first order in
$\mu$ with the back-action tracked at $O(\mu^2)$ by $\varepsilon_G$
(SI Sec.~S1).

\subsubsection*{Measurement model and noise inventory}

The homodyne photocurrent on the output quadrature $X_b^{\vartheta}$ is
$I(t)=\sqrt{\eta_{\rm det}\kappa_b}\,X_b^{\vartheta,\rm out}(t)+\xi_{\rm vac}(t)$.
The per-bin feature is the windowed integral
$f(k,n)=\int_{\rm bin} w(t)I(t)\,dt$, a Gaussian variable whose mean is
given by the accumulated Eq.~\eqref{eq:downconversion} and whose
single-shot variance is $\Sigma_{X_bX_b}^{\vartheta}$. That variance is
reduced by the on-chip PSA gain, applied \emph{before} $\eta_{\rm det}$,
with added noise $\propto(1-\eta_{\rm det})e^{-2r_{\rm psa}}$. The
interleaver alternates $\vartheta\in\{0,\pi/2\}$ on even and odd bins.

\emph{PSA axis selection under fast ellipse rotation.} The PSA axis
tracks the homodyne LO, not the data. Since the homodyne reads a single
fixed quadrature per bin, noiseless gain along that same axis amplifies
the measured component faithfully however fast the signal phase rotates;
the conjugate component is read on the adjacent interleaved bin, at the
Arthurs--Kelly cost already paid by interleaving. Both PSA pump and LO
derive from the master oscillator, so their axes are locked common-mode
by the pilot lock, and the per-bin $X/P$ alternation of the PSA axis is
generated passively by the same one-bin-delay interleaver that switches
the LO, so axis and LO stay locked common-mode. The single-pass,
non-resonant PSA has no linewidth and passes the 20-GHz clock unsmeared.
Gain imbalance between the two interleaved rails
mis-weights the reconstruction of the complex feature from its $X$ and
$P$ samples, mixing $m$ with $\bar m$; it is therefore a charge-mixing
systematic, calibrated against the $|q|\le2$ projector null, and it
joins the falsifier-breaking inventory. A relative gain error
$\epsilon_g$ feeds the $\bar m$ (charge $-1$) component at amplitude
$\epsilon_g/2$, so holding charge mixing below the off-winding
tolerance at the hardware shot floor ($10^{-6}$ amplitude at $B=10^6$)
requires rail matching at the $10^{-2}$ level after calibration---an
electronic, not optical, tolerance.

The twin also carries two stochastic channels. Pump jitter and amplitude
noise enter multiplicatively in $S_k$ and in the source, as
$r_k=r(1+\delta_k)$ and $\mu_k=\mu(1+\gamma_k)$ with
${\rm Var}(\delta),{\rm Var}(\gamma)\lesssim10^{-4}$ at $<50$~fs relative
jitter; their first-order contribution to the feature covariance is
computed analytically alongside the propagation.

Every non-unitary element appears exactly once: loop loss
($\gamma_\ell$), MZ dump ($v$), squeezer intrinsic loss ($\kappa_i$, with
escape efficiency $\eta_{\rm esc}=1-\kappa_i/\kappa_s>0.95$), WDM
insertion loss, detector loss (post-PSA), homodyne shot noise, and gain
noise.

\subsubsection*{Three-tier simulation strategy}

\emph{Tier 1 (closed form):} for frozen inputs the per-mask-period composite $S$
yields stability as $\rho(S)<1$ (distance to the feedback-OPO threshold), the
stationary covariance from the discrete Lyapunov equation
$\bm\Sigma^\star=S\bm\Sigma^\star S^{\top}+\bm{\mathcal N}$, the full cross-time
kernel from powers of $S$, and the entanglement map (log-negativity
\cite{VidalWerner2002}, Duan--Simon \cite{Duan2000,Simon2000}) from $4\times4$
sub-blocks. \emph{Tier 2:} time-domain propagation of
Eq.~\eqref{eq:update} and the moment equations (SI Sec.~S1) with the $\mu^1$ source, jitter term,
and $\varepsilon_G$ gauge --- cost $O(10^4)$ flops per bin on the
$\approx206$-dimensional register (the $\sim3000$-bin memory is a property of the
kernel, never a matrix dimension). \emph{Tier 3:} truncated-Fock integration of
the master equation (SI Sec.~S1) on reduced registers wherever $\varepsilon_G$ exceeds
threshold. The classical simulability of the Gaussian baseline
\cite{MariEisert2012} is thereby the instrument panel of the twin: the machine's
claims live in $C(B)$, and the boundary where the classical description fails is
computed, not conjectured.

\subsection*{Benchmark protocol and baseline constructions}
\label{app:protocol}

\emph{Twin configuration.} Full-register twin (SI Sec.~S2
reduction) with the campaign hardware values: $N=60$ loop bins, braided mask period $N'=61$
(hence $61$ complex per-symbol harvests), $r=0.3$, $\eta_{\rm fb}=0.40$,
$\eta_L=0.89$ (0.5 dB), $\eta_{\rm esc}=0.95$, $\beta=1.0$,
$\varphi_{\rm arm}=\pi/4$. \emph{No hardware parameter was selected on task
performance.} Washout 100, training 1500, test 500 symbols; one symbol per
mask period; drive mapped affinely to $[-1,1]$ and held across the symbol.
Feature tiers: linear (122 real components) and per-unit quadratic
[$(\operatorname{Re}f,\operatorname{Im}f,\operatorname{Re}^2 f,
\operatorname{Im}^2 f,\operatorname{Re}f\operatorname{Im}f)$ per bin, 305].
Features are fit to $y_{t+1}$ by standardized ridge with intercept and
regularization matched to the injected noise variance; three task seeds with
common random numbers; the full pre-registration is bundled with the code archive. The
echo-state property is verified operationally (two initial covariances
converge to relative feature distance $6\times10^{-7}$ at single precision
after 100 circulations).

\emph{Shot-budget accounting: the referred budget.}
$B$ counts homodyne samples per real feature component, \emph{referred to
the fundamental-band feature units} $\avg{a^2}$ in which the twin
reports: the transduction chain's gains are absorbed into the feature
scale, which sets the transduction gain $\mu_T$ to unity in the
simulation. Physically the order-one channel is read \emph{after} the
$\chitwo$ transducer, $\mu_T=\mu=10^{-2}$ at the campaign point [the
same $\mu$ that keeps the Gaussian-closure gauge $\varepsilon_G$ an
order below threshold], so the per-shot variance referred to feature
units is $\sigma_{\rm eff}^2/\mu_T^2$ and the physical shot count is
$B_{\rm phys}=B/\mu_T^2=10^{4}\,B$. Every budget axis in this paper is
the referred budget $B$ (SI Sec.~S5). The convention is exactly neutral for every
\emph{comparison} in which both machines traverse the transducer---the
quantum machine versus its classical-light twin (identical chain,
$\mu_T$ cancels), and versus the digital reservoir at matched
per-feature SNR (matching is performed in referred units)---and for the
$\epsilon^{-3}$ scaling of Corollary~\ref{cor:exchange}, where $\mu_T$
enters the constant as written. It is \emph{not} neutral for the
absolute budget axis, for the energy accounting, or for the moment-readout
comparison, whose baseline homodynes the fundamental directly and pays
no referral; those three are stated under the physical convention where
they appear. Serial head-bin readout means the register does not
divide the budget: each per-bin feature accumulates $B$ samples, with
per-component noise $\sigma_{\rm eff}/\sqrt B$,
$\sigma_{\rm eff}^2=1.25$ (vacuum) or $1.025$ (with pre-detection gain;
audit below). The only division is the quadrature
basis: the interleaved $X$/$P$ readout
(Sec.~\ref{sec:architecture}) supplies the two components at half rate
each, an Arthurs--Kelly factor of two carried into the energy accounting as
$2B_{\rm phys}$ circulations per symbol.

\emph{ESN baseline (matched on readout dimension and data).} A 61-unit echo-state
network, $x_{t+1}=(1-\ell)x_t+\ell\tanh(Wx_t+w_{\rm in}u_t+b)$, hyperparameters
(spectral radius, input scale, leak) selected once on a noiseless validation
block and frozen across rungs, per the pre-registered grid; identical task
instances, splits, and ridge protocol, with the same per-unit quadratic tier
($x$, $x^2$; 122 features). The noise-matched
condition injects observation noise at the per-feature SNR of the
vacuum-readout QRC at each $B$. Three controls are added
(protocols and grids in SI Sec.~S5): a dimension-matched tier
($x$, $x^2$, and lag-$1$--$3$ cross products; $305$ features); an
input-side noise convention, in which the noise enters the ESN input
before the dynamics at matched SNR under two accounting conventions; and
a symmetric tuning protocol, in which the ESN grid is re-selected per
task per reservoir draw (five draws) while the machine is tuned over an
equal-budget $24$-point guard-respecting hardware grid
($r\times\beta\times\eta_{\rm fb}$), both on identical noiseless
validation blocks.

\emph{Classical-light twin (matched on photons and hardware).} Identical machine
with the squeeze map followed by reclassicalizing noise,
$\bm\Sigma \to S\bm\Sigma S^\top + R(\theta/2)\,{\rm diag}(1-e^{-2r},0)\,
R(\theta/2)^\top$: the output on vacuum is ${\rm diag}(1,e^{2r})$ rotated ---
same antisqueezed axis but no quadrature below vacuum, preserving
$P$-representability. Lifting the minor axis to the vacuum floor necessarily adds energy, so
the twin's circulating photon number is not identical but larger---
$\bar n=0.351$ against the machine's $0.156$ ($2.25\times$; settle-state covariances). Hardware and
encoding are matched exactly; flux is matched in the machine's disfavor,
which strengthens rather than weakens every margin reported over the
twin. Its readout carries the same detection chain as the machine,
\emph{including} the pre-detection amplifier: the amplifier's benefit is
loss suppression, which is input-state independent, so denying it to the
twin would misattribute a receiver property to the light. Both machines
therefore run at $\sigma_{\rm eff}^2=1.25$ (vacuum) or $1.025$ (with
gain). What classical light cannot supply is sub-vacuum readout noise~\cite{Caves1981,Tse2019,Lawrie2019}---and, per the audit above, neither can the amplifier: the
residual matched-noise margin between the machines is a
feature-conditioning effect, not a detection-variance one.

\emph{Audit of the amplifier model.} Dividing the input-referred
estimator variance by the gain ($\sigma_{\rm eff}^2=0.1025$) is not
derivable from the measurement model. A phase-sensitive amplifier adds
no excess noise but co-amplifies the arriving vacuum floor with the
mean, so first-principles propagation gives
$\sigma_{\rm eff}^2=1+(1-\eta_{\rm
det})/(\eta_{\rm det}G)=1.025$: a loss-suppression factor of $1.22$.
Three candidate channels for a larger benefit were tested and closed
(SI Sec.~S5).

\emph{Gaussian moment-readout reservoir (matched on total budget, energy, loss).}
The same loop with $\mu=0$ (linear symplectic dynamics) and moment features: per
bin, ${\rm Var}(X)$/${\rm Var}(P)$ interleaved plus covariances at lags 1--3
(four features per bin, granting the scheme its feature-count advantage over the order-one readout), with
per-sample estimator noise ${\rm Var}\approx\Sigma_{kk}\Sigma_{jj}+\Sigma_{kj}^2$
[the $m{=}2$ instance of Eq.~\eqref{eq:factorial}]. This baseline homodynes the
fundamental band directly and therefore pays \emph{no} transduction referral;
under the referred-budget convention the two machines are compared at a
common \emph{physical} budget, mean-readout variance
$\sigma_{\rm eff}^2/(\mu_T^2B_{\rm phys})$ against moment-readout variance
${\rm Var}/B_{\rm phys}$; comparing in referred units
alone would handicap the moment baseline by $\mu_T^{-2}=10^4$.

\subsection*{Real-data campaign protocol}
\label{app:campaign}

\emph{Twin.} Full register, $N=60$ loop bins, braided mask period $N'=N+1=61$,
$r=0.3$, $\eta_{\rm fb}=0.40$, $\eta_L=0.89$, $\eta_{\rm esc}=0.95$,
$\varphi_{\rm arm}=\pi/4$, $\beta=1.0$; round-trip amplitude
$G=e^{r}\sqrt{\eta_{\rm fb}\eta_L\eta_{\rm esc}}=0.785<0.98$, equivalently
covariance guard $\bar\rho=G^{2}=0.616<1$. This guard is quoted under the
hardware-target loss convention ($\eta_L$ per circulation) and is therefore
the tighter of the two readings: the twin of record applies the distributed
loss $\eta_\ell=\eta_L^{1/N}$ once per bin (Subsec.~S1\,D), for which the
per-circulation amplitude is $G=0.831$ and $\bar\rho=0.691$. Both satisfy the
guard; no reported \emph{benchmark} number depends on the choice, and the
memory-capacity curves of Fig.~\ref{fig:t1} are labeled with the
implemented convention. \emph{No hardware
parameter was selected on task performance.} An algebraic reduction is used
throughout: the feedback Mach--Zehnder with per-bin ancilla reset is exactly a
loss channel $\eta_{\rm fb}$ on the head bin (proven in SI Sec.~S2), after which one bin step is a single
$4\times4$ affine map on $(A_h,A_{h-1})$ followed by the loop loss.

\emph{State hygiene.} The register covariance is reset between examples to the
stationary state of the mask orbit, settled over whole joint periods of $N$ and
$N'$, against a fading-memory floor set by the per-bin covariance decay,
$0.907^{256}\approx10^{-11}$.

\emph{Noise inventory.} Per real feature component and per shot: detection
variance $1+(1-\eta_{\rm det})/\eta_{\rm det}=1.25$ at $\eta_{\rm det}=0.8$
without the amplifier, and $1+(1-\eta_{\rm det})/(\eta_{\rm det}G)=1.025$
with $10$~dB of gain (audit below); jitter gain-noise $(r\,\partial f/\partial r)^2{\rm
Var}(\delta)$ with ${\rm Var}(\delta)=10^{-4}$, from dedicated sensitivity runs
at $r(1+10^{-3})$; pump-amplitude noise $|f|^2{\rm Var}(\gamma)$,
${\rm Var}(\gamma)=10^{-4}$; all divided by $B$. A phase-lock residual floor $|f|^2\sigma_\phi^2$ with
$\sigma_\phi=10$~mrad does not average with $B$, and it produces the
observed high-budget saturation.

\emph{Itemized round-trip loss budget.} The campaign transmission
decomposes, per circulation, as: spiral propagation $\approx0.1$~dB;
squeezer-ring insertion and interferometer excess $\approx0.2$~dB;
transducer-segment and WDM insertion $\approx0.2$~dB (together
$\eta_L=0.89$, $0.5$~dB); squeezer escape $\eta_{\rm esc}=0.95$
($0.22$~dB); and the feedback Mach--Zehnder dump $\eta_{\rm fb}=0.40$
($4.0$~dB)---a round-trip total of $4.7$~dB, of which the
\emph{deliberate} MZ dump is $85\%$. The dump, not the waveguide, sets
the campaign memory depth: $\eta_{\rm fb}=0.40$ is chosen so the
stability guard, the antisqueezing budget, and Gaussian closure bind
simultaneously (Table~\ref{tab:params}), making every quoted result a
lower bound. The ultra-low-loss spiral matters in the hardware-target
column, where opening the MZ toward unity moves the machine into the
propagation-loss-limited regime and the depth toward
$(1-G)^{-1}\sim10$.

\emph{Pump extinction at the harmonic readout: the load-bearing
hardware requirement.} The information-bearing signal is a small 775-nm
mean field, homodyned against a pump-derived local oscillator, so any
imperfectly rejected pump light enters $\avg{b}$ coherently. What must
be suppressed is the \emph{fluctuating} leakage residue; the static
residue is removed by dark-reference subtraction. The requirement is
that this residue sit below the shot floor at the largest budget. At
the stated $10$-mW, 20-GHz pump, which is $\approx2\times10^6$ photons
per pulse, and with $1\%$-level pump amplitude and phase fluctuations,
this demands $\lesssim10^{-2}$ leaked photons per bin at the detector,
that is, $\approx85$~dB of pump-to-signal power extinction. We state
plainly that this is the binding requirement of the whole
implementation: demonstrated on-chip pump rejection in comparable
$\chitwo$ platforms is at the $40$--$60$~dB
level~\cite{Lu2019,Nehra2022}, so the balance must come from the
combination of spatial-mode and polarization selectivity, temporal
gating, and, as fallback, off-chip filtering---whose insertion loss
then enters the readout chain after the PSA, where the gain has already
suppressed its penalty. We do not have a line-item allocation for the
remaining $25$--$45$~dB with a demonstrated figure behind each line, and we
state plainly that the architecture is contingent on an advance of that size
over published on-chip rejection. Together with the transduction
amplitude $\mu_T$, which is likewise a design target rather than a measured
device value, this is one of two requirements in this work for which we can
name no existing demonstration. There is a useful bonus. Because the pump
carries phase $2\theta$, its leakage is a charge-$(+2)$ source in the
gauge algebra, and is therefore detected and rejected by the global
charge projector (SI, \emph{Locality of the three kernel laws}). Pump leakage joins the
named falsifier-breaking mechanisms rather than evading them.

\emph{Saturation and closure bounds.} At the campaign point the machine
draws $\bar n=0.156$ signal photons per bin from $\approx2\times10^6$
pump photons per pulse, so fractional pump depletion is
$\approx8\times10^{-8}$ per pass. Gain-saturation and pump-depletion
systematics, including any data-dependent, charge-violating phase they
induce, therefore enter at the $10^{-8}$ level, five orders below the
hardware shot floor at $B=10^6$. The Gaussian-closure gauge is bounded
analytically at the campaign point by
$\varepsilon_G\lesssim\mu^2(2\bar n+1)M\approx7\times10^{-4}$ at
$\mu=10^{-2}$ and memory depth $M=5$, an order below threshold, and
Tier~3 was never triggered. Training uses the matching analytic Tikhonov
term, so $B$ is a parameter rather than a campaign multiplier.

\emph{Ladder consistency.} On NARMA2 the three rungs give NMSE
$0.0039/0.0817/0.0811$ (machine) and $0.0040/0.1017/0.1011$ (classical-light
twin): noiseless parity to $10^{-5}$, a detection-dominated budget axis---the
multiplicative inventory contributes below one percent, which grounds Table~\ref{tab:params}'s
jitter claim in the campaign rather than in assertion---and a $1.10\times$ NMSE
improvement from the amplifier at fixed $B$
. Twenty independent noise realizations ($n=20$) at $B=10^4$ give
$0.0765\pm0.0046$ for vacuum readout and $0.0695\pm0.0042$ with
pre-detection gain (mean $\pm$ s.d.\ across realizations), a dispersion containing the
single-realization full/shot inversion of the ladder. Configuration (the bundled driver regenerates the results file to better
than $10^{-12}$ relative): drive $u_t\sim U(0,0.5)$ mapped affinely to $[-1,1]$ and
held across one mask period per symbol; $2100$ symbols split $100$ washout,
$1500$ training, $500$ test; features are the $61$ complex per-symbol harvests
expanded by the quadratic layer
$(\operatorname{Re}f,\operatorname{Im}f,\operatorname{Re}^2 f,
\operatorname{Im}^2 f,\operatorname{Re}f\operatorname{Im}f)$, $305$ real
components, standardized ridge with intercept and regularization tied to the
injected noise variance. The shot and full rungs inject sampled Gaussian noise
at $B=10^{4}$ shots per real feature component ($\sigma_{\rm eff}^2/B$ per
component); the full rung adds the jitter and phase-lock terms of the
inventory above. Because the rung noise is a sampled realization under fixed
seeds, the full rung falls marginally below the shot rung ($0.0811$ vs
$0.0817$); the expected ordering is monotone, and the twenty-replicate
dispersion reported in Sec.~\ref{sec:narma} contains the inversion.
Session-level time-series diagnostics---the encoding phase, the harvested
readout variables, the output--target traces, and the convergence of test
error with the number of retained harvest bins---are collected in SI Sec.~S9.
\emph{Registration status.} The ladder, W9, and W10 campaigns are
pre-registered (freeze documents bundled verbatim); findings (6) and (7) of SI Sec.~S5, the replicate dispersion, and the power-matched
baseline are \emph{not} pre-registered---they are labeled by their
drivers, use the same protocols and common random numbers as the
registered campaigns, and are deterministic given the disclosed seeds.

\emph{Statistics and reproducibility.} RadioML comparisons: three split seeds with common
random numbers, with $1000$-draw paired bootstrap intervals on the per-seed
accuracy difference as the reported statistic. Per-seed McNemar $p$ values were computed at run time and are archived. Discordant-pair counts are not reported: the archived run retained the $p$ values but not the counts, and counts reconstructed from $p$ values would be a fit rather than observed data; no inference in this paper rests on them. NARMA campaigns: three task seeds under a
frozen-selection, held-out design---every hyperparameter (hardware grid
point, ESN hyperparameters per draw, noise-aware variants) selected on
the first seed only and evaluated unchanged on two fresh seeds---with
five noise realizations and five reservoir draws per cell, scored
against analysis rules frozen before the runs. Headline claims were retained, demoted, or withdrawn
strictly per those rules. Where a paired $t$ statistic is quoted (SI Sec.~S7) it is two-tailed with $n-1=6$ degrees of freedom, $p=9.9\times10^{-5}$ for $t=9.1$ and $8.8\times10^{-4}$ for $t=6.1$. No multiple-comparison correction is applied: each quoted comparison is reported individually with its $n$, and no claim in this paper rests on a significance threshold---decisions follow the pre-registered effect-size rules, and intervals or dispersions are reported throughout, each $\pm$ labeled as a standard deviation across draws or realizations.

\emph{Scope.} The eight-class RadioML acceptance threshold was fixed in the
registered campaign protocol and the eight-class variant fell below it. We do
\emph{not} claim that the four-family scope was pre-registered as the
contingency for that failure: four families appear as the scope of the later
sector-restriction protocol, but no freeze document designates them in advance
as the eight-class fallback. The eight-class accuracy is not quoted here. Both
the unregistered scope choice and the missing number are disclosure gaps, and
we report them as such rather than as favourable selection. A second real-world task (inter-patient arrhythmia
classification) was executed and is \emph{not} reported here: its
parameter search is incomplete, and we decline to report any outcome
from an unconverged configuration---favorable or otherwise. The full
result will be reported only when the registered search completes.

\subsection*{Sector-resolved diagnostics}
\emph{Capacity, defined.} For a target functional $y$ of the input history,
the capacity of a feature set $f$ is
$C_y=1-\min_{w}\mathrm{NMSE}(y;\,w^{\top}\!f)\in[0,1]$: the fraction of the
target's variance captured by the best linear readout $w$ of the features.
The total capacity is $C=\sum_y C_y$ over an orthonormal basis of targets,
bounded above by the number of linearly independent
features~\cite{Dambre2012}. $C(B)$, the figure of merit adopted in the
Introduction, is this quantity evaluated with each feature carrying the
estimator noise of its share of a total budget of $B$ shots, replacing the
infinite-precision idealization by the affordable one.

Sector projection uses the global gauge-shift projector at $K=8$ mask
shifts (harvest at $\theta_{\rm mask}+\chi_k$, DFT in $\chi$).
Capacity decomposition: i.i.d.\ drive; targets are
products of normalized Legendre polynomials of the lagged drive, total
input order $\le4$, lags $\le12$, at most three distinct lags; capacity is
counted only above the 99th percentile of 500 target-permutation
surrogates; ridge with $\lambda=10^{-4}$. Sector-restricted order
probes use the centered $P_m$ targets and the identical 667-feature
post-processing of the displacement-regime protocol, restricted by the
same DFT; the projection is a restriction only for the gauge-equivariant
machine---for the displacement control it grants an eightfold mask
ensemble, making every displacement-side test conservative. The full
pre-registrations, with all disclosed addenda, are archived verbatim in
the code bundle alongside the reported results.

\subsection*{Use of large language models}
A large language model (Claude, Anthropic) was used to check author-developed mathematical derivations and to improve the readability and grammar of the text. All results were reviewed and approved by the author, who takes full responsibility for all content.

\section*{Data availability}
RadioML 2016.10a is an open corpus distributed by DeepSig under CC BY-NC-SA 4.0 and was obtained from its public Zenodo mirror (Methods). The numerical results
underlying every figure and table in this manuscript, the pre-registration
freeze documents, and the number manifest that gates each quoted value are
deposited at Zenodo under DOI~\texttt{10.5281/zenodo.22020535}~\cite{Soh2026code}. Source data
for each display item are provided as JSON files in that deposit.

\section*{Code availability}

The analysis, campaign, figure and gate code is deposited at Zenodo under
DOI~\texttt{10.5281/zenodo.22020535}~\cite{Soh2026code}, released under the MIT license. Every
campaign is deterministic given the disclosed seeds. Results whose generating
driver is not included in the deposit are listed explicitly in the deposit
README, and no such result is quoted in the main text.

\begin{acknowledgments}
D.S. acknowledges support from the U.S. Department of Energy, Office of
Science, under Award No.~DE-SC0025910, and from the National Science
Foundation under Award No.~2529700. The funders played no role in study design, data collection, analysis and interpretation of data, or the writing of this manuscript.
\end{acknowledgments}

\section*{Author contributions}
D.S. conceived the architecture, developed the theory, performed the
numerical campaigns, and wrote the manuscript.

\section*{Competing interests}
The author declares no financial or non-financial competing interests.

\end{document}


\title{Supplementary Information for ``A charge selection rule fixes what a squeezed-light reservoir computer can compute and afford''}
\author{Daniel Soh}
\affiliation{Wyant College of Optical Sciences, University of Arizona, Tucson, AZ, USA}
\maketitle
\medskip\noindent\emph{The argument in a nutshell.} We encode in the
squeeze angle, so histories enter as non-commuting products and the
state carries products of oscillations of the inputs---characters---at
every depth and degree. The $\chitwo$ transducer converts the
covariance holding them into a first moment of the harmonic field, so
the detector only ever measures order-one statistics: single
quadratures, averaged over shots. The measurement is only ever linear
in the quadratures, yet each linear reading naturally contains
polynomials of the encoded input at every degree---the dynamics has
already formed those products in amplitude, before detection. The
readout then simply extracts these input polynomials from the recorded
readings, never paying the factorial price of physically measuring
higher moments; combining features extends the extraction across
charge sectors, one unit per feature, up to $|q|\le D$. Deconvolving
the known one-pole filter and multiplying isolates each character,
whose algebra is dense by Stone--Weierstrass---universality at
polynomial shot cost, with the excluded sectors held off by an
explicit gap. (Terms are defined in Sec.~S3\,A; proofs in Sec.~S3.)

\section{The digital-twin model}
\label{sec:si-model}

This section specifies the dynamical model at the level of detail needed
to build a faithful numerical twin. We give, in order: the Hamiltonians
and Lindblad terms of every element; the time-delayed quantum Langevin
equations that define the non-Markovian loop; their exact Markovian
embedding on the bin register; the resulting Gaussian moment equations;
and the measurement statistics. Throughout we use the conventions
$a=X+iP$, $[X,P]=\tfrac{i}{2}$, and vacuum covariance
$\bm\Sigma_{\rm vac}=\tfrac14\mathds{1}$,
with all fields written in frames rotating at $\omega_0$ (fundamental)
and $2\omega_0$ (harmonic). Numerical covariances --- in Sec.~S2's block
list and in all twin code --- are quoted in \emph{vacuum units},
$\bm\sigma\equiv\bm\Sigma/\tfrac14$, so that the vacuum is the identity;
symplectic blocks are identical in either normalization (Subsec.~S1D).

\subsection{Continuous-time equations of motion}

\emph{Squeezer ring.} The pump-driven intracavity mode $a_s$ obeys the two-photon
Hamiltonian
\begin{equation}
H_s(t)/\hbar = \tfrac{i}{2}\Big[\varepsilon_2(t)\, e^{i\theta(t)}\, a_s^{\dagger 2}
- \varepsilon_2^*(t)\, e^{-i\theta(t)}\, a_s^{2}\Big],
\label{eq:Hsq}
\end{equation}
where $\varepsilon_2(t)=\varepsilon_0\,p(t)\big[1+G_{\rm fb}\!\int\! h(t')\,
Y(t-t')\,dt'\big]$ contains the 20-GHz pulse train $p(t)$ (master clock), the
measured-feedback modulation of the homodyne record $Y$, and
$\theta(t)=\theta_{\rm mask}(t)+\beta s(t)$ carries the mask and data
[Eq.~(2) of the main text]. \emph{Convention:} the
origin of $\theta$ is fixed by the physical pump --- $\theta$ is the
phase of the coefficient $\varepsilon_2e^{i\theta}$ multiplying
$a_s^{\dag2}$ in Eq.~\eqref{eq:Hsq} --- so that Eqs.~(S1)/(S8) and
main-text Eq.~(3) share one orientation. A global shift
$\theta\to\theta+c$ conjugates the Sec.~S3 transfer operator by
$e^{icQ}$, multiplying charge-$q$ kernel coefficients by $e^{iqc}$ and
leaving magnitudes, winding numbers, and every vanishing statement
invariant. With forward-bus, feedback-port, and intrinsic decay
rates $\kappa_f,\kappa_{\rm fb},\kappa_i$
($\kappa_s=\kappa_f+\kappa_{\rm fb}+\kappa_i$), the quantum Langevin equation
\cite{GardinerCollett1985} reads
\begin{align}
\dot a_s &= \varepsilon_2(t)e^{i\theta(t)}\,a_s^\dagger - \tfrac{\kappa_s}{2}a_s
\nonumber\\
&\quad + \sqrt{\kappa_f}\,A_{{\rm in},f}(t) + \sqrt{\kappa_{\rm fb}}\,A_{{\rm
in,fb}}(t) + \sqrt{\kappa_i}\, v_i(t),
\label{eq:langevin}
\end{align}
with input--output relations $A_{{\rm out},f}=\sqrt{\kappa_f}\,a_s-A_{{\rm in},f}$ and $A_{{\rm out},{\rm fb}}=\sqrt{\kappa_{\rm fb}}\,a_s-A_{{\rm in},{\rm fb}}$. Here $A_{{\rm in},f}(t)$ and $A_{{\rm in},{\rm fb}}(t)$ denote the input field
operators entering the forward bus and the feedback port, respectively,
and $A_{{\rm out},f}$, $A_{{\rm out},{\rm fb}}$ the corresponding
output fields; the intrinsic channel $v_i$ is unmonitored and carries no
output relation.

\emph{Coherent time-delayed feedback.} The defining non-Markovian element is the
closure of the loop: the feedback input is the loop-processed, delayed forward
output,
\begin{equation}
A_{\rm in,fb}(t) = e^{i\phi_L}\sqrt{\eta_L}\;
\mathcal U_{\rm loop}\!\big[A_{{\rm out},f}\big](t-\tau)
+ \sqrt{1-\eta_L}\;v_L(t),
\label{eq:delayclosure}
\end{equation}
where $\eta_L$ is the round-trip transmission ($\approx 0.5$~dB loss), $\phi_L$ the
pilot-locked loop phase, $v_L$ vacuum, and $\mathcal U_{\rm loop}$ the composition
of the one-bin interferometer, the $\chitwo$ segment, and the tunable MZ coupler.
Equations~\eqref{eq:langevin}--\eqref{eq:delayclosure} constitute a time-delayed
quantum Langevin system of the class analyzed in
Refs.~\cite{Grimsmo2015,PichlerZoller2016}; the delayed argument makes the reduced
dynamics of $a_s$ non-Markovian by construction.

\emph{Single-pass $\chitwo$ segment.} In the co-moving frame ($t'=t-z/v_g$,
group-velocity matched by design), the quasi-phase-matched traveling-wave
Heisenberg equations for the fundamental $\hat a(z,t')$ and harmonic
$\hat b(z,t')$ envelopes are
\begin{equation}
\partial_z \hat a = 2i\tilde\kappa\, \hat a^\dagger \hat b, \qquad
\partial_z \hat b = i\tilde\kappa\, \hat a^{2},
\label{eq:travelling}
\end{equation}
with $\tilde\kappa$ the poled nonlinear coupling per unit length. Perturbative
integration over the walk-off-limited interaction window $L_{\rm eff}$ (the
temporal aperture, $\approx 2$~ps over 6~mm) gives the per-pass input--output kick
used by the twin,
\begin{equation}
\hat b_{\rm out} = \hat b_{\rm in} + i\mu\, \hat a_{\rm in}^2 + O(\mu^2), \quad
\hat a_{\rm out} = \hat a_{\rm in} + 2i\mu\, \hat a_{\rm in}^\dagger \hat b_{\rm
in},
\label{eq:kick}
\end{equation}
with $\mu=\tilde\kappa L_{\rm eff}\ll 1$; the sign of $\mu$ is a phase
convention, fixed in the main text by the readout phase. Taking
expectations of Eq.~\eqref{eq:kick} on a zero-mean state reproduces
main-text Eq.~(4). Explicitly, with $a=X+iP$ one has
$a^2=X^2-P^2+i(XP+PX)$, the cross terms arriving already
symmetrized, so on a zero-mean state
\begin{equation*}
\avg{a^2}=(\Sigma_{XX}-\Sigma_{PP})+2i\,\Sigma_{XP},
\end{equation*}
which is a linear read-off of three entries of the quadrature covariance;
vacuum ($\Sigma_{XX}=\Sigma_{PP}=\tfrac14$, $\Sigma_{XP}=0$) gives
$\avg{a^2}=0$, consistent with parity. The resonant-transducer variant
obeys $d\avg{b}/dt=-ig\avg{a^2}-(\kappa_b/2)\avg b$ and reduces to the
same kick upon adiabatic elimination for $\kappa_b\Delta\gg1$.

\emph{Parasitics (optional twin channels).} Pump self-phase modulation adds
$H_{\rm SPM}=\hbar\gamma_3 |E_p|^2$-type classical chirp to $p(t)$ (calibrated,
pre-chirp-compensated); cross-phase modulation on the signal,
$H_{\rm XPM}=2\hbar\gamma_3 |E_p|^2 a^\dagger a$ during the walk-off window, is
included as a small input-dependent rotation when peak powers approach the
$B$-integral ceiling.

\subsection{Master equation and exact Markovian embedding}

A delayed equation is awkward to integrate, but it need not stay delayed.
Equations \eqref{eq:langevin}--\eqref{eq:delayclosure} admit an
\emph{exact} Markovian embedding, obtained by promoting the light in
flight to explicit degrees of freedom. The idea is to discretize the loop
into $N=\tau/\Delta$ orthonormal temporal bin modes $A_1,\dots,A_N$, a
conveyor belt, and to identify the continuous field that the ring emits
during one bin period with the bin sitting at the head pointer.

On the enlarged register $\varrho$, comprising modes $a_s$, $b$,
$A_1..A_N$ and ancilla $v$, the dynamics within one bin period is
generated by the Lindblad master equation
\begin{align}
\dot\varrho =&
-\tfrac{i}{\hbar}\big[H_s(t) + H_c(t) + H_g(t),\ \varrho\big]
+ \kappa_i\,\mathcal D[a_s]\varrho
\nonumber\\
&+ \kappa_b\,\mathcal D[b]\varrho
+ \gamma_\ell\, \mathcal D[A_{h}]\varrho ,
\label{eq:master}
\end{align}
with $\mathcal D[c]\varrho = c\varrho c^\dagger - \tfrac12\{c^\dagger c,
\varrho\}$ and the piecewise couplings
\begin{align}
H_c(t)/\hbar &= i\sqrt{\kappa_s^{\rm eff}/\Delta}\;\big(a_s^\dagger A_h - A_h^\dagger a_s\big)
&&\text{(ring--bin exchange)},
\nonumber\\
H_g(t)/\hbar &= g\big(A_h^2\, b^\dagger + A_h^{\dagger 2}\, b\big)
&&\text{(segment transit)},
\nonumber\\
H_{\rm if}/\hbar &= g_{\rm if}\big(A_h^\dagger A_{h-1} + \text{h.c.}\big)
&&\text{(1-bin interferometer)},
\nonumber
\end{align}
each active only during its own stage of the bin period. Here
$\gamma_\ell$ implements the distributed loop loss
($e^{-\gamma_\ell \tau}=\eta_L$), and the MZ coupler is the beamsplitter
unitary between $A_h$ and $v$ with settable transmissivity
$\eta_{\rm fb}$. Advancing the head pointer $h\to h+1 \pmod N$ once per
bin closes the delay exactly: after $N$ steps the bin written by the ring
returns to it, reproducing Eq.~\eqref{eq:delayclosure} without
approximation. This
collision-model form of time-delayed coherent feedback
\cite{Grimsmo2015,PichlerZoller2016} is what the digital twin integrates; in the
strongly over-coupled, broadband limit the ring--bin stage collapses to the
instantaneous oriented squeeze of main-text Eq.~(3), which is the default
twin configuration (the finite-linewidth stage is retained as a switchable
refinement).

\subsection{Gaussian moment equations}

Every generator above is at most quadratic (the trilinear $H_g$ enters
perturbatively), so the state is Gaussian at order $\mu^0$ and the twin
propagates first and second moments exactly. For the quadrature vector
$\bm\xi$ of the register, the continuous stages obey
\begin{equation}
\dot{\bm m} = \bm A(t)\, \bm m, \qquad
\dot{\bm\Sigma} = \bm A(t)\bm\Sigma + \bm\Sigma\bm A(t)^{\!\top} + \bm D(t),
\label{eq:momentODE}
\end{equation}
with drift and diffusion assembled blockwise; e.g., the squeezer block
[from Eq.~\eqref{eq:langevin}]
\begin{equation}
\bm A_s = \begin{pmatrix}
\varepsilon_2\cos\theta - \tfrac{\kappa_s}{2} & \varepsilon_2\sin\theta\\[2pt]
\varepsilon_2\sin\theta & -\varepsilon_2\cos\theta - \tfrac{\kappa_s}{2}
\end{pmatrix},\quad
\bm D_s = \kappa_s\,\bm\Sigma_{\rm vac} = \tfrac{\kappa_s}{4}\,\mathds{1}
\ \ (\kappa_s\,\mathds{1}\ \text{in vacuum units}),
\label{eq:driftsq}
\end{equation}
whose instability threshold $\varepsilon_2=\kappa_s/2$ is the (open-loop) OPO
threshold; with feedback the relevant boundary is the round-trip spectral radius
below. The instantaneous stages act as the symplectic/CP updates
$\bm m\!\to\!S\bm m$, $\bm\Sigma\!\to\!S\bm\Sigma S^{\top}\!+\bm{\mathcal N}$ with the block
matrices listed in Appendix~\ref{app:blocks}; the per-bin composite defines
$S_k(\theta_k)$ of main-text Eq.~(3). The $\mu$-expansion enters as: order
$\mu^0$, Eq.~\eqref{eq:momentODE}; order $\mu^1$, the feature source
main-text Eq.~(4) (a sparse linear read-off of three covariance
entries per bin); order $\mu^2$, the back-action on $\bm\Sigma$ propagated as the
structural-integrity gauge $\varepsilon_G=\|\bm\Sigma^{(\mu^2)}\|/
\|\bm\Sigma^{(\mu^0)}\|$, which simultaneously monitors Gaussian-closure validity
and flags emerging Wigner negativity for the master-equation audit
[direct integration of Eq.~\eqref{eq:master} in a truncated Fock space on 2--3
modes].

\subsection{Operator-level dynamics: Hamiltonian $\to$ Lindbladian $\to$
Heisenberg map}
\label{od:sec:wrapper}

\newcounter{odsaveeq}\setcounter{odsaveeq}{\value{equation}}
\setcounter{equation}{0}\renewcommand{\theequation}{S1D.\arabic{equation}}
\newcounter{odsavetab}\setcounter{odsavetab}{\value{table}}
\setcounter{table}{0}\renewcommand{\thetable}{S1D.\arabic{table}}

\subsubsection{Purpose}

Subsections~S1A--S1C above state the Hamiltonians [Eq.~(S1)], the quantum Langevin
equation [Eq.~(S2)], the delay closure [Eq.~(S3)], the Lindblad master
equation on the embedded register [Eq.~(S6)], and the Gaussian
moment equations [Eq.~(S7)]. They do not exhibit the intermediate
object: the Heisenberg-picture operator evolution, solved as an explicit
Bogoliubov map and composed into the per-bin symplectic $S_k$ of
main-text Eq.~(3). This subsection supplies that step in full, and in doing
so pins one convention that Subsecs.~S1A--S1C leave implicit
(Sec.~\ref{od:sec:pi}).

\emph{Physical picture, before any algebra.} The machine is a lossy
parametric amplifier whose gain axis is steered by the data. A two-photon
drive stretches one quadrature and squeezes the orthogonal one; the pump
phase $\theta$ sets \emph{which} axis is stretched. Because the
interaction is quadratic in the field, rotating the pump by $\theta$
rotates the squeeze ellipse by only $\theta/2$ --- that factor of two is
the whole origin of the half-angle in main-text Eq.~(3). Loss then pulls
the ellipse back toward the vacuum circle, and vacuum noise enters at
exactly the rate needed to keep the commutator equal to one. Everything
below is bookkeeping on those two sentences.

\subsubsection{Conventions}

Fixed once, used throughout, never silently changed:
\begin{equation}
a=X+iP,\quad X=\frac{a+a^\dag}{2},\quad P=\frac{a-a^\dag}{2i},\quad
[X,P]=\frac{i}{2},\quad [a,a^\dag]=1,\quad
\bm\Sigma_{\rm vac}=\tfrac14\mathds{1} .
\label{od:eq:conv}
\end{equation}
These three are not independent choices. Since
$[a,a^\dag]=[X+iP,X-iP]=-2i[X,P]$, demanding $[a,a^\dag]=1$ \emph{forces}
$[X,P]=i/2$, and then
$\avg{X^2}_{\rm vac}=\tfrac14\avg{aa^\dag}=\tfrac14$ follows. Quadrature vector $\bm\xi=(X,P)^\top$; symplectic form
$\bm\Omega=\left(\begin{smallmatrix}0&1\\-1&0\end{smallmatrix}\right)$ with
$[\xi_i,\xi_j]=\tfrac{i}{2}\,\Omega_{ij}$; covariance
$\Sigma_{ij}=\tfrac12\avg{\{\Delta\xi_i,\Delta\xi_j\}}$; uncertainty and
complete positivity take the form
$\bm\Sigma+\tfrac{i}{4}\bm\Omega\succeq0$.

\emph{Convention equivalence.} The entire manuscript and
all code are unified on $a=X+iP$. The alternative convention
$a=(X+iP)/2$, $[X,P]=2i$, $\bm\Sigma_{\rm vac}=\mathds{1}$ is the
uniform rescaling $\bm\xi\to2\bm\xi$ of the present one; equivalently, moving
from it to the present convention rescales
\begin{equation}
\bm\Sigma\to\tfrac14\bm\Sigma,\qquad
\bm D\to\tfrac14\bm D,\qquad
\bm{\mathcal N}\to\tfrac14\bm{\mathcal N},
\label{od:eq:rescale}
\end{equation}
while \emph{every symplectic object is invariant}: $\bm\Phi$, $\bm A_s$,
$S_k$, $R$, $Z$, $\Xi$ and $G$ are unchanged, because a scalar rescaling
commutes with every block,
$(\lambda\mathds{1})S(\lambda\mathds{1})^{-1}=S$. Numerical
covariances in Sec.~S2 and the twin code are quoted in vacuum units and
are unchanged; benchmark invariance holds exactly: the pipeline estimator is
scale-free at fixed regularization. Fields are in
frames rotating at $\omega_0$ and $2\omega_0$. Elementary matrices
$R(\phi)=\left(\begin{smallmatrix}\cos\phi&-\sin\phi\\
\sin\phi&\cos\phi\end{smallmatrix}\right)$,
$Z(r)=\mathrm{diag}(e^{-r},e^{r})$, and the reflection
\begin{equation}
\Xi(\theta)=\begin{pmatrix}\cos\theta&\sin\theta\\
\sin\theta&-\cos\theta\end{pmatrix},\qquad \Xi^2=\mathds{1},\ \det \Xi=-1 .
\label{od:eq:Qdef}
\end{equation}
Symbols introduced later: $A_j$ temporal bin modes and $A_h$ the bin at
the head pointer [Eqs.~\eqref{od:eq:binmodes}--\eqref{od:eq:pointer}];
$\mathfrak u,\mathfrak v$ Bogoliubov coefficients; $\hat F$
integrated noise operator; $\bm\Phi$ quadrature propagator;
$w=t-s$ retarded time; $\eta_{\rm esc}$ escape efficiency; $\eta_{\rm fb}$ feedback Mach--Zehnder power transmissivity retained by the head bin;
$\mathcal L^\dag$ adjoint Lindbladian.

\subsubsection{Hypotheses consumed by the derivation}

Table~\ref{tab:si-hyp} lists every hypothesis the derivation consumes,
with its source and status.

\begin{table}[ht]
\caption{Hypotheses the derivation consumes, with source and status.}
\label{tab:si-hyp}
\begin{ruledtabular}\footnotesize
\begin{tabular}{@{}lp{4.9cm}p{3.5cm}p{6.6cm}@{}}
\# & Hypothesis & Consumed by & Source\\
\colrule
H1 & Rotating-wave approximation & Eq.~\eqref{od:eq:Hs} & SI Eq.~(S1) frame
statement\\
H2 & Undepleted classical pump $\varepsilon_2(t)$ & bilinearity of
Eq.~\eqref{od:eq:Hs} & SI Eq.~(S1)\\
H3 & Flat, $\delta$-correlated reservoirs & Lindblad form
Eq.~\eqref{od:eq:master} & SI Eq.~(S6)\\
H4 & Vacuum (zero-temperature) inputs & Eq.~\eqref{od:eq:Fcomm} &
SI Eq.~(S3) ($v_L$ vacuum)\\
H5 & Group-velocity matching, window $L_{\rm eff}$ & Eq.~\eqref{od:eq:kick}
& SI Eq.~(S4)\\
H6 & $\mu=\tilde\kappa L_{\rm eff}\ll1$ & truncation of
Eq.~\eqref{od:eq:kick} & SI Eq.~(S5)\\
H7 & Broadband, over-coupled ring ($\kappa_s\Delta\gg1$) &
Eq.~\eqref{od:eq:collapse} & SI Sec.~S1B text\\
H8 & $\theta_k$ constant over one bin & Eq.~\eqref{od:eq:bog} &
main-text Eq.~(2)\\
H9 & $N=\tau/\Delta$ integer; bins orthonormal & Sec.~\ref{od:sec:delay} &
SI Eq.~(S6) text\\
H10 & $\kappa_b\Delta\gg1$ & adiabatic elimination of $b$ &
SI Sec.~S1A\\
H11 & Zero-mean state, $\bm m=\bm 0$ & Eq.~\eqref{od:eq:readoff} &
main-text Eq.~(3) text\\
H12 & $G=e^{r}\sqrt\eta<1$ & Eq.~\eqref{od:eq:depth} &
Table~I campaign point\\
H13 & Fundamental/harmonic modes independent, $[a,b^\dag]=0$ &
Eq.~\eqref{od:eq:kickcomm} & WDM band separation, main-text
Fig.~1 station~II\\
H14 & Absolute origin of $\theta$ is free & Sec.~\ref{od:sec:pi} &
pump-phase reference convention\\
\end{tabular}
\end{ruledtabular}
\end{table}

\subsubsection{Generators}

The squeezer ring carries the two-photon Hamiltonian [SI Eq.~(S1)]
\begin{equation}
H_s(t)/\hbar=\frac{i}{2}\Big[\varepsilon_2(t)e^{i\theta(t)}a_s^{\dag2}
-\varepsilon_2^*(t)e^{-i\theta(t)}a_s^2\Big],
\label{od:eq:Hs}
\end{equation}
quadratic, hence Gaussianity-preserving; $\varepsilon_2$ is the
pump-set parametric gain rate (units s$^{-1}$) and
$\theta=\theta_{\rm mask}+\beta s$ the encoded angle. Ring--bin exchange
$H_c$, interferometer $H_{\rm if}$ and the beamsplitter blocks are
likewise quadratic.

\emph{Definition of the bin modes and of $A_h$.} SI Sec.~S1 introduces
``$N=\tau/\Delta$ orthonormal temporal bin modes $A_1,\dots,A_N$'' and
``the bin sitting at the head pointer'' in prose; the symbol $A_h$ is
used from SI Eq.~(S7) onward without a formal definition, which we
supply here. Partition the delay into contiguous, non-overlapping
intervals $I_j=[(j-1)\Delta,\,j\Delta)$ and set
$u_j(t)=\Delta^{-1/2}\mathds{1}_{I_j}(t)$. The bin modes are the
projections of the travelling field $\hat A(t)$, with
$[\hat A(t),\hat A^\dag(t')]=\delta(t-t')$, onto these:
\begin{equation}
A_j=\int\! dt\;u_j(t)\,\hat A(t),\qquad j=1,\dots,N,
\qquad [A_i,A_j^\dag]=\langle u_i|u_j\rangle=\delta_{ij},
\label{od:eq:binmodes}
\end{equation}
the last equality because the $u_j$ have disjoint supports and unit norm
(H9). Each $A_j$ is dimensionless: $\hat A(t)$ carries ${\rm s}^{-1/2}$,
$u_j$ carries ${\rm s}^{-1/2}$, and $dt$ supplies ${\rm s}$.

The subscript $h$ is \emph{not} a mode label but a step-dependent
pointer,
\begin{equation}
h(k)=\big((h_0+k-1)\bmod N\big)+1,\qquad A_h\equiv A_{h(k)},
\label{od:eq:pointer}
\end{equation}
so $A_h$ denotes whichever of the $N$ fixed modes is aligned with the
ring's output port at step $k$, and $A_{h-1}$ in $H_{\rm if}$ is its
predecessor, index $((h(k)-2)\bmod N)+1$. The register
$\{a_s,b,A_1..A_N,v\}$ therefore has $N+3$ modes; $A_h$ is an alias, not
an additional degree of freedom. Two consequences are immediate and are
what the construction buys: $[A_h,A_h^\dag]=1$ at every step
automatically, since $A_h$ is always one of the $A_j$; and
$A_{h(k+N)}=A_{h(k)}$, which is precisely the delay closure of
Sec.~\ref{od:sec:delay}. The factor $\Delta^{-1/2}$ in $u_j$ is also what
fixes the exchange coupling: $\sqrt{\kappa_s^{\rm eff}/\Delta}$ in
$H_c$ carries units of ${\rm s}^{-1}$, as a rate must.

\emph{Why a pointer.} Physically the light in flight moves; one could
shift all $N$ bin amplitudes by one slot each step. Incrementing an
index instead is the same dynamics at $O(1)$ rather than $O(N)$ cost ---
the conveyor belt of SI Sec.~S1, implemented as a rotation of labels. The segment generator
$H_g/\hbar=g(A_h^2b^\dag+A_h^{\dag2}b)$ is \emph{trilinear} and is the
sole non-Gaussian element; it is truncated at order $\mu^1$, discarding
$O(\mu^2)$ back-action which is retained only as the gauge
$\varepsilon_G$ (Sec.~\ref{od:sec:chi2}).

On the enlarged register the dynamics within one bin period is
\begin{equation}
\dot\varrho=-\tfrac{i}{\hbar}[H_s+H_c+H_g,\varrho]
+\kappa_i\mathcal D[a_s]\varrho+\kappa_b\mathcal D[b]\varrho
+\gamma_\ell\mathcal D[A_h]\varrho,
\label{od:eq:master}
\end{equation}
reproducing SI Eq.~(S6), with the stage generators active on disjoint
sub-intervals of the bin period, so they compose as an ordered product
and never overlap.

\subsubsection{The adjoint generator, and why it is not enough}

For $\dot\varrho=-\tfrac{i}{\hbar}[H,\varrho]+\sum_c\gamma_c\mathcal
D[c]\varrho$ the adjoint acting on operators is
\begin{equation}
\mathcal L^\dag[O]=\tfrac{i}{\hbar}[H,O]
+\sum_c\gamma_c\Big(c^\dag Oc-\tfrac12\{c^\dag c,O\}\Big).
\label{od:eq:adjoint}
\end{equation}
Unitality is immediate: $\mathcal L^\dag[\mathds{1}]=c^\dag c-c^\dag c=0$.
Applied to $O=a_s$ with $c=a_s$,
\begin{equation}
a^\dag a a-\tfrac12\{a^\dag a,a\}
=\tfrac12[a^\dag,a]a=-\tfrac12 a ,
\label{od:eq:damp}
\end{equation}
so the dissipator contributes $-\tfrac{\kappa_s}{2}a_s$, while the
Hamiltonian part gives $\tfrac{i}{\hbar}[H_s,a_s]=\varepsilon_2
e^{i\theta}a_s^\dag$. Together they are exactly the drift of
SI Eq.~(S2).

\emph{The catch, stated plainly.} Equation~\eqref{od:eq:adjoint} propagates
\emph{expectation values} correctly, but if one reads it as an operator
equation the solution has $[a_s(t),a_s^\dag(t)]=e^{-\kappa_s t}<1$: the
commutator decays, which is impossible. The resolution is that the
Heisenberg-picture equation is not Eq.~\eqref{od:eq:adjoint} but the
quantum Langevin equation, whose input-noise operators supply precisely
the missing commutator. Equation~\eqref{od:eq:Fcomm} below is that
statement made quantitative.

\subsubsection{Bogoliubov form of the squeezer stage}

With $\theta_k$ constant over the bin (H8) and $\gamma\equiv\kappa_s/2$,
SI Eq.~(S2) reads $\dot a_s=-\gamma a_s+\lambda a_s^\dag+\text{noise}$
with $\lambda=\varepsilon_2e^{i\theta_k}$. Substituting
$a_s=e^{-\gamma t}c$ removes the damping, leaving $\dot c=\lambda
c^\dag$, $\ddot c=|\lambda|^2c$, so
\begin{equation}
a_s(t)=\mathfrak u(t)\,a_s(0)+\mathfrak v(t)\,a_s^\dag(0)+\hat F(t),\qquad
\mathfrak u=e^{-\gamma t}\cosh(\varepsilon_2 t),\quad
\mathfrak v=e^{-\gamma t}e^{i\theta_k}\sinh(\varepsilon_2 t).
\label{od:eq:bog}
\end{equation}
The noise operator is the same Green's function against the input fields,
$\hat F(t)=\int_0^t\!ds\,\big[\mathfrak u(t-s)\big(\sqrt{\kappa_f}\,A_{{\rm in},f}(s)+\sqrt{\kappa_{\rm fb}}\,A_{{\rm in},{\rm fb}}(s)+\sqrt{\kappa_i}\,v_i(s)\big)+\mathfrak v(t-s)\big(\sqrt{\kappa_f}\,A^\dag_{{\rm in},f}(s)+\sqrt{\kappa_{\rm fb}}\,A^\dag_{{\rm in},{\rm fb}}(s)+\sqrt{\kappa_i}\,v_i^\dag(s)\big)\big]$, the three input channels of Eq.~\eqref{eq:langevin}. With
$[A_{\rm in}(s),A^\dag_{\rm in}(s')]=\delta(s-s')$ (H3, H4),
\begin{equation}
|\mathfrak u|^2-|\mathfrak v|^2=e^{-\kappa_s t},\qquad
[\hat F,\hat F^\dag]=\kappa_s\!\int_0^t\! e^{-\kappa_s(t-s)}ds
=1-e^{-\kappa_s t},
\label{od:eq:Fcomm}
\end{equation}
whose sum is unity for every $t$: the vacuum entering through the loss
ports pays exactly the commutator deficit incurred by the decaying
coherent part. \emph{Direction note:} both terms in
Eq.~\eqref{od:eq:Fcomm} are exact equalities, not bounds.

\subsubsection{Quadrature propagator and consistency with SI Eq.~(S8)}

Using $e^{i\theta}a^\dag+e^{-i\theta}a=X\cos\theta+P\sin\theta$ and
$-i(e^{i\theta}a^\dag-e^{-i\theta}a)=X\sin\theta-P\cos\theta$,
Eq.~\eqref{od:eq:bog} becomes $\bm\xi(t)=\bm\Phi(t)\bm\xi(0)+\text{noise}$ with
\begin{equation}
\bm\Phi(t)=e^{-\kappa_s t/2}\Big[\cosh(\varepsilon_2t)\,\mathds{1}
+\sinh(\varepsilon_2t)\,\Xi(\theta_k)\Big].
\label{od:eq:Mprop}
\end{equation}
Differentiating, $\bm\Phi(0)=\mathds{1}$ and
$\dot{\bm\Phi}=\bm A_s\bm\Phi$ for all $t$ with
\begin{equation}
\bm A_s=\begin{pmatrix}
\varepsilon_2\cos\theta-\tfrac{\kappa_s}{2}&\varepsilon_2\sin\theta\\
\varepsilon_2\sin\theta&-\varepsilon_2\cos\theta-\tfrac{\kappa_s}{2}
\end{pmatrix},
\label{od:eq:As}
\end{equation}
which is SI Eq.~(S8) exactly. So
Eq.~\eqref{od:eq:Mprop} \emph{is} the exact propagator of SI Eq.~(S8); no
matrix exponential need be evaluated. Since $\Xi$ has eigenvalues $\pm1$,
$\bm A_s$ has eigenvalues $-\kappa_s/2\pm\varepsilon_2$, recovering the
open-loop OPO threshold $\varepsilon_2=\kappa_s/2$ as the point where the
larger eigenvalue crosses zero.

\subsubsection{The half-angle, and a $\pi$ of convention}
\label{od:sec:pi}

The reflection $\Xi$ factorizes through a \emph{half}-angle rotation,
\begin{equation}
R(\phi)\,\mathrm{diag}(1,-1)\,R(\phi)^\top=\Xi(2\phi),
\label{od:eq:half}
\end{equation} --- the algebraic content of the physical remark that a
quadratic interaction rotates the ellipse at half the rate of the pump.
Combining Eq.~\eqref{od:eq:half} with $\cosh r\pm\sinh r=e^{\pm r}$, in the
broadband limit (H7) with integrated gain $r=\varepsilon_2 t_{\rm int}$
and escape efficiency $\eta_{\rm esc}=e^{-\kappa_s t_{\rm int}}$,
\begin{equation}
\bm\Phi \;\longrightarrow\; \sqrt{\eta_{\rm esc}}\;
R\!\big(\tfrac{\theta_k}{2}\big)\,
\mathrm{diag}(e^{+r},e^{-r})\,
R\!\big(\tfrac{\theta_k}{2}\big)^{\!\top},
\label{od:eq:collapse}
\end{equation}. Main-text Eq.~(3) writes the same object with
$Z(r)=\mathrm{diag}(e^{-r},e^{+r})$ --- the \emph{opposite} orientation.
The two differ by a constant $\pi$ in the definition of $\theta$:
\begin{equation}
R\!\Big(\tfrac{\theta+\pi}{2}\Big)\mathrm{diag}(e^{r},e^{-r})
R\!\Big(\tfrac{\theta+\pi}{2}\Big)^{\!\top}
=R\!\Big(\tfrac{\theta}{2}\Big)Z(r)R\!\Big(\tfrac{\theta}{2}\Big)^{\!\top},
\label{od:eq:pishift}
\end{equation}. \textbf{C5.} With $H_s$ as written in Eq.~\eqref{od:eq:Hs} and
$\varepsilon_2>0$, the angle appearing in main-text Eq.~(3) is the pump
phase \emph{plus} $\pi$. Since $\theta_k=\theta_{\rm mask}(k\bmod
N')+\beta s_n$ and $\theta_{\rm mask}$ is a free pseudorandom mask, a
constant $\pi$ is absorbed into the mask and commutes with the
data term; the map $s_n\mapsto\theta_k$ is unchanged in its
data-dependence. Every twin implementation uses the main-text orientation
consistently, so \emph{no reported number changes}. What is required is
a one-sentence convention statement in SI Sec.~S1 fixing the origin of
$\theta$, so that Eqs.~(S1)/(S8) and Eq.~(3) agree on what $\theta$
means.

\subsubsection{Passive blocks and loss}

Rotations and beamsplitters are symplectic,
$\bm S\bm\Omega\bm S^\top=\bm\Omega$. Loss is written as
a beamsplitter against a vacuum ancilla, so the operator relation is
unitary before tracing; the reduced channel
$\bm S=\sqrt\eta\,\mathds{1}$, $\bm{\mathcal N}=\tfrac{1-\eta}{4}\mathds{1}$
is completely positive, since
\begin{equation}
\bm{\mathcal N}+\tfrac{i}{4}\bm\Omega
-\bm S\big(\tfrac{i}{4}\bm\Omega\big)\bm S^\top
=\tfrac{1-\eta}{4}\big(\mathds{1}+i\bm\Omega\big)\succeq0,
\qquad \mathrm{spec}=\Big\{0,\ \tfrac{1-\eta}{2}\Big\},
\label{od:eq:cp}
\end{equation}. The zero eigenvalue is the statement that the vacuum
saturates the uncertainty relation.

\subsubsection{From operators to the moment equations}
\label{od:sec:lyap}

Writing $\bm\Sigma(t)=\bm\Phi\bm\Sigma(0)\bm\Phi^\top+\bm{\mathcal N}(t)$ with
$\bm{\mathcal N}(t)=\int_0^t \bm\Phi(w)\bm D\bm\Phi(w)^\top dw$, $w=t-s$: the
homogeneous part obeys
$\tfrac{d}{dt}(\bm\Phi\bm\Sigma_0\bm\Phi^\top)=\bm A_s(\cdot)+(\cdot)\bm
A_s^\top$, and the particular part follows by Leibniz from
$\dot{\bm\Phi}=\bm A_s\bm\Phi$ with boundary term $\bm\Phi(0)\bm D\bm
M(0)^\top=\bm D$. Hence
\begin{equation}
\dot{\bm\Sigma}=\bm A_s\bm\Sigma+\bm\Sigma\bm A_s^\top+\bm D_s,
\qquad \bm D_s=\kappa_s\bm\Sigma_{\rm vac}=\tfrac{\kappa_s}{4}\mathds{1},
\label{od:eq:lyap}
\end{equation}
recovering SI Eq.~(S7). Two consistency facts: with $\varepsilon_2=0$
the vacuum is stationary,
$\tfrac14(\bm A_s+\bm A_s^\top)+\bm D_s=0$ --- a
fluctuation--dissipation check that fixes
$\bm D_s=\tfrac{\kappa_s}{4}\mathds{1}$ uniquely in the convention
\eqref{od:eq:conv} --- and the accumulated noise at $\varepsilon_2=0$ is
$\tfrac14(1-e^{-\kappa_s t})\mathds{1}$, i.e.\ exactly the loss channel.

\subsubsection{The $\chitwo$ segment}
\label{od:sec:chi2}

Integrating SI Eq.~(S4) perturbatively (Dyson, first order in
$\mu=\tilde\kappa L_{\rm eff}$, H5--H6),
\begin{equation}
b_{\rm out}=b_{\rm in}+i\mu\,a_{\rm in}^2+O(\mu^2),\qquad
a_{\rm out}=a_{\rm in}+2i\mu\,a_{\rm in}^\dag b_{\rm in},
\label{od:eq:kick}
\end{equation}
which is SI Eq.~(S5). Taking expectations on a zero-mean state (H11) and
using the operator identity
\begin{equation}
a^2=X^2-P^2+i(XP+PX)
\ \Longrightarrow\
\avg{a^2}=(\Sigma_{XX}-\Sigma_{PP})+2i\,\Sigma_{XP},
\label{od:eq:readoff}
\end{equation} gives main-text Eq.~(4): a linear read-off of three
covariance entries. The cross terms arrive already symmetrized, which is
why no ordering ambiguity enters. The $\tfrac14$ prefactor carried by
main-text Eq.~(4) is \emph{absent} here: it has been absorbed by
Eq.~\eqref{od:eq:rescale}, $\bm\Sigma$ itself now being four times smaller.
The physical content --- $\avg{a^2}$ is a linear functional of three
covariance entries, vanishing on vacuum --- is identical. (The sign difference between
Eq.~\eqref{od:eq:kick} and main-text Eq.~(4) is not an inconsistency: SI
Sec.~S1 states that ``the sign of $\mu$ is a phase convention, fixed in
the main text by the readout phase.'')

Two structural properties of Eq.~\eqref{od:eq:kick} deserve statement
because they are exact, not perturbative. First, Manley--Rowe:
\begin{equation}
\partial_z\big(n_a+2n_b\big)=0,
\label{od:eq:MR}
\end{equation}, one harmonic photon per two fundamental photons. Second, the
kick is canonical to first order,
\begin{equation}
[a_{\rm out},a_{\rm out}^\dag]=1+4\mu^2(n_a-n_b)+O(\mu^3),
\label{od:eq:kickcomm}
\end{equation}. The $O(\mu^2)$ residue is precisely the
quantity the structural-integrity gauge $\varepsilon_G$ monitors: it is
the leading departure from a Gaussian symplectic description, and it is
$n$-dependent, which is why $\varepsilon_G$ must be tracked per session
rather than bounded once.

\subsubsection{Delay closure}
\label{od:sec:delay}

The ring writes into the bin $A_h$ at the head pointer
[Eq.~\eqref{od:eq:pointer}]; advancing $h\to
h+1\ (\mathrm{mod}\ N)$ once per bin period means the bin written at
step $k$ is next addressed at step $k+N$, i.e.\ after $N\Delta=\tau$
(H9). In the simulations of record each bin experiences the
distributed loss $\eta_\ell=\eta_L^{1/N}$ \emph{once per circulation}
(round-trip transmission $\eta_L^{1/N}$ per bin, accruing to $\eta_L$
over $N$ circulations); the more pessimistic reading---$\eta_L$ per
round trip, the hardware target of the loss budget---lowers memory
capacity materially, and capacity figures are labeled with the
implemented convention. Under the \emph{hardware-target} convention the
field re-entering the ring at time $t$ is
the loop-processed forward output emitted at $t-\tau$, attenuated by
$\eta_L$ --- which is SI Eq.~(S3) with no approximation; the twin of
record distributes that same round-trip transmission as
$\eta_\ell=\eta_L^{1/N}$ per bin step, and the delay closure is
identical under either reading. ``Exact'' here
means: the reduced dynamics of $a_s$ obtained by tracing the enlarged
register equals the solution of the delayed Langevin system, for every
$\tau$, not merely for $\tau$ small.

\subsubsection{Composition and the stability bound}

One bin step is the ordered product B1--B6 (SI App.~S2); converted to
quadratures it is $\bm\Sigma\to S_k\bm\Sigma S_k^\top+\mathcal N_k$,
main-text Eq.~(3). The map is completely positive and trace preserving: the
Gaussian condition is
$\mathcal N_k+\tfrac{i}{2}\Omega-\tfrac{i}{2}S_k\Omega S_k^\top\succeq0$,
which for symplectic $S_k$ reduces to $\mathcal N_k\succeq0$, satisfied
because every $\mathcal N_k$ here is a physical loss or reset channel. Over one mask period $S=\prod_{k=1}^{N'}S_k$.

The squeeze block $R(\theta/2)Z(r)R(\theta/2)^\top$ is symmetric positive
definite with spectral norm $e^{r}$, so with amplitude loss $\sqrt\eta$
each factor has $\|S_k\|_2=e^{r}\sqrt\eta$ \emph{independently of
$\theta_k$}. Submultiplicativity of the operator norm then
gives
\begin{equation}
\rho\Big(\prod_{k=1}^{N'}S_k\Big)^{1/N'}\;\le\;e^{r}\sqrt\eta\;\equiv\;G .
\label{od:eq:Gbound}
\end{equation}
\emph{Quantity type (load-bearing).} $G$ is a \emph{supremum} over mask
realizations, not a typical or mean value. It is attained only when all
oriented squeezes share an axis; two orthogonal orientations compose to
spectral radius exactly $1$ against $e^{2r}$ for the aligned pair. Since $\theta_{\rm mask}$ randomizes the axis, the realized
contraction is strictly stronger than $G$, which is the precise sense in
which Table~I's campaign point is conservative.

Given $G<1$ (H12), the round-trip series converges and the memory depth
follows from the geometric sum,
\begin{equation}
M\simeq\sum_{m\ge0}G^m=(1-G)^{-1},
\label{od:eq:depth}
\end{equation}
convergent iff $G<1$, divergent at $G=1$ --- the echo-state boundary.

\emph{Why the answer has this shape.} The loop is a contraction composed
with a data-steered rotation-conjugated stretch. The stretch sets how
much a perturbation is amplified per pass; the loss sets how much is
forgotten. Their ratio is the single dial $G$, and everything about
memory depth follows from a geometric series in that dial --- which is
why one number, not a spectrum, controls the reservoir's temporal reach.

\setcounter{equation}{\value{odsaveeq}}\renewcommand{\theequation}{S\arabic{equation}}
\setcounter{table}{\value{odsavetab}}\renewcommand{\thetable}{S\arabic{table}}

\begin{figure}[!t]
\includegraphics[width=0.9\textwidth]{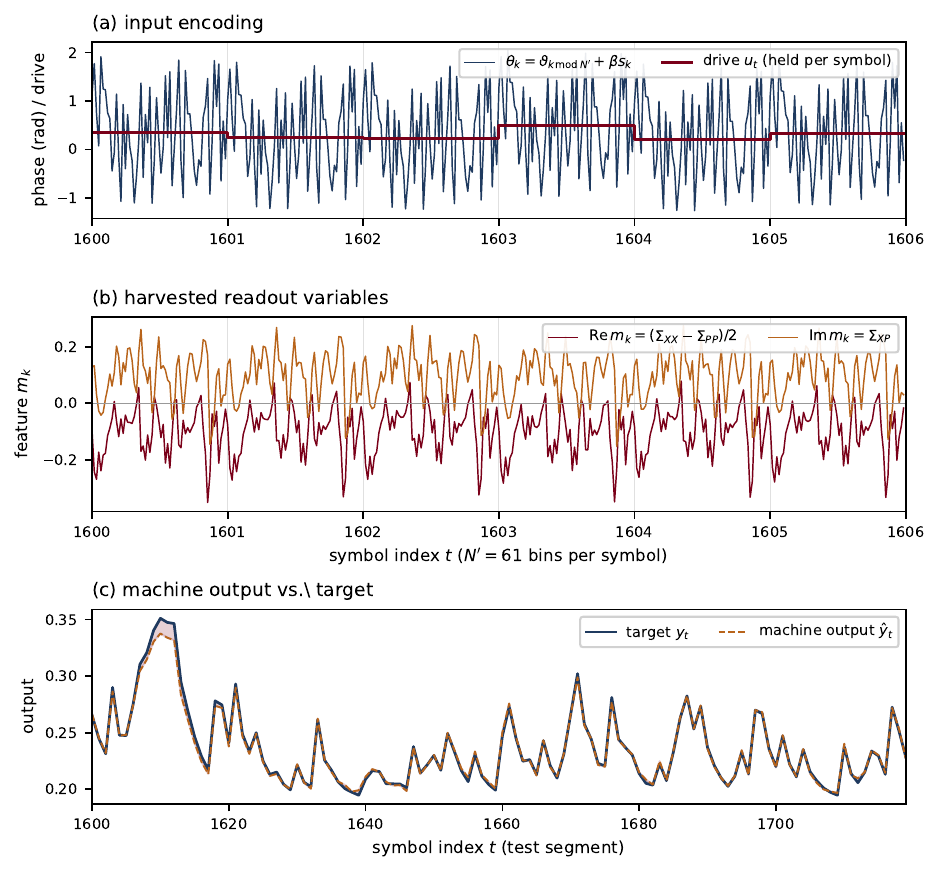}
\caption{Anatomy of one NARMA2 session (noiseless rung, linear-feature
protocol). (a)~Per-bin encoding phase over six test symbols: the frozen
mask orbit repeats each symbol and the held drive $u_t$ (step trace)
enters as a rigid offset. (b)~Harvested head-bin covariance feature
$m_k$ over the same window. (c)~Machine output $\hat y_t$ (dashed)
vs.\ target $y_t$ (solid) over the first $120$ test symbols; the shaded
band is the residual.}
\label{fig:si-session}
\end{figure}

\begin{figure}[!t]
\includegraphics[width=0.95\textwidth]{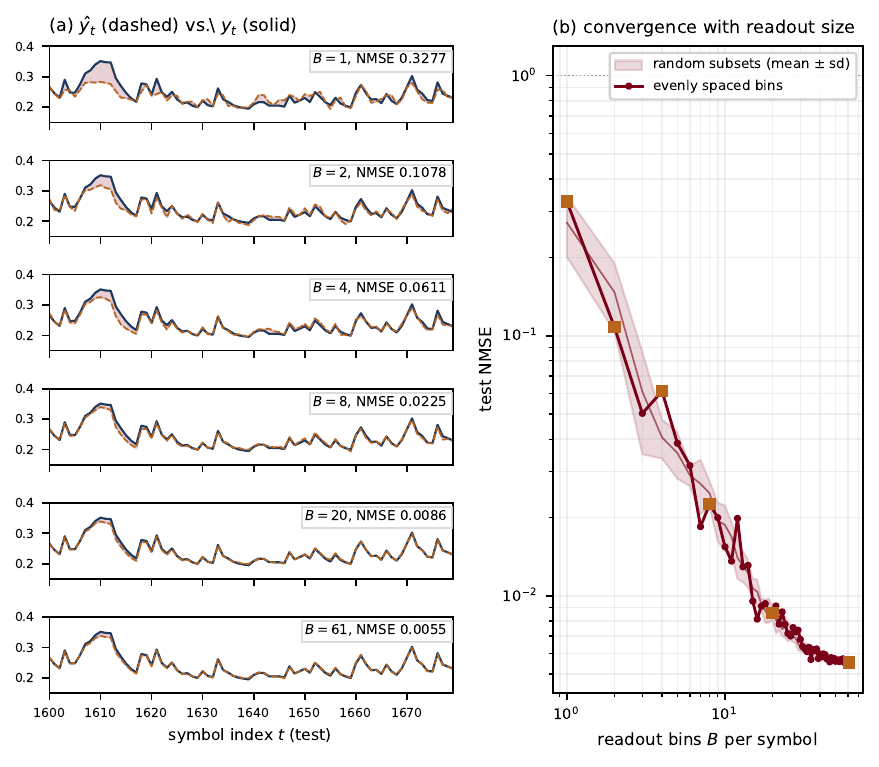}
\caption{Convergence of the session of Fig.~\ref{fig:si-session} with
readout size. (a)~Test-segment traces of $\hat y_t$ vs.\ $y_t$ as the
number of retained harvest bins $B_h$ grows (evenly spaced selection;
NMSE inset per panel). (b)~Test NMSE vs.\ $B_h$: canonical evenly spaced
selection (line with markers; squares mark the traces at left) and the
mean~$\pm$~sd band over ten random $B_h$-subsets per point. The dotted
line marks the mean predictor (NMSE $=1$).}
\label{fig:si-convergence}
\end{figure}

Figures~\ref{fig:si-session} and~\ref{fig:si-convergence} show the
machine of this section in operation on one NARMA2 session, as an
orientation before the formal development: the per-bin encoding phase
carrying the frozen mask orbit and the data offset, the harvested
covariance feature it produces, the ridge readout tracking the target,
and the convergence of test error as harvest bins are added to the
readout. The session-level discussion accompanies them in
Secs.~\ref{sec:si-anatomy} and~\ref{sec:si-readout-convergence}.

\section{Symplectic blocks of the per-bin map}
\label{app:blocks}

{\sloppy Covariances in this section, and in the twin code that
implements it, are quoted in vacuum units ($\bm\sigma=\bm\Sigma/\tfrac14$;
vacuum $=\mathds{1}$), so loss injects $(1-\eta)\mathds{1}$; symplectic
blocks are normalization-invariant. One bin step is the ordered composition B1--B6 acting on the register
$(a_s\ \text{merged into the chip slot } A_h,\ b,\ A_1..A_N,\ v)$; all blocks are
listed with their $\bm m,\bm\Sigma$ action. Elementary matrices:
$R(\phi)=\big(\begin{smallmatrix}\cos\phi&-\sin\phi\\ \sin\phi&\cos\phi
\end{smallmatrix}\big)$, $Z(r)=\mathrm{diag}(e^{-r},e^{r})$,
$BS(\eta)=\big(\begin{smallmatrix}\sqrt\eta\,\mathds{1}_2&\sqrt{1-\eta}\,\mathds{1}_2\\-\sqrt{1-\eta}\,\mathds{1}_2&\sqrt\eta\,\mathds{1}_2\end{smallmatrix}\big)$
on a mode pair; loss $\mathcal L_\eta$: $\bm m\to\sqrt\eta\,\bm m$,
$\bm\Sigma\to\eta\bm\Sigma+(1-\eta)\mathds{1}$ on the addressed mode.\par}

\begin{enumerate}
\item[B1] Feedback MZ and squeezer:
$BS(\eta_{\rm fb})_{(A_h,v)}$, then
$S_{\rm sq}(r,\theta_k)=R(\theta_k/2)Z(r)R(\theta_k/2)^{\top}$ on $A_h$, then
$\mathcal L_{\eta_{\rm esc}}$ on $A_h$ (intrinsic squeezer loss). The input enters
only through $\theta_k=\theta_{\rm mask}(k\bmod N')+\beta s_n$.
\item[B2] One-bin interferometer:
$BS(\tfrac12)\big[R(\varphi_{\rm arm})\oplus\mathds{1}\big]BS(\tfrac12)$ on
$(A_h,A_{h-1})$.
\item[B3] $\chitwo$ segment [Eq.~\eqref{eq:kick}]: order $\mu^0$ identity;
order $\mu^1$ mean source on $b$,
$\Delta\bm m_b=\mu\big(2\Sigma_{X_hP_h},\ -(\Sigma_{X_hX_h}-\Sigma_{P_hP_h})
\big)^{\top}$ (plus $-i\mu\avg{a}^2$ and the linear $a\!\leftrightarrow\!b$ mean
coupling when the displacement toggle is on); order $\mu^2$ covariance
back-action, propagated into $\varepsilon_G$.
\item[B4] Output chain on $b$: WDM loss $\mathcal L_{\eta_{\rm wdm}}$; PSA
$Z(-r_{\rm psa})$ aligned to the read quadrature; detector loss
$\mathcal L_{\eta_{\rm det}}$; feature harvest
$f=m_{X_b^\vartheta}+\zeta$, $\zeta\sim\mathcal N(0,\Sigma_{X_bX_b}^\vartheta)$,
$\vartheta$ interleaved; then reset of $b$ toward vacuum at rate
$\kappa_b\Delta\gg1$.
\item[B5] Loop: $\mathcal L_{\eta_\ell}$ on the outgoing slot
($\eta_\ell=\eta_L^{1/N}$), head advance $h\to h+1\pmod N$.
\item[B6] Ancilla reset: trace and re-vacuum $v$.
\end{enumerate}

The composite over one mask period, $S=\prod_{k=1}^{N'}S_k$ with accumulated noise
$\bm{\mathcal N}$, feeds Tier 1: stability $\rho(S)<1$; stationary state
$\bm\Sigma^\star=S\bm\Sigma^\star S^{\top}+\bm{\mathcal N}$; two-time covariance
${\rm Cov}(\bm\xi_{t+m},\bm\xi_t)=S^m\bm\Sigma^\star$; log-negativity of a bin pair
from the partially transposed symplectic eigenvalue $\tilde\nu_-$ of its
$4\times4$ sub-covariance, $E_{\mathcal N}=\max(0,-\log\tilde\nu_-)$.

\medskip
\noindent\textbf{Lemma (the feedback MZ with per-bin ancilla reset is
exactly the loss channel $\mathcal L_{\eta_{\rm fb}}$).}
\emph{Let the tunable Mach--Zehnder act on the pair (head bin $A$,
ancilla $v$) as the passive coupler $BS(\eta_{\rm fb})$ above, with the
ancilla prepared in vacuum, uncorrelated with the register, and reset
each bin (block B6), and with the dumped port traced out. Then the
reduced map on $A$ is exactly the loss channel
$\mathcal L_{\eta_{\rm fb}}$: in vacuum units,
$\bm m_A\to\sqrt{\eta_{\rm fb}}\,\bm m_A$,
$\bm\Sigma_A\to\eta_{\rm fb}\bm\Sigma_A+(1-\eta_{\rm fb})\mathds 1$,
and every cross-covariance block $C$ between $A$ and the rest of the
register scales as $C\to\sqrt{\eta_{\rm fb}}\,C$ with the rest of the
register untouched. Any output phase of the coupler is absorbed by the
pilot-locked in-loop phase tuner.}

\smallskip
\emph{Proof, by two independent routes.} \emph{Route 1 (covariance
level).} Order the quadratures $(X_A,P_A,X_v,P_v)$ and append the
remaining register modes. The coupler $S=BS(\eta_{\rm fb})$ is
symplectic ($S\Omega S^{\top}=\Omega$ blockwise, from the matrix
displayed above), and the joint input covariance is block
$\mathrm{diag}(\bm\Sigma_A,\mathds 1)$ against the register with
cross block $C$ in the $A$ rows and zero in the ancilla rows (the
ancilla is fresh vacuum). Carrying out $S\,\bm\Sigma_{\rm in}
S^{\top}$ and deleting the ancilla rows and columns (the partial
trace at covariance level) gives, exactly,
\begin{equation*}
\bm\Sigma_A'=\eta_{\rm fb}\,\bm\Sigma_A+(1-\eta_{\rm fb})\,\mathds 1,
\qquad
C'=\sqrt{\eta_{\rm fb}}\,C,
\qquad
\bm m_A'=\sqrt{\eta_{\rm fb}}\,\bm m_A,
\end{equation*}
the register block unchanged. \emph{Route 2 (characteristic-function
level; all moments, arbitrary input state).} With $U$ the coupler
unitary and $D_A(\bm\xi)=e^{i\bm\xi^{\top}\Omega\bm R_A}$ the Weyl
operator, the linear Heisenberg action
$U^{\dagger}\bm R_AU=\sqrt{\eta_{\rm fb}}\,\bm R_A
+\sqrt{1-\eta_{\rm fb}}\,\bm R_v$ lifts to Weyl operators without
correction terms, because $A$ and $v$ quadratures commute:
\begin{equation*}
U^{\dagger}D_A(\bm\xi)U
=D_A\big(\sqrt{\eta_{\rm fb}}\,\bm\xi\big)\,
D_v\big(\sqrt{1-\eta_{\rm fb}}\,\bm\xi\big).
\end{equation*}
Evaluating the output characteristic function
$\chi_{\rm out}(\bm\xi)=\operatorname{Tr}\big[(\rho\otimes
|0\rangle\langle0|_v)\,U^{\dagger}D_A(\bm\xi)U\big]$, the trace
factorizes over the product state:
\begin{equation*}
\chi_{\rm out}(\bm\xi)
=\chi_{\rho}\big(\sqrt{\eta_{\rm fb}}\,\bm\xi\big)\;
\chi_{\rm vac}\big(\sqrt{1-\eta_{\rm fb}}\,\bm\xi\big),
\end{equation*}
which is the defining characteristic-function composition of the
Gaussian loss channel of transmissivity $\eta_{\rm fb}$, at the level
of the complete state --- Gaussian or not. The two routes share no
intermediate quantity; both import only the coupler's Heisenberg
action. Two remarks fix the scope. The per-bin reset (B6) is
load-bearing: a correlated or un-reset ancilla would return memory to
the dump port and the reduced map would no longer be a (Markovian)
loss channel. And the map is charge-neutral: it commutes with the
gauge rotation ($\mathds 1$ is rotation invariant), consistent with
register neutrality --- only the oriented squeezers carry pump
charge.\hfill$\blacksquare$

\section{Proofs of the reachability theorem}
\label{sec:si-proofs}

This section proves main-text Lemma~1, Corollary~1, Theorem~1,
Proposition~1, and Proposition~2. It is written to be read without a background in quantum
optics. Every symbol is defined at its first appearance; each proof is
preceded by a plain-language account of why it works; and the only
external tools are three standard ones: a perturbation (Neumann/Dyson)
series, inverse filtering of a one-pole exponential-decay filter, and the
discrete Fourier transform. A reader who knows recurrent networks and
Fourier analysis has everything needed. Numbering convention: unprefixed
environment numbers (Lemma~1, Corollary~1, Theorem~1) refer to the main
text; S-prefixed numbers (Lemma~S1, Theorem~S1, Remark~S1) refer to
environments of this Supplement.

\subsection{Reader's guide: the setting, the claim, and the notation}
\label{sec:si-guide}

\subsubsection*{The machine, stripped of optics}

Mathematically, the reduced single-loop model of main-text Sec.~IIC is a
recurrent system with a three-component hidden state and
\emph{input-steered weights}. That is the whole of it, and a reader who
holds onto that sentence can follow every proof below.

The three components are expectation values of the circulating light
field. Writing $a$ for the annihilation operator of the loop mode and
$\avg{\cdot}$ for the quantum expectation value in the current state, the
state vector at time step $t$ is
\begin{equation}
v_t=(m_t,\ \bar m_t,\ J_t),\qquad
m_t:=\avg{a^2}_t,\qquad
J_t:=\avg{a^\dagger a}_t+\tfrac12 .
\label{eq:si-state}
\end{equation}
For the purposes of the proofs, the reader may forget where these numbers
come from and retain only their roles. $m_t$ is a single
\emph{complex-valued feature per time step}, and it is the only quantity
the detector reads; main-text Eq.~(4) shows that the measured signal is
proportional to it. $\bar m_t$ is its complex conjugate, carried along so
that the update is linear. $J_t$ is a real, energy-like channel, the mean
photon number plus the vacuum's half quantum, which is never read
directly but which continually feeds the other two. The vacuum state,
meaning the machine with no input history, has $(m,J)=(0,\tfrac12)$.

Inputs are one-sided real sequences $s=(s_t)_{t\le0}$ with
$s_t\in[-s_{\max},s_{\max}]$; the space of all such histories,
$\mathcal U=[-s_{\max},s_{\max}]^{\mathbb Z_{\le 0}}$, is compact in the
product topology, and all function-approximation statements below are
uniform statements on the space $C(\mathcal U)$ of continuous functions
of the input history. Each input symbol enters the dynamics through a
single angle,
\begin{equation}
\theta_t=\varphi_t+\beta s_t ,
\label{eq:si-angle}
\end{equation}
where $\varphi_t$ is a fixed, input-independent \emph{mask} phase --- the per-slot value of the hardware mask $\theta_{\rm mask}$ --- (a
per-step constant chosen by the experimenter, playing the role of a
positional encoding) and $\beta>0$ is the encoding gain. The state
update is the exact covariance recursion of main-text Eq.~(6),
\begin{equation}
v_t=T(\theta_t)\,v_{t-1}+b_{\rm vac},\qquad
b_{\rm vac}=\big(0,\,0,\,\tfrac{1-\eta}{2}\big)^{\!\top},
\label{eq:si-recursion}
\end{equation}
with $T(\theta)$ the $3\times3$ matrix displayed in main-text Eq.~(6),
built from two hardware constants: the per-step squeeze strength $r>0$
(how strongly the parametric amplifier acts each step) and the per-step
transmission $\eta\in(0,1)$ (the fraction of the field surviving one
circulation; the constant vector $b_{\rm vac}$ is the vacuum noise injected in
place of what is lost). The essential structural point is that the input
never enters additively: $s_t$ appears only \emph{inside the weight
matrix}, through the angle $\theta_t$. The machine is a linear recurrence
whose weights are rotated by the data --- the continuous-variable
analogue of a recurrent network with input-dependent weights --- and this
is the sole source of its nonlinearity in the input history ---
unrolling the recurrence produces products of $T(\theta_t)$ across time
steps, hence products of trigonometric functions of distinct inputs,
which no additive drive could generate.

Two standing hardware assumptions are in force throughout:
the \emph{stability guard} $\bar\rho:=\eta e^{2r}<1$ (gain does not
outrun loss; this will be shown to make $T$ a contraction) and
\emph{encoding injectivity} $0<\beta s_{\max}<\pi$ (two different input
values never encode to the same angle configuration).

\subsubsection*{The question, and the basis in which it is answered}

Train a readout on the features: collect the values
$m_{t_1},\dots,m_{t_k}$ (and their conjugates) at finitely many readout
slots, possibly across a finite ensemble of operating settings (different
masks $\varphi$ and guard-compatible squeeze strengths $r$, the standard
input-mask multiplexing of delay-line reservoir computing, at fixed
$\beta,\eta$), and apply a complex polynomial $p$ of polynomial order (total degree) at
most $D$. The \emph{reachable class} $\mathcal A_D$ is the closure, in the
uniform norm on $C(\mathcal U)$, of every functional of the input history
obtainable this way. The theorem characterizes $\mathcal A_D$ exactly.

The natural basis for the answer is the Fourier basis on input
histories. For each finitely supported integer pattern
$n:\mathbb Z_{\le0}\to\mathbb Z$ (i.e., $n_t\in\mathbb Z$ with only
finitely many $n_t\ne0$), define the \emph{character}
\begin{equation}
E_n(s)=\prod_{t}e^{i\beta n_t s_t},
\label{eq:si-character}
\end{equation}
a product of pure oscillations, one per engaged slot, with $n_t$ the
integer frequency at slot $t$. Two integers summarize a pattern: its
\emph{charge} $q(n)=\sum_t n_t$ (the net winding number: shift all inputs
together by $\tau$ and $E_n$ responds by the phase $e^{iq(n)\beta\tau}$,
absorbing $q(n)$ units of the applied shift---the shift is continuous, the
charge is the integer multiplying it) and its
\emph{weight} $\|n\|_1=\sum_t|n_t|$ (the total frequency content in the
inputs). A convention that holds throughout this section: every
``frequency'' is a winding number with respect to a phase angle---the
common shift $\chi$, the pump phase $\theta$, or a single input value
$s_t$---and never a frequency in time. We
write $\mathcal V_q$ for the closed span of all charge-$q$ characters.
In this language, main-text Theorem~1 says:
\begin{itemize}
\item[\textbf{(I)}] $\mathcal A_D$ is exactly the closed span of the
\emph{$D$-assembleable} characters, $n\in\mathfrak N_D$: those splitting
into at most $D$ admissible or co-admissible blocks (the patterns of
main-text Corollary~1 and their negatives). In particular order-$D$ polynomial
readout is confined to the Fourier sectors with net winding number
(charge) at most $D$,
$\mathcal A_D\subseteq\overline{\operatorname{span}}
\bigcup_{|q|\le D}\mathcal V_q$, and the containment is strict
(Sec.~\ref{sec:si-prfI}).
\item[\textbf{(II)}] The union over all $D$ is dense in
$C(\mathcal U)$: the machine plus polynomial post-processing is a
universal approximator of fading-memory functionals.
\item[\textbf{(III)}] The exclusion in (I) is quantitative: any single
character of charge $D{+}1$ stays at least the explicit distance
$\delta(D,\beta s_{\max})>0$ of main-text Eq.~(10) away from
\emph{everything} polynomial order $D$ can build --- an
inapproximability margin,
not merely a non-membership statement.
\end{itemize}

\subsubsection*{The conservation law that runs all three proofs}

One symmetry does all the work, and it is worth dwelling on, because
everything else in this section is bookkeeping around it.

Shift every input symbol by the same constant, $s\mapsto s+\tau\mathbf 1$.
By Eq.~\eqref{eq:si-angle} every angle then shifts by the same amount,
$\theta_t\mapsto\theta_t+\chi$ with $\chi=\beta\tau$. Lemma~1(a), proved
below, states that the dynamics responds \emph{exactly} by a phase
rotation of the state: $m_t\mapsto e^{i\chi}m_t$,
$\bar m_t\mapsto e^{-i\chi}\bar m_t$, and $J_t\mapsto J_t$. In other
words the readable feature carries charge $+1$, its conjugate charge
$-1$, and the energy channel charge $0$; and these assignments are
conserved by the exact dynamics, loss and all.

Why should that be so? The physical picture is pump-phase bookkeeping. A
squeezer does not create photons singly; it creates and destroys them in
pairs, drawing each pair from the pump laser, whose phase is the common
reference. The readable harmonic signal is generated by $a^2$, which
consumes two signal photons, that is, exactly one pump quantum. So $a^2$
carries precisely one net unit of the reference, while $a^\dagger a$
consumes and returns one and therefore carries none. Loss, beamsplitters
and delays never touch the pump, so they cannot alter anyone's count.

Now compare the two sides. A character $E_n$ responds to the same shift
with the phase $e^{iq(n)\chi}$. Matching that response against the
$e^{i\chi}$ response of $m_t$ forces every character appearing in the
expansion of $m_t$ to have charge exactly $+1$. This is the selection
rule.

From the selection rule the three clauses follow by three separate
routes, in increasing order of machinery.
\begin{itemize}
\item \textbf{Clause (III)} uses the symmetry and nothing else about the
machine. The shift direction slices every order-$\le D$ polynomial functional
into
at most $2D{+}1$ frequencies; a charge-$(D{+}1)$ target oscillates at a
frequency outside that band; and a classical polynomial-interpolation
argument turns ``outside the band'' into the explicit distance $\delta$.
(Sec.~\ref{sec:si-prfIII}.)
\item \textbf{Clause (II)} works at leading order in the input coupling.
The weakest-order content of $m_t$ is a known exponential-decay filter of
the sequence $e^{i\theta_t}$. Inverting that filter, by a two-tap
deconvolution, recovers each single-slot oscillation $e^{i\beta s_j}$ up
to a controlled error; products of these approximate every character; and
density of trigonometric polynomials (Stone--Weierstrass) finishes the
argument. (Sec.~\ref{sec:si-prfII}.)
\item \textbf{Clause (I)} is constructive, and exact at every order. A
designed sweep of the mask over $5^L$ settings makes every candidate
pattern announce its in-window frequency content as a distinct tone of a
discrete Fourier transform. Because the per-slot frequencies are capped
at $|n_t|\le 2$ (Corollary~1), base-$5$ digits suffice and the tones
never collide, so a DFT projects out any prescribed pattern exactly, at
every perturbative order simultaneously. (Sec.~\ref{sec:si-prfI}.)
\end{itemize}

\begin{figure*}[!htb]
\centering
\includegraphics[width=\textwidth]{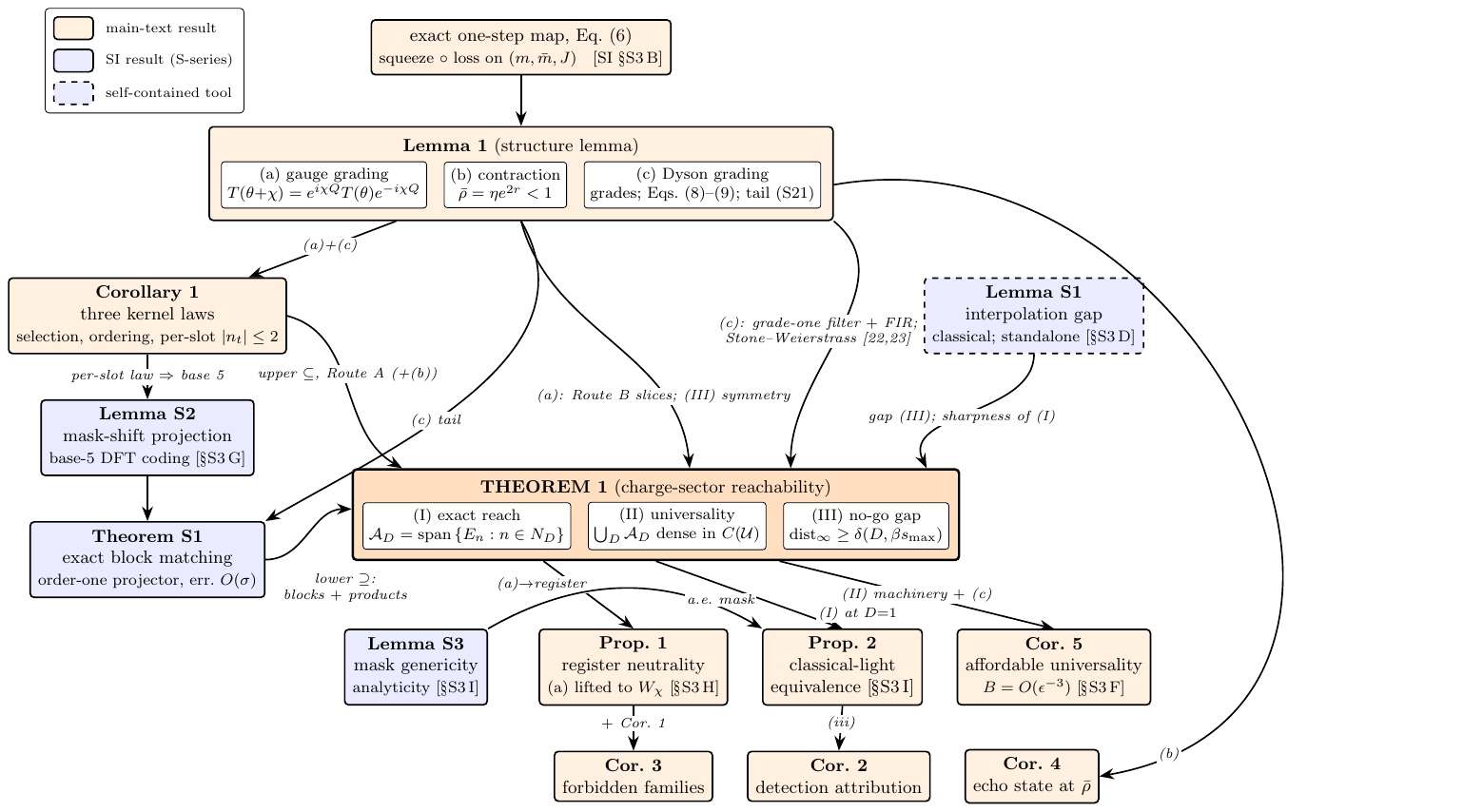}
\caption{\textbf{Proof architecture of main-text Theorem~1.}
Dependency graph of the results proved in this section (blue,
S-series) and the main-text statements they establish (orange). Each
arrow is labeled by the specific clause it imports. Clause (III)
consumes only the gauge grading, Lemma~1(a), plus the standalone
interpolation gap (Lemma~S1); clause (II) consumes only the Dyson
grading, Lemma~1(c); clause (I) consumes both, via the two
upper-inclusion routes and the projector pair Lemma~S2 $\to$
Theorem~S1. Lemma~S3 enters only through the genericity hypothesis of
Proposition~2(iii).}
\label{fig:si-proofmap}
\end{figure*}

Figure~\ref{fig:si-proofmap} charts how the results of this section
assemble the main theorem. One input is load-bearing: the gauge
grading, Lemma~1(a). Clause (III) consumes it alone, together with the
self-contained interpolation gap (Lemma~\ref{lem:gap}); clause (II)
consumes only its Dyson-grading consequence, Lemma~1(c), through the
FIR deconvolution and Stone--Weierstrass; clause (I) consumes both, via
the two upper-inclusion routes and the projector pair
Lemma~\ref{lem:projection} $\to$ Theorem~\ref{thm:matching}. The three
S-series lemmas are the self-contained technical tools ---
interpolation, DFT coding, analyticity --- each imported exactly once.
Downstream, Propositions~1--2 transport the law to the register and to
classical light, with Lemma~\ref{lem:genericity} entering only through
the genericity hypothesis of Proposition~2(iii).

Table~\ref{tab:si-notation} collects every recurring symbol with its
meaning and provenance; the reader may treat it as a glossary for this
section.

\begin{table*}[!htb]
\caption{Glossary of symbols used in the proofs
(Sec.~\ref{sec:si-proofs}). ``MT'' = main text.}
\label{tab:si-notation}
\begin{ruledtabular}
\begin{tabular}{lp{0.62\textwidth}l}
Symbol & Meaning & Defined \\
\colrule
$s=(s_t)_{t\le0}$,\ $s_{\max}$ & input history; $s_t\in[-s_{\max},s_{\max}]$ & Eq.~\eqref{eq:si-state} ff.\\
$\mathcal U$,\ $C(\mathcal U)$ & compact space of input histories; continuous functionals on it, sup norm & Sec.~\ref{sec:si-guide}\\
$a$,\ $\avg{\cdot}$ & loop-mode annihilation operator; quantum expectation value & Eq.~\eqref{eq:si-state}\\
$m_t=\avg{a^2}_t$ & the readable complex feature (one per step); carries charge $+1$ & Eq.~\eqref{eq:si-state}\\
$J_t=\avg{a^\dagger a}_t+\frac12$ & energy channel (photon number $+$ vacuum half); charge $0$; vacuum value $\frac12$ & Eq.~\eqref{eq:si-state}\\
$v_t=(m_t,\bar m_t,J_t)$ & hidden state of the recursion & Eq.~\eqref{eq:si-state}\\
$\theta_t=\varphi_t+\beta s_t$ & squeeze angle: mask phase $\varphi_t$ (fixed) $+$ encoding gain $\beta$ $\times$ input & Eq.~\eqref{eq:si-angle}\\
$r$,\ $\eta$ & per-step squeeze strength; per-step transmission (survival fraction) & Eq.~\eqref{eq:si-recursion} ff.\\
$T(\theta)$,\ $b_{\rm vac}$ & one-step transfer matrix (MT Eq.~(6)); vacuum-injection vector & Eq.~\eqref{eq:si-recursion}\\
$\bar\rho=\eta e^{2r}<1$ & contraction rate of $T$; the stability guard and fading-memory rate & Lemma~1(b)\\
$Q=\mathrm{diag}(1,-1,0)$ & charge assignment of the channels $(m,\bar m,J)$ & Eq.~\eqref{eq:si-grading}\\
$T_d$ ($|d|\le2$) & Laurent component of $T(\theta)$ at pump-phase frequency $d$ & Eq.~\eqref{eq:si-laurent}\\
$e_m=(1,0,0)^{\!\top}$ & coordinate vector selecting the readable channel & Sec.~\ref{sec:si-prf-lemma}\\
$x=\eta\cosh^2\!r$ & input-free per-step survival gain of the $m$ channel (the filter pole) & Lemma~1(a)\\
$T_0$,\ $T_{\rm od}$ & input-free diagonal part of $T$ (equal to the frequency-zero Laurent component); input-carrying off-diagonal part & Lemma~1(c)\\
$\sigma=\eta(\sinh2r+\sinh^2\!r)$ & size of the input coupling, $\||T_{\rm od}|\|_\infty$; small when $r$ is small & Lemma~1(c)\\
$\bar\rho_0=\eta\cosh 2r$ & contraction rate of the input-free part, $\||T_0|\|_\infty$ & Lemma~1(c)\\
grade $k$; engagement & order in the expansion in $T_{\rm od}$; one step at which the input acts & Lemma~1(c)\\
$J_{\rm ss}=\frac{1-\eta}{2(1-\eta\cosh2r)}$ & stationary value of $J$ under the input-free dynamics & Eq.~\eqref{eq:si-Jss}\\
$\varkappa=-\eta\sinh(2r)J_{\rm ss}$ & amplitude of a single engagement into $m$ (grade-one prefactor) & MT Eq.~(8)\\
$E_n(s)=\prod_te^{i\beta n_ts_t}$ & character (Fourier basis function); pattern $n$ has $n_t\in\mathbb Z$, finite support & Eq.~\eqref{eq:si-character}\\
$q(n)=\sum_tn_t$,\ $\|n\|_1$ & charge (winding number under a uniform shift of all inputs) and weight (total frequency) & Eq.~\eqref{eq:si-character} ff.\\
$\mathcal V_q$,\ $\mathcal A_D$ & closed span of charge-$q$ characters; reachable class at polynomial
order $D$ & Sec.~\ref{sec:si-guide}\\
$\mathfrak N_D$ & $D$-assembleable patterns: sums of $\le D$ admissible or co-admissible blocks & MT, Thm.~1(I)\\
$v_\infty=\frac{1-\eta}{2(1-\bar\rho)}$ & uniform bound on the stationary state & Lemma~1(b)\\
$a=\beta s_{\max}$,\ $\chi$,\ $\mathcal E_D$ & encoding depth; global-shift/slice variable $\chi=\beta\tau$ (one symbol, both roles); span of $e^{iq\chi}$, $|q|\le D$ & Sec.~\ref{sec:si-prfIII}\\
$g_j$ & deconvolved (FIR-filtered) order-one feature at slot $j$ & Eq.~\eqref{eq:si-fir}\\
$L$; window & number of lags addressed by a projector; the slots $j,\dots,j-L+1$ & Lemma~\ref{lem:projection}\\
$\varphi^{(k)}$,\ $\psi$,\ $R(\nu)$,\ $w_k$ & mask ray of $5^L$ epochs; its step; the digit code; DFT weights & Lemma~\ref{lem:projection}\\
$\nu=n-e_j$;\ $C_{n^*}$ & window exponents after FIR normalization; walk coefficient of pattern $n^*$ & Lemma~\ref{lem:projection}, Thm.~\ref{thm:matching}\\
$k^*$ & grade (engagement count) of the matched pattern $n^*$ & Thm.~\ref{thm:matching}\\
$\mathrm{gr}(n)$ & grade of pattern $n$: engagement count of its unique walk & Thm.~\ref{thm:matching}\\
$\zeta_j$;\ $\rho_j^{\rm res}$;\ $\Upsilon$ & per-feature detector noise; FIR residual; noise-gain abbreviation $C_{\rm stat}\sigma_{\rm eff}/(\mu_T\sqrt B)$ & Sec.~\ref{sec:si-budget}\\
$\omega$,\ $d_\omega$ & fading-memory weighting sequence and its metric & Sec.~\ref{sec:si-prfII}\\
\multicolumn{3}{l}{\emph{Notation notes.} $\epsilon$ (accuracy) vs.\ $\varepsilon_2,\varepsilon_G$ (physical); bare $\varkappa$ (walk amplitude) vs.\ $\kappa_s,\kappa_b,\tilde\kappa$ (rates); $a$ dual-use (operator / depth $\beta s_{\max}$, Sec.~\ref{sec:si-prfIII} is optics-free);}\\
\multicolumn{3}{l}{$\sigma$ (coupling) vs.\ $\bm\sigma=\bm\Sigma/\tfrac14$ (bold, vacuum units); $p$ (readout polynomial / monic $p(z)$ / pump $p(t)$, by argument); $L$ (window) vs.\ $L_{\rm eff}$, $\mathcal L_\eta$; $m$ as moment order or lag where stated.}\\
\end{tabular}
\end{ruledtabular}
\end{table*}

\subsection{Proof of the structure lemma (main-text Lemma 1)}
\label{sec:si-prf-lemma}

Throughout, $e_m=(1,0,0)^{\!\top}$, $e_{\bar m}=(0,1,0)^{\!\top}$,
$e_J=(0,0,1)^{\!\top}$ denote the coordinate vectors of the three
channels, and $\|\cdot\|_\infty$ on matrices is the maximum absolute row
sum (the operator norm on sup-normed vectors). Main-text Eq.~(6) is
obtained by composing the one-step squeeze
$a\mapsto\cosh r\,a-e^{i\theta}\sinh r\,a^\dagger$ with the loss channel
$\bm\Sigma\mapsto\eta\bm\Sigma+(1-\eta)\mathds{1}$ (vacuum units;
absolutely, the injected term is $(1-\eta)\bm\Sigma_{\rm vac}$) and
evaluating the
three moments \eqref{eq:si-state}.

Because the whole section rests on this matrix, we derive it in full.
Write $a'=\cosh r\,a-e^{i\theta}\sinh r\,a^\dagger$ for the field after
the squeeze. Squaring and taking expectations, using
$[a,a^\dagger]=1$ so that
$\avg{aa^\dagger+a^\dagger a}=2\avg{a^\dagger a}+1=2J$,
\begin{equation}
\avg{a'^2}
=\cosh^2\!r\,\avg{a^2}
-e^{i\theta}\sinh r\cosh r\,\avg{aa^\dagger+a^\dagger a}
+e^{2i\theta}\sinh^2\!r\,\avg{a^{\dagger2}}
=\cosh^2\!r\,m
+e^{2i\theta}\sinh^2\!r\,\bar m
-e^{i\theta}\sinh(2r)\,J,
\label{eq:si-derive-m}
\end{equation}
where $2\sinh r\cosh r=\sinh 2r$ collected the middle term. Similarly,
using $\avg{aa^\dagger}=\avg{a^\dagger a}+1=J+\tfrac12$ and
$\cosh^2\!r+\sinh^2\!r=\cosh 2r$,
\begin{equation}
\avg{a'^\dagger a'}+\tfrac12
=\cosh^2\!r\,\big(J-\tfrac12\big)+\sinh^2\!r\,\big(J+\tfrac12\big)
-\sinh r\cosh r\,\big(e^{-i\theta}m+e^{i\theta}\bar m\big)+\tfrac12
=\cosh(2r)\,J-\tfrac12\sinh(2r)\big(e^{-i\theta}m+e^{i\theta}\bar m\big).
\label{eq:si-derive-J}
\end{equation}
The loss channel then rescales the fluctuation moments,
$\avg{a^2}\mapsto\eta\avg{a^2}$ and
$\avg{a^\dagger a}\mapsto\eta\avg{a^\dagger a}$, so
$m\mapsto\eta m$ and
$J\mapsto\eta\big(J-\tfrac12\big)+\tfrac12=\eta J+\tfrac{1-\eta}2$:
multiplying \eqref{eq:si-derive-m}--\eqref{eq:si-derive-J} through by
$\eta$ and adding the vacuum-injection constant $\tfrac{1-\eta}2$ to the
$J$ row reproduces main-text Eq.~(6) entry by entry --- every entry of
the table below, and the source vector $b_{\rm vac}$, can be read off these two
displays.

\begin{proof}
\textbf{(a) Gauge grading.} Assign to the channels $(m,\bar m,J)$ the
charges $(+1,-1,0)$, collected in $Q=\mathrm{diag}(1,-1,0)$. The claim
\begin{equation}
T(\theta+\chi)=e^{i\chi Q}\,T(\theta)\,e^{-i\chi Q},\qquad
e^{i\chi Q}b_{\rm vac}=b_{\rm vac}
\label{eq:si-grading}
\end{equation}
is an entrywise statement: since $Q$ is diagonal with entries $q_c$,
$e^{i\chi Q}$ is diagonal with entries $e^{i\chi q_c}$, and a diagonal
conjugation acts on each entry separately,
$\big(e^{i\chi Q}\,M\,e^{-i\chi Q}\big)_{cc'}
=e^{i\chi q_c}\,M_{cc'}\,e^{-i\chi q_{c'}}
=e^{i\chi(q_c-q_{c'})}M_{cc'}$. So
\eqref{eq:si-grading} holds if and only if every entry of $T(\theta)$
depends on $\theta$ through the phase $e^{i(q_c-q_{c'})\theta}$ exactly.
Reading main-text Eq.~(6) entry by entry:
\begin{center}
\begin{tabular}{lccc}
entry $(c\leftarrow c')$ & value & $\theta$-phase & $q_c-q_{c'}$\\
\colrule
$m\leftarrow m$ & $\eta\cosh^2\!r$ & $e^{0}$ & $0$\\
$m\leftarrow\bar m$ & $\eta e^{2i\theta}\sinh^2\!r$ & $e^{+2i\theta}$ & $+2$\\
$m\leftarrow J$ & $-\eta e^{i\theta}\sinh2r$ & $e^{+i\theta}$ & $+1$\\
$J\leftarrow m$ & $-\tfrac\eta2 e^{-i\theta}\sinh2r$ & $e^{-i\theta}$ & $-1$\\
$J\leftarrow\bar m$ & $-\tfrac\eta2 e^{+i\theta}\sinh2r$ & $e^{+i\theta}$ & $+1$\\
$J\leftarrow J$ & $\eta\cosh2r$ & $e^{0}$ & $0$\\
\end{tabular}
\end{center}
with the $\bar m$ row the complex conjugate of the $m$ row (the phases
match in every case; $b_{\rm vac}$ is supported on the charge-$0$ channel, so
$e^{i\chi Q}b_{\rm vac}=b_{\rm vac}$). The physical reading of the phase column: each entry
transfers content between channels while exchanging photon pairs with the
pump, and its phase factor $e^{id\theta}$ counts the net number $d$ of
pump-phase units absorbed ($d>0$) or returned ($d<0$) in that transfer.
Collecting entries by this frequency defines the Laurent decomposition
\begin{equation}
T(\theta)=\sum_{|d|\le2}T_d\,e^{id\theta},\qquad [Q,T_d]=d\,T_d.
\label{eq:si-laurent}
\end{equation}
Explicitly, sorting the entries of main-text Eq.~(6) by frequency,
\begin{align}
T_0&=\eta\,\mathrm{diag}\!\big(\cosh^2\!r,\ \cosh^2\!r,\ \cosh2r\big),
\nonumber\\
T_{+1}&=\begin{pmatrix}0&0&-\eta\sinh2r\\ 0&0&0\\ 0&-\tfrac\eta2\sinh2r&0\end{pmatrix},
\qquad
T_{-1}=\begin{pmatrix}0&0&0\\ 0&0&-\eta\sinh2r\\ -\tfrac\eta2\sinh2r&0&0\end{pmatrix},
\nonumber\\
T_{+2}&=\begin{pmatrix}0&\eta\sinh^2\!r&0\\ 0&0&0\\ 0&0&0\end{pmatrix},
\qquad\qquad\;\;
T_{-2}=\begin{pmatrix}0&0&0\\ \eta\sinh^2\!r&0&0\\ 0&0&0\end{pmatrix},
\qquad
T_d=0\ \ (|d|\ge3),
\label{eq:si-laurent-explicit}
\end{align}
in the channel basis $(m,\bar m,J)$; the grading $[Q,T_d]=d\,T_d$ is
checkable entrywise, since each nonzero entry sits at a position with
charge difference $q_c-q_{c'}=d$. The table shows directly that (i) no entry carries $|d|\ge3$ --- one
squeeze exchanges at most two pump-phase units per step, which is the
per-slot law in embryo; (ii) the readable row is
\emph{holomorphic in the pump phase}: the three $m$-row entries carry
frequencies $0,+2,+1$ and never a negative one, so
$e_m^\dagger T_{-1}=e_m^\dagger T_{-2}=0$; and (iii) the frequency-$0$
part of the readable row is pure survival,
$e_m^\dagger T_0=x\,e_m^\dagger$ with $x:=\eta\cosh^2\!r$ --- with no
input engagement, the feature just decays geometrically by the factor
$x$ per step. Equivalently, in one identity:
$T(\theta)=e^{i\theta Q}\,T(0)\,e^{-i\theta Q}$ [set $\theta\to0$,
$\chi\to\theta$ in \eqref{eq:si-grading}] --- squeezing at pump angle
$\theta$ is squeezing at angle zero conjugated by the pump-frame
rotation, and loss commutes with that rotation, so the \emph{entire}
$\theta$-dependence of the dynamics is a frame rotation; the Laurent
support $|d|\le2$ then just reads off the charge differences
$|q_c-q_{c'}|\le2$ available among three channels of charges
$(+1,-1,0)$ under one pair exchange.

\smallskip
\textbf{(b) Contraction.} Sum the entries along each row of the
entrywise absolute value $|T(\theta)|$; the phases drop out. Row $m$ (and its conjugate row):
$\eta(\cosh^2\!r+\sinh^2\!r+\sinh2r)=\eta(\cosh2r+\sinh2r)=\eta e^{2r}$.
Row $J$:
$\eta(\tfrac12\sinh2r+\tfrac12\sinh2r+\cosh2r)=\eta e^{2r}$.
Hence $\||T(\theta)|\|_\infty=\bar\rho:=\eta e^{2r}$ for every $\theta$
--- the guard $\bar\rho<1$ makes every step a strict contraction in the
sup norm, uniformly in the input. Three standard consequences follow,
and each is one line once the recursion is unrolled. Iterating
\eqref{eq:si-recursion} backward from the infinite past gives
\begin{equation}
v_t=b_{\rm vac}+\sum_{k\ge1}\Big[\prod_{j=0}^{k-1}T(\theta_{t-j})\Big]b_{\rm vac},
\label{eq:si-unrolled}
\end{equation}
the leading term being the injection with zero propagation steps
(equivalently, the $k=0$ term of the unsplit series under the
empty-product convention $\prod_{j=0}^{-1}:=\mathds 1$). \emph{Existence:} the $k$-th term
is a product of $k$ matrices each of norm $\le\bar\rho$ applied to $b_{\rm vac}$,
hence of size $\le\bar\rho^{\,k}\|b_{\rm vac}\|_\infty$, and the geometric series
converges absolutely; summing it bounds the stationary state by
$v_\infty=\|b_{\rm vac}\|_\infty/(1-\bar\rho)=\tfrac{1-\eta}{2(1-\bar\rho)}$.
\emph{Uniqueness:} two bounded solutions of \eqref{eq:si-recursion}
driven by the same input differ, after $t'$ steps, by a product of $t'$
transfer matrices applied to their initial difference, which is
$\le\bar\rho^{\,t'}\times\mathrm{const}\to0$; so the difference is zero.
\emph{Fading memory:} changing the input at lag $\ell$ changes only the
matrices $T(\theta_{t-j})$ with $j\ge\ell$, which enter only the terms
$k>\ell$ of \eqref{eq:si-unrolled}; each such term is bounded by
$\bar\rho^{\,k}\|b_{\rm vac}\|_\infty$ before and after the change, so
$\|\Delta v_t\|_\infty\le2\sum_{k>\ell}\bar\rho^{\,k}\|b_{\rm vac}\|_\infty
=O(\bar\rho^{\,\ell})$ --- the echo-state property, at the hardware rate
$\bar\rho$, independent of the input statistics.

\smallskip
\textbf{(c) Dyson grading.} Split the transfer matrix into its
input-free and input-carrying parts,
\begin{equation}
T=T_0+T_{\rm od},\qquad
T_0=\eta\,\mathrm{diag}(\cosh^2\!r,\,\cosh^2\!r,\,\cosh2r),
\label{eq:si-split}
\end{equation}
so that $T_0$ (the diagonal --- exactly the frequency-zero Laurent
component of Eq.~\eqref{eq:si-laurent}) contains no $\theta$ and all input
dependence sits in the off-diagonal $T_{\rm od}$, of size
$\||T_{\rm od}|\|_\infty=\sigma:=\eta(\sinh2r+\sinh^2\!r)$; note
$\sigma\to0$ as $r\to0$, so $\sigma$ is the small parameter of a
weak-coupling expansion. Substitute \eqref{eq:si-split} into each
$n$-fold product of the unrolled recursion \eqref{eq:si-unrolled} and
distribute: the product becomes a sum of $2^n$ \emph{words} in the two
letters $T_0$ and $T_{\rm od}$ --- the same algebra as a Neumann series,
or as the Taylor expansion of a recurrent network in its input-coupling
weights. Classify each word by the set of time slots at which the
letter $T_{\rm od}$ appears (the \emph{engagements}); the rearrangement
from the $n$-indexed series to the engagement-indexed double sum below
is licensed by absolute convergence: replacing every letter by its
entrywise absolute value dominates the doubly indexed family
term-by-term by $\sum_n\bar\rho^{\,n}\|b_{\rm vac}\|_\infty<\infty$,
so the series may be regrouped freely. This organizes $v_t$ by the
number $k$ of engagements:
\begin{equation}
v_t=\sum_{k\ge0}v_t^{(k)},\qquad
v_t^{(k)}=\!\!\sum_{t\ge j_k>\cdots>j_1}\!\!
T_0^{\,t-j_k}\,T_{\rm od}(\theta_{j_k})\,
T_0^{\,j_k-j_{k-1}-1}\cdots
T_{\rm od}(\theta_{j_1})\,T_0^{\,*}\,b_{\rm vac},
\label{eq:si-dyson}
\end{equation}
where the starred factor is the resummed input-free run-up before the
first engagement,
\begin{equation*}
T_0^{\,*}\,b_{\rm vac}:=\sum_{\ell\ge0}T_0^{\,\ell}\,b_{\rm vac}
=(\mathds 1-T_0)^{-1}\,b_{\rm vac}
=\Big(0,\;0,\;\tfrac{1-\eta}{2(1-\eta\cosh2r)}\Big)^{\!\top}
=(0,\,0,\,J_{\rm ss})^{\top},
\end{equation*}
convergent since $\bar\rho_0:=\||T_0|\|_\infty=\eta\cosh2r<\bar\rho<1$:
$T_0$ is diagonal and $b_{\rm vac}$ is supported on the $J$ channel, so
the resummation is the scalar geometric series whose value is the
stationary energy $J_{\rm ss}$ of Eq.~\eqref{eq:si-Jss} below ---
this identity is what anchors the prefactor of main-text Eq.~(8).
To bound grade $k$, count what a term contains: exactly $k$
off-diagonal factors, each of norm $\le\sigma$, separated by $k{+}1$
dwell stretches (before the first engagement, between consecutive
engagements, and after the last). Summing over the position of each
engagement is the same as summing each dwell length independently, and
each dwell sum is the geometric series
$\sum_{d\ge0}\bar\rho_0^{\,d}=(1-\bar\rho_0)^{-1}$ with
$\bar\rho_0=\||T_0|\|_\infty=\eta\cosh2r$. Hence
$\|v^{(k)}\|_\infty\le\sigma^k(1-\bar\rho_0)^{-(k+1)}
\|b_{\rm vac}\|_\infty$: grade $k$ costs $k$ powers of the small coupling and one
dwell resummation per stretch. Summing the tail over $k\ge2$ with the
substitution $u:=\sigma/(1-\bar\rho_0)$,
\begin{equation*}
\sum_{k\ge2}\frac{\sigma^k\,\|b_{\rm vac}\|_\infty}{(1-\bar\rho_0)^{k+1}}
=\frac{\|b_{\rm vac}\|_\infty}{1-\bar\rho_0}\sum_{k\ge2}u^k
=\frac{\|b_{\rm vac}\|_\infty}{1-\bar\rho_0}\cdot\frac{u^2}{1-u},
\end{equation*}
and clearing $u$ gives
\begin{equation}
\Big\|\sum_{k\ge2}v^{(k)}\Big\|_\infty
\le\frac{\sigma^2\,\|b_{\rm vac}\|_\infty}
{(1-\bar\rho_0)^2\,\big(1-\bar\rho_0-\sigma\big)}
=O(\sigma^2)
\label{eq:si-tail}
\end{equation}
along any family of guard-compatible operating points with $r\to0$
(the denominator is positive for $\sigma<1-\bar\rho_0$, which
holds for $r$ small at fixed $\eta<1$); this is the error bookkeeping
that clause (II) will consume.

The two lowest grades can now be written in closed form, and the cleanest
way to get them is to \emph{decode walks backward from the readable
channel}. Think of a grade-$k$ term as a walk on the three channels. It
starts in $J$, because the source $b_{\rm vac}$ feeds only $J$. It dwells under
$T_0$, which is diagonal, so dwelling never changes channel. It
engages the input $k$ times, each engagement being one off-diagonal
entry, that is, one channel change carrying a definite pump-phase
frequency from the table above. And it must terminate in $m$ in order to
be read.

\emph{Grade one.} A single engagement from $J$ into $m$: the only such
entry is $(m\leftarrow J)$, frequency $+1$, amplitude $-\eta\sinh2r$.
Before it, the state dwells in $J$; the input-free $J$ dynamics
$J\mapsto\eta\cosh(2r)J+\tfrac{1-\eta}2$ relaxes geometrically to
\begin{equation}
J_{\rm ss}=\frac{1-\eta}{2\,(1-\eta\cosh2r)},
\label{eq:si-Jss}
\end{equation}
which is the resummed value the engagement acts on --- explicitly,
$\big(\sum_{\ell\ge0}T_0^{\,\ell}b_{\rm vac}\big)_J
=\sum_{\ell\ge0}(\eta\cosh2r)^\ell\,\tfrac{1-\eta}2=J_{\rm ss}$, the
same geometric dwell sum as above, evaluated on the $J$ diagonal.
After it, the state dwells in $m$, decaying by $x=\eta\cosh^2 r$ per
step. Assembling the three pieces --- the readout dwell
$e_m^\dagger T_0^{\,t-j}=x^{t-j}e_m^\dagger$, the engagement
entry $-\eta e^{i\theta_j}\sinh2r$, and the run-up value $J_{\rm ss}$
--- the walk that engages at slot $j$ contributes
$x^{t-j}\cdot\big(-\eta\sinh2r\big)e^{i\theta_j}\cdot J_{\rm ss}$, and
summing over the engagement slot:
\begin{equation}
m^{(1)}_t=\sum_{j\le t}\varkappa\,x^{t-j}\,e^{i\theta_j},
\qquad \varkappa=-\eta\sinh(2r)\,J_{\rm ss},
\label{eq:si-grade1}
\end{equation}
which is main-text Eq.~(8): in signal-processing terms, the leading
readable content is a \emph{one-pole filter} --- an exponentially
decaying moving average with known, input-independent pole $x$ ---
applied to the phase-encoded sequence $e^{i\theta_j}$.

\emph{Grade two.} Two engagements, ending in $m$. Work backward. The
last engagement must land in $m$ with a positive frequency (holomorphy
of the readable row, part (a)): its options are $(m\leftarrow J)$ at
frequency $+1$ or $(m\leftarrow\bar m)$ at frequency $+2$. The first
engagement starts from $J$. If the last engagement is
$(m\leftarrow J)$, the first must return the walk to $J$, but every
entry leaving $J$ leaves it --- there is no $J\to J$ engagement (the
$J\to J$ entry is diagonal, hence part of $T_0$, not an
engagement) --- so the walk would need a third engagement to return:
grade three, not two. The unique grade-two walk is therefore
$(\bar m\leftarrow J)$ at frequency $-1$ (amplitude
$-\tfrac\eta2\cdot 2\sinh 2r$ acting on $J_{\rm ss}$, i.e., the
conjugate-channel copy of $\varkappa$) at some step $i$, a dwell in
$\bar m$ (factor $x$ per step, since the $\bar m$ diagonal equals the
$m$ diagonal), then $(m\leftarrow\bar m)$ at frequency $+2$ (amplitude
$\eta\sinh^2\!r$) at step $j>i$, then the $m$ dwell to the readout:
\begin{equation}
m^{(2)}_t=\eta\sinh^2\!r\;\varkappa
\sum_{i<j\le t}x^{t-j}\,x^{j-1-i}\,e^{i(2\theta_j-\theta_i)},
\label{eq:si-grade2}
\end{equation}
which is main-text Eq.~(9). Note the charges check out at every grade:
$+1$ (grade one) and $-1+2=+1$ (grade two).
\end{proof}

\subsection{Proof of the three kernel laws (main-text Corollary 1)}
\label{sec:si-prf-laws}

The corollary states three combinatorial constraints on which characters
$E_n$ can appear in the expansion of the readable feature $m_t$:
\emph{(selection)} $q(n)=+1$; \emph{(ordering)} the latest engaged slot
carries $n_{t^*}\in\{+1,+2\}$, never a negative frequency;
\emph{(per-slot)} $|n_t|\le2$ at every slot. Each is one structural
property of Lemma~1(a) read off at the level of the series.

\begin{proof}
\emph{Selection.} By the equivariance \eqref{eq:si-grading}, a uniform
angle shift $\theta\mapsto\theta+\chi\mathbf 1$ rotates the exact state:
$v_t\mapsto e^{i\chi Q}v_t$, hence $m_t\mapsto e^{i\chi}m_t$. Write
$m_t(s;\chi)$ for the feature at the shifted angles and expand it in
characters; the expansion is absolutely convergent, uniformly in
$\chi$, by the contraction bound of Lemma~1(b), and under the shift
each character $E_n$ acquires the phase $e^{iq(n)\chi}$. Grouping the
characters by charge,
\begin{equation*}
m_t(s;\chi)=\sum_{q\in\mathbb Z}e^{iq\chi}\,c_q(s),
\qquad
c_q(s):=\!\!\sum_{n:\,q(n)=q}\!\!a_n\,E_n(s),
\end{equation*}
with $a_n$ the series coefficients, while the conservation law states
\begin{equation*}
m_t(s;\chi)=e^{i\chi}\,m_t(s).
\end{equation*}
Equate the two expressions, multiply by $e^{-iq'\chi}/2\pi$, and
integrate $\chi$ over $[0,2\pi]$; absolute convergence licenses
term-by-term integration, and the orthogonality
\begin{equation*}
\frac{1}{2\pi}\int_0^{2\pi}e^{i(q-q')\chi}\,d\chi=\delta_{qq'}
\end{equation*}
isolates one coefficient per charge:
\begin{equation*}
c_{q'}(s)=m_t(s)\,\delta_{q'1}
\qquad\text{for every }s\in\mathcal U,
\end{equation*}
so the readable functional carries no content outside the
charge-$(+1)$ sector. At the level of individual characters the same
law is constructive: in the Dyson representation \eqref{eq:si-dyson}
every walk starts on the charge-$0$ channel ($b_{\rm vac}$ is supported
on $J$), ends on the charge-$(+1)$ channel (the readout
$e_m^\dagger$), and dwells are diagonal while each engagement shifts
the channel charge by exactly its Laurent grade
[Eq.~\eqref{eq:si-grading}], so the windings deposited along any walk
telescope to $q(n)=(+1)-0=+1$. Hence $q(n)=+1$ for every character
present.

\emph{Ordering.} Let $t^*$ be the latest engaged slot of a pattern $n$
appearing in $m_t$. In the Dyson representation \eqref{eq:si-dyson}
every engagement of a contributing walk sits at a slot $\le t^*$, and
the strict ordering $j_k>\cdots>j_1$ means each slot hosts at most one
engagement per walk, so the walk's readable coefficient factors as
\begin{equation*}
e_m^\dagger\,T_0^{\,t-t^*}\,T_{d^*}\,W\,b_{\rm vac},
\end{equation*}
where $T_{d^*}$ ($d^*\in\{\pm1,\pm2\}$) is the Laurent component
deposited by the final engagement at $t^*$ and $W$ collects every
earlier factor. By the dwell identity the readout covector is a left
eigencovector of the input-free part,
\begin{equation*}
e_m^\dagger\,T_0=x\,e_m^\dagger
\quad\Longrightarrow\quad
e_m^\dagger\,T_0^{\,t-t^*}=x^{\,t-t^*}\,e_m^\dagger,
\end{equation*}
so it slides through the final dwell at the cost of the strictly
positive scalar $x^{\,t-t^*}$ and lands directly on the final
engagement. There, holomorphy of the readable row [Lemma~1(a)],
\begin{equation*}
e_m^\dagger\,T_{-1}=e_m^\dagger\,T_{-2}=0,
\end{equation*}
annihilates the negative grades:
\begin{equation*}
x^{\,t-t^*}\,e_m^\dagger\,T_{d^*}\,W\,b_{\rm vac}=0
\qquad\text{for }d^*\in\{-1,-2\},
\end{equation*}
for every $W$, i.e., regardless of the walk's earlier history. A
nonvanishing coefficient therefore requires $d^*\in\{+1,+2\}$ (the
$d=0$ component is diagonal --- a dwell, not an engagement), and since
$t^*$ is the latest engaged slot, no later engagement adds winding
there: $n_{t^*}=d^*\in\{+1,+2\}$, never a negative frequency.

\emph{Per-slot.} Two facts combine. $T_d=0$ for $|d|\ge3$
[Eq.~\eqref{eq:si-laurent}]: a single step exchanges at most two
pump-phase units. And the strict ordering $j_k>\cdots>j_1$ in
\eqref{eq:si-dyson} means each slot is engaged at most once per walk,
so the winding deposited at any slot is a single Laurent grade:
$|n_t|\le2$ at every slot.
\end{proof}

\subsection{Proof of clause (III): symmetry meets polynomial interpolation}
\label{sec:si-prfIII}

\emph{What is proved.} Every functional in $\mathcal A_D$ stays at least
the explicit distance $\delta(D,\beta s_{\max})$ of main-text Eq.~(10)
from any character of charge $D{+}1$.

\emph{Why it is true.} Shifting all inputs together is a one-parameter
probe direction. Along it, every order-$\le D$ polynomial readout can oscillate
only at the $2D{+}1$ frequencies $-D,\dots,D$, whereas the target
oscillates at frequency $D{+}1$. A function of one variable confined to a
$(2D{+}1)$-dimensional space of exponentials cannot track an exponential
lying outside that band, and classical polynomial interpolation converts
that obstruction into an explicit, dimension-free lower bound. Notice
what the argument does \emph{not} use: no perturbation series, and no
property of the machine beyond the symmetry \eqref{eq:si-grading}. The
bound therefore survives any hardware imperfection that preserves the
symmetry.

\smallskip\noindent\emph{Step 1: the shift direction slices
$\mathcal A_D$ into $2D{+}1$ frequencies.} Under
$s\mapsto s+\tau\mathbf 1$ every angle shifts by $\chi=\beta\tau$, so by
\eqref{eq:si-grading} each feature responds exactly as
$m_t\mapsto e^{i\beta\tau}m_t$ and
$\bar m_t\mapsto e^{-i\beta\tau}\bar m_t$ --- for every operating
setting, hence for every member of an ensemble simultaneously. A
monomial in the features with $\alpha$ unconjugated and $\bar\alpha$
conjugated factors ($\alpha,\bar\alpha\in\mathbb Z_{\ge0}$ the two
factor counts) therefore transforms as a product of its factors'
phases: each unconjugated factor contributes $e^{+i\beta\tau}$, each
conjugated factor $e^{-i\beta\tau}$, so the monomial as a whole picks up
$\big(e^{i\beta\tau}\big)^{\alpha}\big(e^{-i\beta\tau}\big)^{\bar\alpha}
=e^{i(\alpha-\bar\alpha)\beta\tau}$ --- a single frequency
$q=\alpha-\bar\alpha$, confined to the band by the triangle inequality
on non-negative integers:
\begin{equation*}
|q|\;=\;|\alpha-\bar\alpha|\;\le\;\alpha+\bar\alpha\;\le\;D .
\end{equation*}
Collecting terms of an order-$\le D$
polynomial $g$ by this frequency,
\begin{equation}
g(s+\tau\mathbf 1)=\sum_{|q|\le D}e^{iq\beta\tau}\,g_q(s):
\label{eq:si-slice}
\end{equation}
every \emph{slice} $\tau\mapsto g(s+\tau\mathbf 1)$ lies in the
$(2D{+}1)$-dimensional space
$\mathcal E_D=\operatorname{span}\{e^{iq\chi}:|q|\le D\}$ of the
slice variable $\chi=\beta\tau$. The space $\mathcal E_D$ is finite
dimensional, hence closed, so the same confinement holds for every
element of the closure $\mathcal A_D$, not just for finite readouts.

\smallskip\noindent\emph{Step 2: the target slices outside the band.}
For a pattern $n$ with $q(n)=D{+}1$,
$E_n(s+\tau\mathbf 1)=e^{i\beta n\cdot s}\,e^{i(D+1)\chi}$: along
every slice the target is a unimodular constant times the frequency
$D{+}1$ exponential.

\smallskip\noindent\emph{Step 3: a one-variable inapproximability bound.}
The problem is now classical: how far is $e^{i(D+1)\chi}$ from
$\mathcal E_D$ in the sup norm on an interval? The standard tool is
duality: for any function $f$, any subspace $V$, and any bounded linear
functional $\Lambda$ that annihilates $V$,
\begin{equation}
\operatorname{dist}_\infty(f,V)\;\ge\;\frac{|\Lambda(f)|}{\|\Lambda\|},
\qquad
\|\Lambda\|:=\sup_{\|g\|_\infty\le1}|\Lambda(g)|<\infty .
\label{eq:si-duality}
\end{equation}
The proof is three attributed lines. Fix any $v\in V$. Linearity and
the annihilation hypothesis $\Lambda(v)=0$ give
\begin{equation*}
\Lambda(f)=\Lambda(f)-\Lambda(v)=\Lambda(f-v).
\end{equation*}
The operator norm bounds any value by the norm times the argument's
size --- for $g\ne0$ the rescaled $g/\|g\|_\infty$ is admissible in the
supremum defining $\|\Lambda\|$, and homogeneity restores the scale ---
so
\begin{equation*}
|\Lambda(f)|=|\Lambda(f-v)|\;\le\;\|\Lambda\|\,\|f-v\|_\infty .
\end{equation*}
The left-hand side does not depend on $v$; taking the infimum over
$v\in V$ on the right,
\begin{equation*}
|\Lambda(f)|\;\le\;\|\Lambda\|\,\inf_{v\in V}\|f-v\|_\infty
=\|\Lambda\|\operatorname{dist}_\infty(f,V),
\end{equation*}
and dividing by $\|\Lambda\|>0$ yields \eqref{eq:si-duality} (for
$\Lambda\equiv0$ the bound is vacuous). Note the direction of use: the
norm sits in the denominator, so any \emph{upper} bound on
$\|\Lambda\|$ suffices --- overestimating the gain only weakens, never
invalidates, the lower bound on the distance. It remains to build a
good annihilating functional, which interpolation supplies:

\begin{lemma}[Interpolation gap]
\label{lem:gap}
On $C\big[-\tfrac a2,\tfrac a2\big]$ with $0<a<2\pi$,
\begin{equation}
\operatorname{dist}_\infty\!\big(e^{i(D+1)\chi},\,\mathcal E_D\big)
\;\ge\;\prod_{j=1}^{2D+1}\sin\!\Big(\frac{j\,a}{4(2D+1)}\Big)\;>\;0 .
\end{equation}
\end{lemma}

\begin{proof}
Take the $2D{+}2$ equally spaced points $\chi_\ell=\ell h$,
$\ell=0,\dots,2D{+}1$, with spacing $h=\tfrac a{2(2D+1)}$ (all inside
$[0,\tfrac a2]$), and let
$p(z)=\prod_{|q|\le D}\big(z-e^{iqh}\big)=\sum_{\ell=0}^{2D+1}w_\ell
z^\ell$ be the monic polynomial whose roots are the band frequencies
placed on the unit circle. Define the functional as the weighted sum of
point evaluations $\Lambda(f)=\sum_\ell w_\ell f(\chi_\ell)$. Then:

\emph{(i) It annihilates the band.} For $|q|\le D$,
$\Lambda(e^{iq\chi})=\sum_\ell w_\ell\,(e^{iqh})^\ell=p(e^{iqh})=0$,
since $e^{iqh}$ is a root of $p$.

\emph{(ii) It sees the target.} Using
$|e^{ix}-e^{iy}|=2\,|\sin\tfrac{x-y}2|$,
\begin{equation*}
\big|\Lambda(e^{i(D+1)\chi})\big|=\big|p(e^{i(D+1)h})\big|
=\prod_{|q|\le D}\Big|2\sin\tfrac{(D+1-q)h}{2}\Big|
=\prod_{j=1}^{2D+1}2\sin\!\Big(\frac{jh}{2}\Big),
\end{equation*}
where $j=D{+}1{-}q$ runs over $1,\dots,2D{+}1$; every factor is
strictly positive because the largest argument is
$(2D{+}1)h/2=a/4<\pi/2$.

\emph{(iii) It is bounded.} Since $|f(\chi_\ell)|\le\|f\|_\infty$
at every node, $\|\Lambda\|\le\sum_\ell|w_\ell|$, the coefficient
$\ell_1$-norm of $p$. That norm is submultiplicative under polynomial
multiplication --- the coefficients of a product are convolutions of
the factors' coefficients, and the triangle inequality gives
$\|pq\|_{\ell_1}\le\|p\|_{\ell_1}\|q\|_{\ell_1}$ --- and each linear
factor $(z-e^{iqh})$ has $\ell_1$-norm $1+|e^{iqh}|=2$ because its root
is unimodular. Multiplying the $2D{+}1$ factors,
$\|\Lambda\|\le2^{2D+1}$.

Insert (i)--(iii) into the duality bound \eqref{eq:si-duality}; the
factors of $2$ cancel against $2^{2D+1}$, leaving
$\prod_{j=1}^{2D+1}\sin(jh/2)$ with $h/2=\tfrac a{4(2D+1)}$.
\end{proof}

\smallskip\noindent\emph{Step 4: assembly.} Restrict base points $s$ to
the half-box $\|s\|_\infty\le s_{\max}/2$, so that
$s+\tau\mathbf 1\in\mathcal U$ for all $|\chi|\le a/2$ with
$a=\beta s_{\max}$ --- the slice stays inside the input domain. Fix any
such $s$ and any $g\in\mathcal A_D$, and chain the three steps:
\begin{equation*}
\|E_n-g\|_{\infty,\mathcal U}
\;\ge\;\sup_{|\chi|\le a/2}
\big|E_n(s+\tau\mathbf 1)-g(s+\tau\mathbf 1)\big|
\;=\;\sup_{|\chi|\le a/2}
\Big|e^{i\beta n\cdot s}e^{i(D+1)\chi}
-\sum_{|q|\le D}e^{iq\chi}g_q(s)\Big|
\;=\;\operatorname{dist}_\infty\!\big(e^{i(D+1)\chi},
\mathcal E_D\big),
\end{equation*}
where the first inequality holds because the slice points lie inside
$\mathcal U$, the middle equality substitutes Step 2 for the target and
Step 1 for $g$, and the last equality divides through by the unimodular
constant $e^{i\beta n\cdot s}$ --- which changes no absolute value and
maps $\mathcal E_D$ onto itself, since $\mathcal E_D$ is a subspace.
Lemma~\ref{lem:gap} then gives
$\|E_n-g\|_\infty\ge\delta(D,a)$ with $\delta$ as in main-text
Eq.~(10).\hfill$\blacksquare$

\subsection{Proof of clause (II): deconvolve the leading filter, then Fourier}
\label{sec:si-prfII}

\emph{What is proved.} $\bigcup_D\mathcal A_D$ is dense in
$C(\mathcal U)$.

\emph{Why it is true.} At leading order in the input coupling $\sigma$,
the readable feature is a known exponential-decay filter of the
phase-encoded input [Eq.~\eqref{eq:si-grade1}]. A one-pole filter is
inverted exactly by a two-tap finite-impulse-response (FIR) filter, so a
fixed linear combination of two adjacent features isolates the
single-slot oscillation $e^{i\beta s_j}$ up to an $O(\sigma)$ error.
Products of these, and of their conjugates, approximate every character
$E_n$. Characters form exactly the sort of algebra to which the
Stone--Weierstrass theorem applies, and density follows. By this point
the quantum optics has already done its work; what remains is
deconvolution plus Fourier approximation.

\smallskip\noindent\emph{Step 1: deconvolution.} The grade-one content
\eqref{eq:si-grade1} is the one-pole filter
$m^{(1)}_j=\varkappa\sum_{i\le j}x^{\,j-i}e^{i\theta_i}$ with pole
$x=\eta\cosh^2\!r$ --- crucially, $x$ and $\varkappa$ are hardware
constants, independent of the input. The inverse of a one-pole filter is
the two-tap FIR filter with taps $(1,-x)$: define, for any slot $j$ and
\emph{any} mask,
\begin{equation}
g_j:=\frac{m_j-x\,m_{j-1}}{\varkappa\,e^{i\varphi_j}} .
\label{eq:si-fir}
\end{equation}
On the grade-one content the telescoping is exact at all depths ---
write out the two sums and subtract:
\begin{equation*}
m^{(1)}_j-x\,m^{(1)}_{j-1}
=\varkappa\sum_{i\le j}x^{\,j-i}e^{i\theta_i}
-\varkappa\sum_{i\le j-1}x^{\,j-i}e^{i\theta_i}
=\varkappa\,e^{i\theta_j},
\end{equation*}
since every slot $i<j$ appears in both sums with the identical weight
$x^{\,j-i}$ and cancels (dividing a geometric lag structure by its
transfer function).
The mask
normalization $e^{i\varphi_j}$ then strips the known phase from
$e^{i\theta_j}=e^{i\varphi_j}e^{i\beta s_j}$, leaving $e^{i\beta s_j}$
exactly. The residual of \eqref{eq:si-fir} is
the same FIR filter applied to the grades $\ge2$: writing
$h_j:=m_j-m^{(1)}_j=\sum_{k\ge2}m^{(k)}_j$ for the higher-grade
remainder, the residual is
$(h_j-x\,h_{j-1})/(\varkappa e^{i\varphi_j})$, bounded by the triangle
inequality as $\le(1+x)\,\sup_j\|h_j\|_\infty\,/\,|\varkappa|$. The numerator is
$O(\sigma^2)$ by Eq.~\eqref{eq:si-tail}, and the filter's gain is
$(1+x)/|\varkappa|=O(1/\sigma)$ (note
$\varkappa=O(\sigma)$); hence [this working bound is sharpened to $O(\sigma^{2})$ on the deconvolved feature in Subsec.~\ref{sec:si-budget}, Step~1, where the grade-two amplitude is bounded explicitly rather than through the Dyson tail; the weaker exponent here suffices for density]
\begin{equation}
\big\|g_j-e^{i\beta s_j}\big\|_\infty=O(\sigma)\;\longrightarrow\;0
\quad\text{as }r\to0,
\label{eq:si-firerr}
\end{equation}
uniformly on $\mathcal U$, along guard-compatible operating points.

\smallskip\noindent\emph{Step 2: products build every character.}
Conjugated factors $e^{-i\beta s_j}$ come from the same construction on
$\bar m$. A product of $k$ such factors at slots $j_1,\dots,j_k$
approximates $\prod_ie^{\pm i\beta s_{j_i}}$; since each factor is
bounded by $1+v_\infty$ [by \eqref{eq:si-firerr},
$|g_j|\le1+O(\sigma)$, and $O(\sigma)\le v_\infty$ for $\sigma$ small
along the family, $v_\infty\to\tfrac12$] and carries error $O(\sigma)$, the product
carries error at most $k\,(1+v_\infty)^{k-1}O(\sigma)$, by the standard
factor-by-factor telescoping of a difference of products,
\begin{equation*}
\prod_{i=1}^k a_i-\prod_{i=1}^k b_i
=\sum_{i=1}^k\Big(\prod_{l<i}a_l\Big)(a_i-b_i)\Big(\prod_{l>i}b_l\Big):
\end{equation*}
each of the $k$ summands replaces one factor by its error, at most
$O(\sigma)$, and pads the rest with factors bounded by $1+v_\infty$
(the approximants $a_l$) or by $1$ (the exact unimodular targets
$b_l$). Every character $E_n$ with
integer exponents is such a product (repeat slots $|n_t|$ times), so
every character lies in the closure of the reachable class at the
appropriate polynomial order.

\smallskip\noindent\emph{Step 3: Stone--Weierstrass.} The set of finite
linear combinations of characters is an algebra of continuous functions
on the compact space $\mathcal U$: it contains the constants
($E_0=1$), is closed under products
($E_nE_{n'}=E_{n+n'}$) and complex conjugation
($\bar E_n=E_{-n}$), and separates points --- if $s\ne s'$ differ at
slot $t$, then $|\beta(s_t-s'_t)|\le 2\beta s_{\max}<2\pi$ by encoding
injectivity, so $e^{i\beta s_t}\ne e^{i\beta s'_t}$. By the complex
Stone--Weierstrass theorem this algebra is dense in $C(\mathcal U)$;
this is Fourier approximation on the input torus. Since every character
is reachable (Step 2), $\bigcup_D\mathcal A_D$ is dense in
$C(\mathcal U)$. \emph{Topology and the fading-memory class} (completing
the embedding claim): equip $\mathcal U=\prod_{t\le0}[-s_{\max},s_{\max}]$
with the product topology, compact by Tychonoff and metrized by
$d_\omega(s,s')=\sup_{t\le0}\omega_{|t|}\,|s_t-s'_t|$ for any weighting sequence
$\omega:\mathbb N\to(0,1]$ that is decreasing with $\omega_k\to0$. The reservoir
filter is $d_\omega$-continuous for every admissible $\omega$ dominating the state
contraction, $\omega_k\ge c\,\bar\rho^{\,k}$: a perturbation of the input $k$
slots in the past enters every readable feature through walk coefficients
bounded by $C\bar\rho^{\,k}$ [Dyson bound, Eq.~\eqref{eq:si-tail} and the
causal ordering law], so the induced functional has fading memory with
respect to $\omega$ in the sense of Boyd--Chua~\cite{BoydChua1985}, and the
weighted-norm and product topologies coincide on the compact
$\mathcal U$ (Grigoryeva--Ortega~\cite{GrigoryevaOrtega2018}). Uniform density in $C(\mathcal U)$ is therefore
equivalent to fading-memory approximation for every such $\omega$, and the
sup-norm gap of clause (III) is a distance in the same topology,
completing clause (II).\hfill$\blacksquare$

\subsection{Budgeted universality: proof of the affordable-universality
corollary}
\label{sec:si-budget}

\emph{What is proved.} Main-text Corollary (Affordable universality): a
weight-$k$ character is estimated to RMS accuracy $\epsilon$, uniformly
over inputs, at per-feature measurement shot budget
$B=O\big((\sigma_{\rm eff}/\mu_T)^2\,k^3(1+v_\infty)^{3(k-1)}\,
\epsilon^{-3}\big)$.

\emph{Why it is true.} The clause-(II) construction has two error
channels, and they pull in opposite directions in the coupling $\sigma$.
One is a deterministic FIR residual, of size $O(\sigma^2)$. This is
sharper than the $O(\sigma)$ working bound of clause (II), because the
grade-two amplitude carries $\sinh^2\!r$. The other is a statistical
noise gain, of size $O(1/\sigma)$. Because one grows with $\sigma$ and
the other shrinks, there is an optimum: balancing them against a budget
$B$ puts $\sigma^\star\propto B^{-1/6}$ and error $\propto B^{-1/3}$.
Inverting gives the cubic budget law, and the product telescoping of
clause (II), Step 2, carries it up to weight $k$.

\smallskip\noindent\emph{Step 1: the sharpened residual.} The FIR output
decomposes as
\begin{equation*}
g_j=e^{i\beta s_j}+\rho_j^{\rm res},\qquad
\rho_j^{\rm res}=\frac{\big(m_j-m_j^{(1)}\big)-x\big(m_{j-1}-m^{(1)}_{j-1}\big)}
{\varkappa\,e^{i\varphi_j}},
\end{equation*}
the filter applied to the grades $\ge2$. Grade two is bounded
\emph{explicitly} from Eq.~\eqref{eq:si-grade2}: summing the two dwell
factors independently,
\begin{equation*}
\big|m^{(2)}_j\big|\;\le\;\eta\sinh^2\!r\,|\varkappa|
\sum_{i<j}x^{\,j-1-i}\,x^{\,t-j}
\;=\;\frac{\eta\sinh^2\!r\,|\varkappa|}{(1-x)^2},
\end{equation*}
so its FIR image divided by $\varkappa$ obeys
\begin{equation*}
\frac{(1+x)\,\big|m^{(2)}\big|}{|\varkappa|}\;\le\;
\frac{(1+x)\,\eta\sinh^2\!r}{(1-x)^2}\;=\;O(\sigma^2),\qquad
\sinh^2\!r=O(\sigma^2)\ \text{as }r\to0\text{ at fixed }\eta.
\end{equation*}
Grades $\ge3$ are bounded by the Dyson tail [Eq.~\eqref{eq:si-tail}],
\begin{equation*}
\sum_{k\ge3}\big\|v^{(k)}\big\|\;\le\;C_3\,\sigma^3
\quad\Longrightarrow\quad
\frac{(1+x)}{|\varkappa|}\sum_{k\ge3}\big\|v^{(k)}\big\|\;=\;O(\sigma^2),
\qquad \varkappa=\Theta(\sigma),
\end{equation*}
along guard-compatible settings. Hence
\begin{equation}
\sup_j\big\|\rho_j^{\rm res}\big\|_\infty\;\le\;C_{\rm res}\,\sigma^2,
\label{eq:si-res2}
\end{equation}
with $C_{\rm res}$ depending only on $(\eta,\bar\rho)$-margins.

\smallskip\noindent\emph{Step 2: the statistical channel.} The detector
returns
\begin{equation*}
\hat m_j=m_j+\zeta_j,\qquad
\mathbb E|\zeta_j|^2=\frac{\sigma_{\rm eff}^2}{\mu_T^2\,B}
\end{equation*}
at per-feature measurement shot budget $B$ (Methods conventions; both
quadratures interleaved). The FIR filter maps $\zeta$ through gain
$(1+x)/|\varkappa|$, and
\begin{equation*}
|\varkappa|=\eta\sinh(2r)\,J_{\rm ss}\;\ge\;c_\varkappa\,\sigma
\end{equation*}
along the same settings, so the per-factor statistical RMS is
\begin{equation}
\varepsilon_{\rm stat}\;\le\;
\frac{C_{\rm stat}}{\sigma}\,\frac{\sigma_{\rm eff}}{\mu_T\sqrt B}.
\label{eq:si-stat}
\end{equation}

\smallskip\noindent\emph{Step 3: balance and inversion.} Abbreviate
$\Upsilon:=C_{\rm stat}\,\sigma_{\rm eff}/(\mu_T\sqrt B)$, so the
per-factor error is
\begin{equation*}
f(\sigma)=C_{\rm res}\,\sigma^2+\frac{\Upsilon}{\sigma}.
\end{equation*}
Its stationarity condition,
\begin{equation*}
f'(\sigma)=2C_{\rm res}\,\sigma-\frac{\Upsilon}{\sigma^2}=0
\quad\Longrightarrow\quad
\sigma^{\star3}=\frac{\Upsilon}{2C_{\rm res}},\qquad
\sigma^\star=\Big(\frac{C_{\rm stat}\,\sigma_{\rm eff}}
{2C_{\rm res}\,\mu_T\sqrt B}\Big)^{\!1/3},
\end{equation*}
is a minimum since $f''(\sigma)=2C_{\rm res}+2\Upsilon/\sigma^3>0$.
Substituting back, both terms scale identically,
\begin{equation*}
\sigma^{\star2}\propto B^{-1/3},\qquad
\frac{\Upsilon}{\sigma^\star}\propto B^{-1/2}\cdot B^{1/6}=B^{-1/3},
\end{equation*}
so the optimized error is
\begin{equation*}
\varepsilon^\star=c'\,\Big(\frac{\sigma_{\rm eff}}{\mu_T}\Big)^{\!2/3}
B^{-1/3},
\end{equation*}
with $c'$ collecting the constants. Requiring the product-telescoping
budget of clause (II), Step 2, with per-factor bound $1+v_\infty$,
\begin{equation*}
\varepsilon^\star\;\le\;\frac{\epsilon}{k\,(1+v_\infty)^{k-1}}
\quad\Longleftrightarrow\quad
c'\Big(\frac{\sigma_{\rm eff}}{\mu_T}\Big)^{\!2/3}B^{-1/3}\;\le\;
\frac{\epsilon}{k\,(1+v_\infty)^{k-1}},
\end{equation*}
and cubing both sides and solving for $B$ yields
\begin{equation}
B\;\le\;c\,\Big(\frac{\sigma_{\rm eff}}{\mu_T}\Big)^{\!2}
k^{3}\,(1+v_\infty)^{3(k-1)}\;\epsilon^{-3},
\end{equation}
which is the corollary; the estimator uses $\le2k$ distinct features, so
the total sample count is $O(kB)$ under the per-feature convention.
Because $\sigma_{\rm eff}$ enters squared, any reduction of the
per-shot deviation shifts $B$ proportionally at every accuracy; with
the corrected amplifier model the available
reduction is the loss-suppression factor $1.25/1.025=1.22$, shared by
the classical twin.\hfill$\blacksquare$

\begin{remark}[Polynomial in accuracy, exponential in weight, factorial
for moments]
At fixed target the budget is polynomial in $1/\epsilon$ (order three);
at fixed accuracy the constant grows exponentially in the character
weight $k$ with base $(1+v_\infty)^3$---bounded, with
$v_\infty\to\tfrac12$ in the small-$\sigma$ limit and $v_\infty=0.862$,
base $\approx6.5$, at the campaign point---whereas order-$m$
moment readout pays the super-exponential $(2m-1)!!$ of
main-text Eq.~(1) at every accuracy. The comparison is therefore
polynomial-versus-factorial in the accuracy variable and
exponential-versus-factorial in the order variable; both favor
in-dynamics placement.
\end{remark}

\subsection{Proof of clause (I): exact character projectors from a
designed mask sweep}
\label{sec:si-prfI}

\emph{What is proved.} The exact characterization
$\mathcal A_D=\overline{\operatorname{span}}\{E_n:n\in\mathfrak N_D\}$, with $\mathfrak N_D$ the $D$-assembleable patterns of the
main text: sums of at most $D$ admissible or co-admissible blocks.
\emph{The two directions differ in depth.} The \emph{upper} inclusion
($\subseteq$) is the selection rule plus arithmetic. The \emph{lower}
inclusion ($\supseteq$) is the substantial one: it must show that every
charge-allowed character is actually reachable, including multi-slot
patterns such as $e^{i\beta(2s_{t-3}-s_{t-7})}$.

Here the per-slot law $|n_t|\le2$ of Corollary~1 turns from a restriction
into the enabling tool. Because each in-window slot can carry only the
five frequencies $\{-2,\dots,2\}$, a pattern's in-window frequency
content is a word in a five-letter alphabet. We then sweep the mask
through $5^L$ designed settings, chosen so that each such word announces
itself as a \emph{distinct tone} of a discrete Fourier transform. A DFT
projects out any prescribed word exactly, at every perturbative grade
simultaneously, with no truncation anywhere.

\smallskip\noindent\emph{Upper inclusion, Route A (character
algebra).} By the three kernel laws, each feature $m_t$ ($\bar m_t$)
is an absolutely convergent series of characters indexed by admissible
(co-admissible) patterns [convergence from the contraction bound,
Lemma~1(b)]. A monomial of total polynomial order $D'\le D$ in such series
expands, by the Cauchy product of absolutely convergent series, into
characters $E_{n^{(1)}+\cdots+n^{(D')}}$ with every $n^{(i)}$
admissible or co-admissible---$D$-assembleable characters, with
absolutely summable coefficients. Every finite readout therefore lies
in $W_D:=\overline{\operatorname{span}}\{E_n:n\in\mathfrak N_D\}$,
and $W_D$ is closed, so $\mathcal A_D\subseteq W_D$. Imported
inputs: Corollary~1 (admissible index set), Lemma~1(b) (absolute
convergence).

\smallskip\noindent\emph{Upper inclusion, Route B (symmetry
slices; independent, consequence-level).} Without any character
expansion of the features, the gauge equivariance
\eqref{eq:si-grading} alone re-derives the two falsifiable
consequences of Route A's inclusion: the global slice
$s\mapsto s+\tau\mathbf 1$ confines every element of
$\mathcal A_D$ to frequencies $|q|\le D$ (the Step-1 argument of the
clause-(III) proof), and the single-slot slice
$s_j\mapsto s_j+\tau$ confines it to per-slot frequencies
$|k|\le2D$, since each of $\le D$ factors responds through its slot
exponent of modulus $\le2$ and the containing span of exponentials is
finite dimensional, hence closed. The full
block-form inclusion is carried by Route A;
Route B independently confirms its sector and per-slot consequences
from the symmetry alone.

\smallskip\noindent\emph{The projector.} Fix an \emph{anchor} slot $j$
and a \emph{window} of $L$ lags, the slots $j,j-1,\dots,j-L+1$. We
perturb the mask along a designed ray and record the deconvolved feature
$g_j$ of Eq.~\eqref{eq:si-fir} at each setting.

\begin{lemma}[Mask-shift projection]
\label{lem:projection}
Define the mask ray of $5^L$ \emph{epochs}
\begin{align}
\varphi^{(k)}&=\varphi+k\,\psi,\qquad k=0,\dots,5^L-1,\nonumber\\
\psi_{j-\ell}&=2\pi\,5^{-(\ell+1)}\ \ (0\le\ell<L),\qquad
\psi_s=0\ \ \text{otherwise},
\label{eq:si-ray}
\end{align}
i.e., the mask phase at lag $\ell$ inside the window advances by
$2\pi/5^{\ell+1}$ per epoch, and the mask outside the window is not
touched. Let $g_j^{(k)}$ be the deconvolved feature \eqref{eq:si-fir}
recorded at epoch $k$. Then a pattern $n$ contributes to $g_j^{(k)}$
with the epoch dependence $e^{2\pi i\,kR(\nu)/5^L}$, where
$\nu:=n-e_j$ ($e_j$ = the unit pattern at the anchor, removed by the
FIR normalization) and
\begin{equation}
R(\nu)=\sum_{\ell<L}\nu_{j-\ell}\,5^{\,L-1-\ell}
\label{eq:si-digitcode}
\end{equation}
is a function of the \emph{window exponents only}. Because the per-slot
law confines $\nu_s\in\{-2,\dots,2\}$, the code $R$ is injective on
window exponent vectors modulo $5^L$, and consequently the DFT weights
$w_k=5^{-L}e^{-2\pi i\,kR(\nu^*)/5^L}$ satisfy: the average
$\sum_kw_k\,g_j^{(k)}$ annihilates --- exactly, and at every grade ---
every pattern whose window exponents differ from the prescribed
$\nu^*$, and preserves, with unit weight, every pattern that agrees.
\end{lemma}

\begin{proof}
A pattern $n$ enters the feature through the product
$\prod_t e^{in_t\theta_t}$; along the ray, the mask at epoch $k$ is
$\varphi^{(k)}=\varphi+k\psi$, so the pattern's epoch dependence and
the FIR normalization of \eqref{eq:si-fir} combine as
\begin{equation*}
\prod_t e^{in_t\theta^{(k)}_t}
=e^{ik\langle n,\psi\rangle}\prod_t e^{in_t\theta_t},\qquad
\frac{1}{e^{i\varphi^{(k)}_j}}
=e^{-ik\psi_j}\,\frac{1}{e^{i\varphi_j}},
\end{equation*}
for a net epoch phase
\begin{equation*}
e^{ik\langle\nu,\psi\rangle},\qquad \nu=n-e_j.
\end{equation*}
Substituting the ray \eqref{eq:si-ray},
\begin{equation*}
\langle\nu,\psi\rangle=\sum_{\ell<L}\nu_{j-\ell}\,2\pi\,5^{-(\ell+1)}
=\frac{2\pi R(\nu)}{5^L},
\end{equation*}
with $R$ as in \eqref{eq:si-digitcode} --- the window exponents read
out as signed digits of a base-$5$ integer, the deeper lags in the
less significant digits.

\emph{Injectivity of $R$ modulo $5^L$} is the uniqueness of signed
base-$5$ representations with digit modulus $\le2$ (``balanced
base~$5$''). Suppose
\begin{equation*}
R(\nu)\equiv R(\nu')\pmod{5^L},\qquad \nu_s,\nu'_s\in\{-2,\dots,2\};
\end{equation*}
the difference $\Delta=\nu-\nu'$ has digits
$\Delta_s\in\{-4,\dots,4\}$ and satisfies
\begin{equation*}
\sum_{\ell<L}\Delta_{j-\ell}\,5^{\,L-1-\ell}\equiv0\pmod{5^L}.
\end{equation*}
Reducing modulo $5$ isolates the least significant digit:
\begin{equation*}
\Delta_{j-L+1}\equiv0\pmod5,\qquad
|\Delta_{j-L+1}|\le4\;\Longrightarrow\;\Delta_{j-L+1}=0;
\end{equation*}
dividing by $5$ and repeating peels off the digits one by one, so
$\Delta=0$. Equivalently, the digit bound gives
\begin{equation*}
|R(\nu)|\le 2\sum_{\ell<L}5^{\ell}=\frac{5^L-1}{2}<\frac{5^L}{2},
\end{equation*}
so two codes cannot differ by a nonzero multiple of $5^L$.

\emph{Annihilation/preservation} is then character orthogonality on
the cyclic group $\mathbb Z_{5^L}$: with $\Delta R:=R(\nu)-R(\nu^*)$,
\begin{equation*}
\sum_{k=0}^{5^L-1}w_k\,e^{2\pi ikR(\nu)/5^L}
=\frac1{5^L}\sum_{k=0}^{5^L-1}\big(e^{2\pi i\,\Delta R/5^L}\big)^{k}
=\mathds{1}\big[\Delta R\equiv0\ (\mathrm{mod}\ 5^L)\big]
=\mathds{1}\big[\nu=\nu^*\ \text{on the window}\big],
\end{equation*}
the middle step the finite geometric series --- its ratio is a
$5^L$-th root of unity,
\begin{equation*}
\sum_{k=0}^{5^L-1}\zeta^{\,k}=5^L\,\mathds{1}[\zeta=1],
\qquad \zeta^{5^L}=1,
\end{equation*}
--- and the final step the injectivity just proved. No expansion was
truncated anywhere in this argument --- the epoch phase of a pattern
is exact at every grade --- so the projection acts exactly on the full
nonlinear feature.
\end{proof}

\smallskip\noindent\emph{Why base $5$.} The base of the sweep is the
size of the per-slot frequency alphabet, $5=2\cdot2+1$: a balanced digit
set $\{-2,\dots,+2\}$ fixed by the per-slot law, itself the statement
that one squeeze exchanges at most two pump quanta per step. Base $5$ is
the smallest base for which the signed-digit representation is unique:
injectivity of $R$ requires every digit difference
$\Delta_s\in\{-4,\dots,4\}$ to satisfy
$\Delta_s\equiv0\pmod b\Rightarrow\Delta_s=0$, which holds iff $b\ge5$.
At $b=4$ the projector fails concretely: single-slot exponents $+2$ and
$-2$ differ by $\Delta R=4\cdot4^{L-1}=4^L\equiv0\pmod{4^L}$, so their
DFT tones coincide and the projector cannot separate a pattern from its
conjugate-doubled partner. Any $b>5$ is also injective but spends
$2\,b^L$ epochs against $2\cdot5^L$, a $(b/5)^L$ overhead with nothing
bought. A hardware process exchanging $d_{\max}$ pump quanta per slot
would set the base at $2d_{\max}+1$.

\begin{theorem}[Exact block matching]
\label{thm:matching}
Let $n^*$ be any admissible charge-one pattern of grade $k^*$ (its number
of engagements) and span $\le L$ (all engaged slots inside the window),
anchored at $j$ (its latest engaged slot), and let $\varepsilon>0$. The
\emph{order-one} functional
\begin{equation}
f=\frac1{C_{n^*}}\sum_{k=0}^{5^L-1}w_k\,g_j^{(k)},
\label{eq:si-certificate}
\end{equation}
with $w_k$ the projection weights of Lemma~\ref{lem:projection} at
$\nu^*=n^*-e_j$ and $C_{n^*}\ne0$ the closed-form walk coefficient of
$n^*$ (the product of its engagement amplitudes, dwell factors, and
$J_{\rm ss}$, as in Eqs.~\eqref{eq:si-grade1}--\eqref{eq:si-grade2}),
satisfies $\|f-E_{n^*}\|_\infty\le\varepsilon$ for $\sigma$ small
enough along guard-compatible operating points. Arbitrary
prescriptions on all window patterns follow by linearity, the same
$5^L$ epochs furnishing every projector.
\end{theorem}

\begin{proof}
Three filters act in sequence, and the first two are exact. Expand the
deconvolved feature at epoch $k$ in characters, using the Dyson series
\eqref{eq:si-dyson} and the epoch phase of Lemma~\ref{lem:projection}:
\begin{equation}
g_j^{(k)}=\sum_{n}a_n\,E_n(s)\,e^{2\pi i\,kR(\nu)/5^L},\qquad
\nu:=n-e_j,
\label{eq:si-thmpf-expand}
\end{equation}
absolutely convergent by Lemma~1(b), with $a_n$ the walk coefficient of
pattern $n$ and $\mathrm{gr}(n)$ its grade (engagement count); the FIR
step \eqref{eq:si-fir} has already restricted the sum to patterns
anchored at $j$ (its own $O(\sigma^2)$ residual is priced separately in
Sec.~\ref{sec:si-budget}). Applying the DFT weights
$w_k=5^{-L}e^{-2\pi i\,kR(\nu^*)/5^L}$ and exchanging the finite epoch
sum with the absolutely convergent character sum,
\begin{equation}
\sum_{k=0}^{5^L-1}w_k\,g_j^{(k)}
=\sum_n a_n\,E_n(s)\;
\mathds 1\big[\nu=\nu^*\ \text{on the window}\big],
\label{eq:si-thmpf-proj}
\end{equation}
by the orthogonality relation of Lemma~\ref{lem:projection} --- exact at
every grade, no expansion truncated. Splitting off the matched pattern
and normalizing,
\begin{equation}
f=\frac1{C_{n^*}}\sum_kw_k\,g_j^{(k)}
=E_{n^*}(s)
+\frac1{C_{n^*}}\!\!\sum_{\substack{n\ne n^*\\ \nu|_{\rm win}=\nu^*}}
\!\!a_n\,E_n(s).
\label{eq:si-thmpf-split}
\end{equation}
Every survivor in the residual sum agrees with $n^*$ on the entire
window; since a pattern is supported exactly where its engagements are,
it must carry at least one additional engagement \emph{outside} the
window, so $\mathrm{gr}(n)\ge k^*+1$. Its coefficient is controlled by
the walk decoding of Lemma~1(c) (unique per pattern): the coefficient
factors into engagement amplitudes, dwell factors, and $J_{\rm ss}$,
\begin{equation}
a_n=J_{\rm ss}\Big[\prod_{s\in\operatorname{supp}\nu^*}
A_{\nu^*_s}\!(s)\Big]
\Big[\prod_{\text{outside engagements}}A_d\Big]
\times(\text{dwell factors}),
\label{eq:si-thmpf-decode}
\end{equation}
with $|A_{\pm1}|\in\{\eta\sinh2r,\tfrac\eta2\sinh2r\}=\Theta(\sigma)$
and $|A_{\pm2}|=\eta\sinh^2\!r=\Theta(\sigma^2)$ [entries of
Eq.~\eqref{eq:si-laurent-explicit}]. Because the survivor's window
engagements carry the \emph{same} Laurent grades as the unique walk of
$n^*$ --- the exponents agree slot by slot, and each slot hosts one
engagement of a single grade --- the window product in
\eqref{eq:si-thmpf-decode} matches $|C_{n^*}|$ up to a constant
depending only on $n^*$, while each outside engagement contributes a
factor $\le\sigma$ and each dwell resummation a factor
$\le(1-\bar\rho_0)^{-1}$. Summing over survivors,
\begin{equation}
\sum_{\substack{n\ne n^*\\ \nu|_{\rm win}=\nu^*}}|a_n|
\;\le\;C_{n^*\!,\rm loc}\;|C_{n^*}|\;\sigma,
\label{eq:si-thmpf-tail}
\end{equation}
with $C_{n^*\!,\rm loc}$ depending only on $n^*$ and the guard margins.
(This refines the cruder pairing
$O(\sigma^{k^*+1})\times O(\sigma^{-k^*})$: for patterns carrying a
$\pm2$ window exponent the walk coefficient is smaller than
$\sigma^{k^*}$ --- the double-charge amplitude is $\Theta(\sigma^2)$,
cf.\ Eq.~\eqref{eq:si-grade2}, whose grade-two coefficient is
$\Theta(\sigma^3)$ --- and the per-winding bookkeeping above is what
delivers the bound uniformly over admissible $n^*$.) Since $\|w\|_1=1$
(the DFT average is unimodular) and $|E_n|=1$,
\begin{equation}
\big\|f-E_{n^*}\big\|_\infty
\;\le\;\frac1{|C_{n^*}|}
\sum_{\substack{n\ne n^*\\ \nu|_{\rm win}=\nu^*}}|a_n|
\;\le\;C_{n^*\!,\rm loc}\,\sigma\;\longrightarrow\;0
\label{eq:si-thmpf-error}
\end{equation}
as $r\to0$ along guard-compatible operating points. Finally
$C_{n^*}\ne0$: the walk decoding of Lemma~1(c) is unique, and each of
its factors --- engagement amplitudes, dwells, $J_{\rm ss}$ --- is
individually nonzero.
\end{proof}

\smallskip\noindent\emph{Lower inclusion, assembled.} Let
$n\in\mathfrak N_D$: by definition
\begin{equation}
n=\sum_{i=1}^{D'}n^{(i)},\qquad D'\le D,\quad
\text{each }n^{(i)}\text{ admissible or co-admissible}.
\label{eq:si-lower-decomp}
\end{equation}
Theorem~\ref{thm:matching} supplies order-one functionals $f_i$ with
$\|f_i-E_{n^{(i)}}\|_\infty\le\varepsilon'$ (co-admissible blocks via
the conjugate features), each bounded by $\|f_i\|_\infty\le1+v_\infty$.
Since characters multiply, $E_n=\prod_iE_{n^{(i)}}$, the telescoping of
clause (II), Step~2 [the factor-by-factor product-difference identity of
Sec.~\ref{sec:si-prfII}] gives
\begin{equation}
\Big\|\prod_{i=1}^{D'}f_i-E_n\Big\|_\infty
\;\le\;\sum_{i=1}^{D'}\Big[\prod_{l<i}\|f_l\|_\infty\Big]
\big\|f_i-E_{n^{(i)}}\big\|_\infty
\;\le\;D'\,(1+v_\infty)^{D'-1}\,\varepsilon'
\;\xrightarrow[\varepsilon'\to0]{}\;0,
\label{eq:si-lower-telescope}
\end{equation}
a polynomial-order-$D'$ readout, so $E_n\in\mathcal A_D$. Together with
Route A this proves $\mathcal A_D=\overline{\operatorname{span}}
\{E_n:n\in\mathfrak N_D\}$. \emph{(Caution: it is not the case that
every pattern with $|q(n)|\le D$ factors into $D'\le D$
charge-$\pm1$ blocks; it does not: blocks of charge $\pm1$ summing to
$q$ force
\begin{equation}
D'\ge|q|,\qquad D'\equiv q\ (\mathrm{mod}\ 2),
\label{eq:si-lower-parity}
\end{equation}
so nontrivial charge-$0$ patterns need $D'\ge2$; per-slot exponents cap
at $2D'$; and block admissibility constrains further.)}

\smallskip\noindent\emph{Two obstructions, making the refinement
sharp.} (i)~\emph{Parity.} Take $n=e_a-e_b$ (exponent $+1$ at slot $a$,
$-1$ at slot $b$): $q(n)=0$, yet $n\notin\mathcal N_1$, a single block
having charge $\pm1$. Quantitatively $E_n\notin\mathcal A_1$. Under the
global slice $s\mapsto s+\tau\mathbf 1$, $\chi=\beta\tau$, every
$g\in\mathcal A_1$ decomposes while the target is invariant:
\begin{equation}
g(s+\tau\mathbf 1)=c+e^{i\chi}A(s)+e^{-i\chi}B(s),\qquad
E_n(s+\tau\mathbf 1)=E_n(s),
\label{eq:si-obstr-slice}
\end{equation}
with frequency-$0$ component the \emph{constant} $c$ (the constant term
of the readout, base-point independent). Apply the three-point instance
of Lemma~\ref{lem:gap} --- nodes $\chi_\ell=\ell h$, $\ell=0,1,2$,
spacing $h=a/4$ inside the half-box slice range, $a=\beta s_{\max}$;
weights the coefficients of $p(z)=(z-e^{ih})(z-e^{-ih})$, which
annihilates the frequencies $\pm1$ and sees the frequency-$0$ component
with $|p(1)|=4\sin^2(a/8)$, $\|\Lambda\|\le4$:
\begin{equation}
\operatorname{dist}_\infty(E_n,\mathcal A_1)
\;\ge\;\sup_s\frac{|E_n(s)-c|\cdot4\sin^2(a/8)}{4}
\;\ge\;\sin(a)\,\sin^2\!\Big(\frac a8\Big),
\label{eq:si-obstr-parity}
\end{equation}
the last step because on the half-box $E_n$ sweeps the arc
$\{e^{i\varphi}:|\varphi|\le a\}$, whose endpoints lie $2\sin a$ apart,
so by the triangle inequality any fixed $c$ sits at distance $\ge\sin a$
from one of them. At the campaign point $a=1$ the bound is
$1.3\times10^{-2}$. (ii)~\emph{Per-slot.} For any $D$, a pattern with
one slot exponent $n_{t_0}=2D{+}1$, compensated elsewhere so
$|q(n)|\le D$, satisfies the sector cap but is not $D$-assembleable.
Under the single-slot slice $s_{t_0}\mapsto s_{t_0}+\tau$, Route B
confines every element of $\mathcal A_D$ while the target escapes:
\begin{equation}
g\in\operatorname{span}\{e^{ik\beta\tau}:|k|\le2D\},\qquad
E_n\propto e^{i(2D+1)\beta\tau},
\label{eq:si-obstr-perslot}
\end{equation}
and Lemma~\ref{lem:gap} with band $|k|\le2D$ and target frequency
$2D{+}1$ yields
\begin{equation}
\operatorname{dist}_\infty(E_n,\mathcal A_D)
\;\ge\;\delta(2D,a)
\;=\;\prod_{j=1}^{4D+1}\sin\!\Big(\frac{j\,a}{4(4D+1)}\Big)\;>\;0.
\label{eq:si-obstr-perslot-gap}
\end{equation}
Hence $\overline{\operatorname{span}}\bigcup_{|q|\le D}\mathcal V_q
\not\subseteq\mathcal A_D$ for every $D$: the sector equality is
false, not merely unproven.\hfill$\blacksquare$

\begin{remark}[Confinement enables matching]
The projection is finite and exact \emph{because} the per-slot law caps
the window exponents at modulus two: five characters per slot suffice,
and the digit base is a constant --- independent of the grade, the
order, and the depth of the matched block: charge confinement thins the
combinatorics to a fixed alphabet. That the alphabet cannot be smaller
--- base $4$ collides --- is the \emph{Why base $5$} note following the
proof of Lemma~\ref{lem:projection}. The same algebra that forbids in clause (III) is what makes
the projector finite in clause (I).
\end{remark}

\begin{remark}[Designed versus generic operation]
The projector spends $2\cdot5^L$ runs (two readout slots per epoch),
priced in shots and amortized over the entire window block. A
generically operated device does not pay it: at a generic operating
point the projector is a diagnostic instrument, not a runtime cost.
\end{remark}

\subsection{Proof of register neutrality (main-text Proposition 1)}
\label{sec:si-prf-register}

\emph{What is proved.} On the full register, with
$\mathsf M_{ij}=\avg{a_ia_j}$, $\bar{\mathsf M}=\overline{\mathsf M}$,
and $\mathsf J_{ij}=\avg{a_i^\dagger a_j}+\tfrac12\delta_{ij}$, the
gauge action of Lemma~1(a) survives every added hardware element:
advancing all pump-referenced squeeze angles by a common $\chi$ maps the
exact state by $\mathsf M\mapsto e^{i\chi}\mathsf M$,
$\bar{\mathsf M}\mapsto e^{-i\chi}\bar{\mathsf M}$,
$\mathsf J\mapsto\mathsf J$; consequently the selection rule, the
ordering law, the per-slot bound, the upper inclusion of clause~(I), and
clause~(III) hold verbatim on the register.

\emph{Why it is true.} The single-mode conservation law was a frame
statement: squeezing at angle $\theta$ is squeezing at angle zero
conjugated by a pump-frame rotation. Every element added by the register
--- beamsplitters, static phases, delays, per-bin loss, ancilla reset,
detection --- is \emph{phase covariant}: it commutes with rotating all
modes together, because it never references the pump. Only the oriented
squeezers know the pump phase, and they respond to a frame rotation
exactly by the angle advance. So a global advance of the squeeze angles
is, for the entire register dynamics, a change of frame.

\begin{proof}
Let $W_\chi$ be the \emph{band-weighted} frame rotation:
\begin{equation*}
a_i\mapsto e^{i\chi/2}a_i\quad[R(\chi/2)]\ \ \text{on every
fundamental-band mode }(a_s,\,A_1..A_N,\,v),\qquad
b\mapsto e^{i\chi}b\quad[R(\chi)],
\end{equation*}
the harmonic carrying two fundamental quanta, hence twice the frame
phase. Classify the elements of one bin step
(Sec.~\ref{app:blocks}, B1--B6):

\emph{(i) Oriented squeezers.} With $R(\phi)$ the quadrature-plane
rotation and $Z(r)=\mathrm{diag}(e^{r},e^{-r})$ the axis-aligned
squeeze, the oriented squeezer is by definition
\begin{equation*}
S_{\rm sq}(r,\theta)=R\big(\tfrac{\theta}{2}\big)\,Z(r)\,
R\big(\tfrac{\theta}{2}\big)^{\!\top}
\end{equation*}
--- squeeze axes at half the pump angle. Composing rotations, both
sides of the conjugation identity equal
$R(\tfrac{\theta}{2}+\phi)\,Z(r)\,R(\tfrac{\theta}{2}+\phi)^{\top}$:
\begin{equation*}
R(\phi)\,S_{\rm sq}(r,\theta)\,R(\phi)^{\top}
=S_{\rm sq}(r,\theta+2\phi),
\end{equation*}
and at $\phi=\chi/2$,
\begin{equation*}
S_{\rm sq}(r,\theta+\chi)=W_\chi\,S_{\rm sq}(r,\theta)\,W_\chi^{\top}
\end{equation*}
on the squeezed mode.

\emph{(ii) Passive elements.} Beamsplitters, static phases, arm
rotations, and delays act \emph{within} one wavelength band (the WDM
separates fundamental from harmonic), as a mode-space unitary $U$ on
modes sharing a common $W_\chi$ rotation; hence
\begin{equation*}
[\,U,\,W_\chi\,]=0\qquad\text{bandwise.}
\end{equation*}

\emph{(iii) Loss, reset, and injection.} Per-mode loss,
\begin{equation*}
\bm\Sigma\mapsto\eta\,\bm\Sigma+(1-\eta)\,\bm\Sigma_{\rm vac},\qquad
\bm\Sigma_{\rm vac}\propto\mathds{1}\ \text{rotation invariant},
\end{equation*}
has a scalar transmission factor; ancilla reset replaces a mode by the
rotation-invariant vacuum. The $\chitwo$ mean source of block B3
drives the harmonic mean linearly by the fundamental
$\mathsf M$-channel,
\begin{equation*}
\Delta\beta_b\propto-i\mu\,m_h
\qquad\Longrightarrow\qquad
W_\chi:\ \Delta\beta_b\mapsto e^{i\chi}\Delta\beta_b,\quad
m_h\mapsto e^{i\chi}m_h,
\end{equation*}
both sides gaining the same phase --- covariant, a source
\emph{inside} the graded algebra, as in Proposition~2(i). More
compactly, the entire $\chitwo$ segment at every order is generated by
\begin{equation*}
H\propto\mu\big(a_h^2b^\dagger+a_h^{\dagger2}b\big),\qquad
a_h^2b^\dagger\mapsto\big(e^{i\chi/2}\big)^{2}e^{-i\chi}\,
a_h^2b^\dagger=a_h^2b^\dagger,
\end{equation*}
invariant under the band-weighted rotation, so its $\mu^1$ mean source
and $\mu^2$ covariance back-action are covariant together. One scoping
note: the optional linear $a\!\leftrightarrow\!b$ mean coupling of
block B3 (``displacement toggle on'') is covariant only if the
displacement field is itself pump referenced, rotating with the
fundamental band; the operating mode treated here is zero-mean
(toggle off), and the neutrality claim is scoped to it. The
fixed-frame output chain (PSA and homodyne local oscillator, block B4)
is deliberately outside $W_\chi$: it is detection, not dynamics, and
its interaction with the gauge phase is precisely what the
gauge-projector measurement reads out (Remark on locality) and what
the detection-attribution corollary prices.

Composing (i)--(iii): with $\Phi_{\bm\theta}$ the full one-bin map at
squeeze angles $\bm\theta$,
\begin{equation*}
\Phi_{\bm\theta+\chi\mathbf 1}=W_\chi\,\Phi_{\bm\theta}\,W_\chi^{\top},
\end{equation*}
and the fixed sources (vacuum injections) are $W_\chi$-invariant.
Iterating over the input history and evaluating the three channel
blocks yields
\begin{equation*}
\mathsf M\mapsto e^{i\chi}\,\mathsf M,\qquad
\mathsf J\mapsto\mathsf J\qquad\text{exactly}
\end{equation*}
--- the register conservation law.

Transport of the laws is now bookkeeping. \emph{Selection:} the
coefficient-matching argument of Sec.~\ref{sec:si-prf-laws} used only
the equivariance and absolute convergence; both hold on the register
(contraction from the Tier-1 stability $\rho(S)<1$).
\emph{Per-slot and ordering, by sector confinement.} These two laws
follow from the charge grading alone, with no count of how many
squeezers a symbol addresses --- important, because under input-hold
mask multiplexing one symbol may drive several bins. The register
channels carry charges $(+1,-1,0)$, so a walk's in-flight charge is
confined to $\{-1,0,+1\}$; the net exponent a symbol $t$ contributes
to a readable character is the charge change across that symbol,
\begin{equation*}
n_t=q_{\rm out}(t)-q_{\rm in}(t),\qquad
q_{\rm in},q_{\rm out}\in\{-1,0,+1\}
\;\Longrightarrow\;|n_t|\le2
\end{equation*}
--- the per-slot law, for any assignment of bins to symbols. For the
latest symbol with $n_{t}\ne0$: symbols engaged with net zero preserve
the sector, and the walk must end in the readable $\mathsf M$ sector,
so
\begin{equation*}
q_{\rm out}(t^*)=+1,\qquad
n_{t^*}=1-q_{\rm in}(t^*)\in\{+1,+2\},
\end{equation*}
never negative --- the ordering law. (Blockwise holomorphy of the
$\mathsf M$ row, inherited entrywise from the single-mode table
tensored with charge-$0$ passive factors, gives the same conclusion at
bin resolution.) The upper inclusion of clause~(I) is Route~A run on
the register series, and clause~(III) used nothing but the
equivariance. What does \emph{not} transport by this argument is
single-setting richness (the lower-inclusion diversity), which remains
an open hypothesis --- exactly as stated in the main text.
\end{proof}

\subsection{Proof of the classical-light equivalence (main-text
Proposition 2)}
\label{sec:si-prf-classical}

The classical-light twin (main-text Proposition~2 and Methods) is the
same machine with one alteration. After each squeeze, we add just enough
classical noise $\bm{\mathcal N}_{\rm cl}(\theta)$ to restore the squeezed,
sub-vacuum quadrature to the vacuum level, while preserving the
antisqueezed axis and its orientation. The state is thereby pushed back
to the boundary of what a classical, P-representable light field can be.
Crucially, the data are encoded in the orientation of the noise ellipse,
and the orientation is exactly what this operation leaves alone. That is
the intuition behind the equivalence: the only thing removed is the
sub-vacuum axis, which never carried any data.

\begin{proof}
\emph{(i) The source lives inside the graded algebra.} With
$\Xi(\theta)=\big(\begin{smallmatrix}\cos\theta&\sin\theta\\
\sin\theta&-\cos\theta\end{smallmatrix}\big)$ the reflection, the
projector onto the antisqueezed axis decomposes as
\begin{equation}
R\big(\tfrac{\theta}{2}\big)\,\mathrm{diag}(1,0)\,
R\big(\tfrac{\theta}{2}\big)^{\!\top}
=\tfrac12\,\mathds1+\tfrac12\,\Xi(\theta),
\label{eq:si-cl-proj}
\end{equation}
isotropic vacuum plus a traceless part along the antisqueezed axis
[expand $R(\varphi)\,\mathrm{diag}(1,0)\,R(\varphi)^{\top}
=\tfrac12\mathds1+\tfrac12\Xi(2\varphi)$ at $\varphi=\theta/2$].
Translate any symmetric quadrature source $\bm V$ into the channel
basis \eqref{eq:si-state} --- for a mean-zero mode,
$a^2=X^2-P^2+i(XP+PX)$ and $a^\dagger a=X^2+P^2-\tfrac12$
[using $[X,P]=\tfrac i2$], so an added covariance $\bm V$ shifts the
channels by
\begin{equation}
\bm V\;\longmapsto\;
b[\bm V]=\big(\,V_{XX}-V_{PP}+2iV_{XP},\;\;
\overline{(\cdot)},\;\;V_{XX}+V_{PP}\,\big)^{\!\top}.
\label{eq:si-cl-translate}
\end{equation}
Applied to $\bm{\mathcal N}_{\rm cl}(\theta)=\nu_{\rm cl}
\big(\tfrac12\mathds1+\tfrac12\Xi(\theta)\big)$: the isotropic part
contributes only to $J$; the $\Xi$ part gives
$\Xi_{XX}-\Xi_{PP}=2\cos\theta$ and $2i\,\Xi_{XP}=2i\sin\theta$; hence
\begin{equation}
b_{\rm cl}(\theta)=s_{\rm cl}\,
\big(e^{i\theta},\,e^{-i\theta},\,1\big)^{\!\top},\qquad
e^{i\chi Q}\,b_{\rm cl}(\theta)=b_{\rm cl}(\theta+\chi),
\label{eq:si-cl-source}
\end{equation}
with $s_{\rm cl}>0$ the common strength (at the reduced-model
conventions of \eqref{eq:si-state}, $s_{\rm cl}=\tfrac\eta4(1-e^{-2r})$:
the twin code injects the noise between squeeze and loss, so one factor
of the per-step transmission multiplies it; on the register the
corresponding prefactor is the $\eta_{\rm esc}(1-e^{-2r})$ of the
main-text statement, the unit choices being the invariant rescalings of
Subsec.~S1\,D\,2). The three components sit at charges $(+1,-1,0)$ ---
Laurent support $|d|\le1$ --- so the twin recursion
\begin{equation}
v_t^{\rm cl}=T(\theta_t)\,v_{t-1}^{\rm cl}+b_{\rm vac}
+b_{\rm cl}(\theta_t)
\label{eq:si-cl-recursion}
\end{equation}
is generated by the same gauge-graded transfer operator plus a source
inside the graded algebra: under $\theta\mapsto\theta+\chi\mathbf1$
every term of the unrolled series transforms covariantly [by
\eqref{eq:si-grading} for the $T$-factors, invariance of $b_{\rm vac}$,
and \eqref{eq:si-cl-source}], so $v_t^{\rm cl}\mapsto e^{i\chi Q}
v_t^{\rm cl}$ --- the selection rule; the readable-row holomorphy
$e_m^\dagger T_{-1}=e_m^\dagger T_{-2}=0$ is a property of $T$ alone ---
the ordering law; and each slot receives one engagement grade
$|d|\le2$ or one source grade $|d|\le1$ --- the per-slot bound.
Corollary~1 holds for the classical twin \emph{unchanged}.

\emph{(ii) Superposition and containment.} The recursion
\eqref{eq:si-cl-recursion} is affine, so its bounded solution splits
exactly:
\begin{equation}
v_t^{\rm cl}=v_t+\delta v_t,\qquad
\delta v_t=\sum_{k\ge0}\Big[\prod_{j=0}^{k-1}T(\theta_{t-j})\Big]
b_{\rm cl}(\theta_{t-k}),\qquad
\delta_{\rm cl}:=e_m^\dagger\,\delta v_t,
\label{eq:si-cl-split}
\end{equation}
absolutely convergent by the contraction bound,
$\|\delta v_t\|_\infty\le s_{\rm cl}\,\sqrt2/(1-\bar\rho)$.
Dyson-grading $\delta v_t$ in $T_{\rm od}$ as in
Eq.~\eqref{eq:si-dyson}: each $\delta$-walk is a quantum walk with its
initial $b_{\rm vac}$ emission replaced by a $b_{\rm cl}$ emission at
slot $i$, depositing winding $q_c\in\{+1,-1,0\}$ there and launching
the walk at in-flight charge $q_c$; the deposited windings telescope to
\begin{equation}
q(n)=q_c+\big(q_{\rm end}-q_{\rm start}\big)
=q_c+\big({+1}-q_c\big)=+1
\label{eq:si-cl-charge}
\end{equation}
for every readable $\delta$-character [the in-flight bookkeeping of
Sec.~\ref{sec:si-prf-register}], and the ordering and per-slot checks
run as in (i). Hence $\delta_{\rm cl}$ is an absolutely convergent
series of admissible characters, so $\delta_{\rm cl}\in\mathcal A_1$
by Theorem~1(I) at $D=1$; and any order-$D$ polynomial in the twin
features $m_{\rm cl}=m+\delta_{\rm cl}$, $\bar m_{\rm cl}$ expands by
the Cauchy product into $D$-assembleable characters: every twin feature
lies in $\mathcal A_D$.

\emph{(iii) Span equality.} Fix $D$, a finite readout window, and the
finite basis $\{E_n:n\in\mathfrak N_D^{\rm win}\}$ of $D$-assembleable
characters supported there. By (ii) and Theorem~1(I) both machines'
noiseless order-$D$ monomial families are finite combinations of this
one basis: feature vectors $F(\varphi)=C(\varphi)\,E$ and
$F^{\rm cl}(\varphi)=C^{\rm cl}(\varphi)\,E$ with coefficient matrices
depending on the mask $\varphi$. Span equality at mask $\varphi$ is
$\operatorname{row\,span}C(\varphi)=\operatorname{row\,span}
C^{\rm cl}(\varphi)$, implied by both Gram determinants
$\det(CC^\dagger)$, $\det(C^{\rm cl}C^{\rm cl\dagger})$ attaining full
rank on the shared basis; these are real-analytic on the mask torus by
the following lemma, so nonvanishing at one mask implies nonvanishing
for almost every mask.

\begin{lemma}[Mask genericity]\label{lem:genericity}
On the guard region $e^{r}\sqrt{\eta_{\rm fb}\eta_L\eta_{\rm esc}}<1$,
the stationary settle state, and hence every noiseless feature, is
real-analytic in the mask phases $(\theta_1,\dots,\theta_{N'})$; the
order-$D$ feature Gram determinant is therefore real-analytic on the
torus $\mathbb T^{N'}$, and if it is nonzero at one mask it is nonzero
for almost every mask.
\end{lemma}

\begin{proof}
Each per-bin operation (squeeze at angle $\theta$, loss, interferometer,
feedback) is an entire function of the mask phases acting on covariance
blocks; the settle state is the limit of the geometrically convergent
composition (contraction rate $\bar\rho<1$ under the guard), and a
uniform limit of analytic maps with geometric error is analytic on the
open guard region. Determinants and Gram matrices of analytic families
are analytic; the zero set of a nonzero real-analytic function on
$\mathbb T^{N'}$ has measure zero.\hfill$\blacksquare$
\end{proof}

Clause (iii) is therefore span
equality for almost every mask; the operating masks are assumed
generic, and ``computes the same functions'' in the main text carries
this status.
\end{proof}
\section{Feature statistics and calibrations}
\label{app:features}

Per-bin single-shot features are Gaussian with the mean and variance of
Appendix~\ref{app:blocks}(B4); the finite shot budget $B$ enters as the number of
harvested bins. The optional quadratic layer uses products $f(k,n)f(k',n')$
restricted to lags with Tier-1 kernel/entanglement support; same-bin squares carry
the shot-noise bias $\Sigma_{X_bX_b}^\vartheta$, measured once with the input dark
and subtracted. Jitter enters as $r_k=r(1+\delta_k)$, $\mu_k=\mu(1+\gamma_k)$; to
first order the feature covariance acquires
$\bm\Sigma_{\rm feat}\mathrel{+}= {\rm Var}(\gamma)\,\partial_\gamma\bm
m_b\,\partial_\gamma\bm m_b^{\top}+{\rm Var}(\delta)\,\partial_\delta\bm
m_b\,\partial_\delta\bm m_b^{\top}$, evaluated along the propagation. Readout
training is ridge regression with regularization matched to the measured feature
noise floor; the echo-state property is verified operationally (two initial
conditions, Frobenius convergence) after a washout of $\gtrsim5/(1-\rho(S))$
circulations.

\section{Benchmark validation and diagnostics}
\label{sec:si-narma}

We exercise the full-register twin on the NARMA2 benchmark,
$y_{t+1}=0.4y_t+0.4y_ty_{t-1}+0.6u_t^3+0.1$ with $u_t\sim U(0,0.5)$,
using the pre-registered protocol and baseline constructions detailed in
the Methods of the main text. There are three task seeds with common
random numbers, and all numbers below are three-seed means; the
single-seed arm reproduces the ladder of the main-text
Methods exactly.

One methodological commitment governs everything here: ``fair''
comparison means \emph{matched on the currency of the claim being
tested}. Concretely, that is readout dimension and training data for the
software baseline; photons and hardware for the classical-light control;
and total circulation count for the moment-readout reservoir.

\begin{figure*}[t]
\centering
\includegraphics[width=\textwidth]{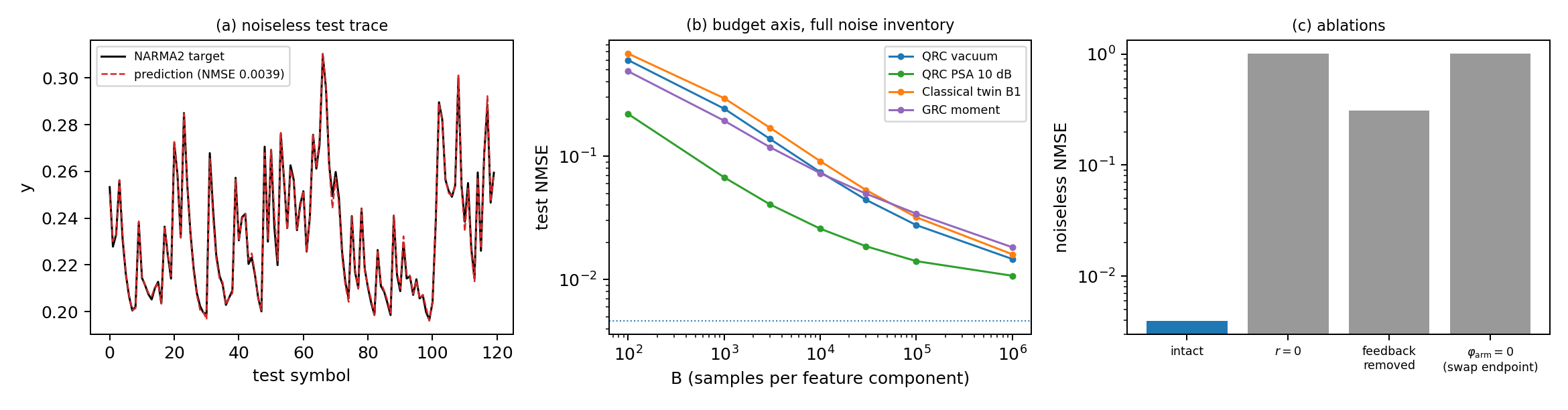}
\caption{\textbf{NARMA2 performance of the full-register twin.} (a) Test-set
trace, NMSE $=0.0039$ with noiseless features (all features are covariance
read-offs of a strictly zero-mean state; echo-state property verified
operationally to $6\times10^{-7}$ at single precision). (b) NMSE versus $B$
(samples per feature component) under the full noise inventory, for
vacuum-limited and PSA-squeezed readout, the classical-light twin, and the
moment-readout reservoir; 10~dB of readout squeezing is worth approximately a
tenfold budget (horizontal shift between curves). (c) Ablations --- removing
squeezing ($r{=}0$) destroys the machine entirely (the covariance then carries
no imprint of the encoding); removing the coherent feedback path degrades NMSE
by nearly two orders of magnitude (the loop supplies the memory NARMA2
requires); the interferometer arm phase is structural, not incidental: at the
swap endpoint $\varphi_{\rm arm}{=}0$ the encoded bin is routed entirely into
the delay and the readout port carries no input dependence.}
\label{fig:narma2}
\end{figure*}

\begin{figure*}[t]
\centering
\includegraphics[width=0.92\textwidth]{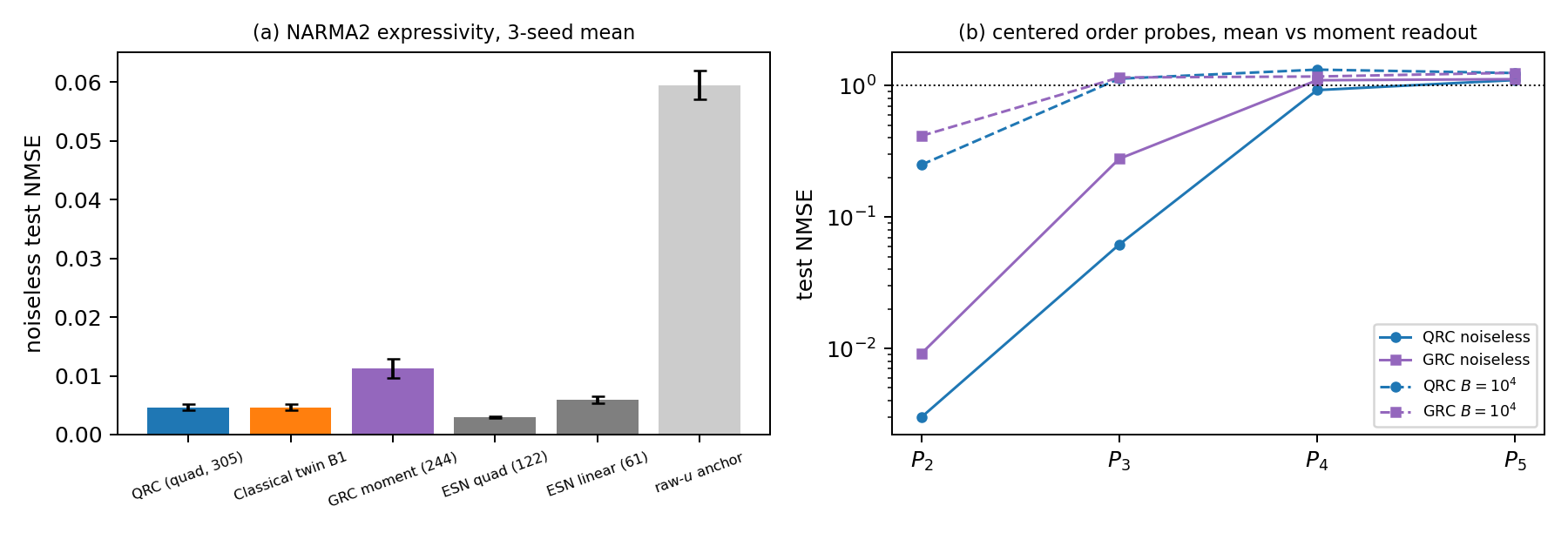}
\caption{\textbf{Matched baselines and order probes.} (a) Noiseless
(expressivity) comparison on NARMA2. The classical-light twin matches the
quantum machine to $8\times10^{-6}$ --- the classical simulability of the
Gaussian tier \cite{MariEisert2012} manifest in the data --- and a tuned
61-unit echo-state network with the same per-unit quadratic tier wins the
noiseless comparison outright. (b) Centered order probes
$P_m=\prod_{i=1}^m s_{t-i}$ (zero-mean drive; pure order-$m$ content):
order-one mean readout with the quadratic tier beats the moment readout
within reach ($P_2$, $P_3$), and both Gaussian readouts saturate by
$P_4$--$P_5$ at the register's loss budget.}
\label{fig:si-baselines}
\end{figure*}

Seven findings follow (Fig.~\ref{fig:narma2}; Fig.~\ref{fig:si-baselines};
main-text Table~I), each of which sharpens rather than merely supports the
thesis. Findings (1)--(5) are from the pre-registered campaign; findings (6)
and (7) are post-review additions, labeled by their tranche drivers and
deterministic given the disclosed seeds (see the main-text Methods).

\emph{(1) No quantum expressivity advantage, as the theory demands.}
With noiseless features, the classical-light control matches the quantum
machine to $8\times10^{-6}$ in NMSE. The reason is the one identified in
Proposition~2: the data-bearing structure lives in the orientation of the
noise ellipse, which classical light reproduces exactly, and the
sub-vacuum axis contributes nothing to noiseless feature content. This is
the simulability theorem \cite{MariEisert2012} observed empirically in
our own machine, and it validates the framing chosen throughout the
paper: the claims live in economics, not expressivity.

The tuned echo-state network wins the noiseless comparison outright at
the same quadratic tier (0.0029 versus 0.0046). A physical reservoir
should not be expected to beat float-precision software on feature
quality, and we report the result plainly. All machines clear the
raw-input anchor (0.060) by an order of magnitude or more.

\emph{(2) At low task order, the surviving readout advantage is loss
suppression, shared with the twin.}
The amplifier's benefit is the loss-suppression factor $1.22$ in
variance ($1.10\times$ in NMSE at $B=10^4$), and it accrues to the
classical-light twin equally.
Under finite budgets, what separates the
machines at matched detection noise is the conditioning margin of
main-text Table~I, not a detection-variance option: the amplifier is
granted to both. The noise-matched ESN, granted the QRC's per-feature
SNR, degrades faster than every optical machine (0.112 at $B=10^4$),
which tells us the reservoir's features are the better-conditioned basis
at fixed readout SNR.

\emph{(3) Moment readout is competitive at low budget, and is overtaken
on the budget axis.} The Gaussian moment-readout reservoir, granted the
same circulation count, edges the vacuum-readout machine at $B\le10^4$
(0.073 versus 0.074). That is not surprising: the factorial penalty
$(2m{-}1)!!$ is only a factor of 3 at $m{=}2$, and its features have
$O(\Sigma)$ amplitude.

But the penalty is a \emph{noise} penalty, and noise penalties tell on
the budget axis. By $B=10^5$ the order-one machine leads (0.028 versus
0.034), and the gap widens through $B=10^6$ (0.0147 versus 0.0183).
Moreover the moment estimator's sampling noise $\propto\Sigma^2$ cannot
be improved by phase-sensitive gain, so the PSA option, available only to
order-one readout, beats the moment machine at \emph{every} budget
tested. Our cost model further predicts that the residual low-budget
competitiveness must vanish as required task order grows, and we tested
that prediction directly on order-isolating probes.

\emph{(4) The order sweep: the reduced-model result falsified our
crossing prediction, and the full register locates its boundary.} On the
reduced (Tier-2) model we had predicted that the moment readout's
factorial cost must reverse the NARMA2 gap as required order grows. The
prediction failed: the moment readout matched or beat the order-one
machine at every order tested there.

The mechanism was identifiable, and we retain it as a correction to the
framing of continuous-variable QRC, our own included. The factorial cost
[main-text Eq.~(1)] taxes schemes that must reach input order $m$ by
\emph{climbing the moment ladder}. But the ordered-product richness of
bilinear encoding lives in the \emph{state}, and both readouts inherit
the state.

On the hardware-faithful full register the same probe, run on
order-isolating centered targets $P_m=\prod_{i=1}^m s_{t-i}$ (main-text
Fig.~4b), sharpens this into a boundary statement---corrected under the
C3 referral. \emph{Noiselessly} (referral-invariant) order-one mean
readout with the per-unit quadratic tier beats the moment readout at
$P_2$ (0.003 versus 0.009) and decisively at $P_3$ (0.062 versus
0.277): the informational lead of bilinear encoding. \emph{Under a
common physical budget}, however, the moment readout leads at every
$B_{\rm phys}$ tested, because at moment orders $2$--$3$ the factorial
penalty
[$3!!$, $5!!$] is far below the transduction penalty
$\mu_T^{-2}=10^4$---the crossover law $m^\ast(\mu_T)=6$ of the main
text, reproduced on the register.

The reach itself, however, is finite. At the register's loss budget (measured per-bin covariance decay $0.907$)
both Gaussian readouts saturate by $P_4$--$P_5$. So
the reduced model's high-order regime, where the moment readout had
appeared to express deep parity essentially exactly, is a property of the
lossless idealization and not of the machine.

The measurement-economics thesis survives, in a sharpened and narrower
form, with two clauses. (i) The factorial argument applies to
displacement-encoded or otherwise encoding-limited Gaussian schemes,
which must estimate genuinely high-order moments. (ii) Where strong
encoding nonlinearity is available, the operative quantum resource at any
reachable input order is squeezed detection, which order-one readout can
exploit and moment estimators cannot, their sampling noise
$\propto\Sigma^2$ amplifying under phase-sensitive gain.

\emph{(5) The encoding-limited regime: the crossing appears.} We repeat
the comparison with data encoded in a \emph{displacement}, at fixed
squeeze angle. This is the regime where covariances are input-independent
and all signal lives in the means. The Gaussian machine's features are
then strictly linear in input history, so order $m$ is reachable only by
post-processing products of noisy estimates, whereas the bilinear machine
forms its products before detection, through the non-commuting symplectic
composition. Both machines are granted identical post-processing (per-bin
monomials to polynomial order 3 plus adjacent triple products);
Table~\ref{tab:disp} reports centered targets and three-seed means.

At orders 2 and 3 the displacement machine's $O(1)$-amplitude means cube
affordably and win under budget (0.076 versus 0.474 at $P_3$,
$B{=}10^5$). At order 4 it fails at every budget (NMSE $\ge1$), while the
bilinear machine retains genuine order-4 content (0.70 noiseless) before
the loss saturation of finding~(4) closes the window at $P_5$.

The demonstrated law is that pre-detection products act as an \emph{order
multiplier}. At matched polynomial order $p$, order-one readout of
a linear-mean machine reaches task order $p$ and no further, while the
bilinearly encoded machine reaches beyond $p$, up to the register's
loss-limited depth. In-dynamics nonlinearity is decisive exactly when the
required order exceeds what post-processing can affordably supply. This
is the precise, and now demonstrated, form of the measurement-economics
thesis.

\emph{(6) The coherent arm's value over measured feedback is
measurement-economic, not expressive.}
The architecture provides an
electronic measured-feedback arm (Sec.~S6), in which coherent
re-injection is replaced by a programmable exponential kernel of the
homodyne record modulating the pump,
$r_k=r[1+G_{\rm fb}\sum_jh_j\hat Y_{k-jN}]$. The gain was selected once
on the seed-11 training block over the six-point grid
$G_{\rm fb}\in\{0.05,\dots,1.2\}$, on which the training objective
decreases monotonically to the grid boundary $G_{\rm fb}=1.2$, where the
$\pm50\%$ actuator-excursion bound saturates the modulation. That is an
edge selection, and we disclose it as such.

Granted a \emph{noiseless} record, the electronic arm nearly emulates the
coherent memory ($0.0071\pm0.0005$ versus $0.0046\pm0.0006$ noiseless
NMSE, three task seeds; the no-memory control sits at $0.32$). At the
matched record budget $B=10^4$, however, the emulation pays its
structural price: measurement noise recirculates through the dynamics,
and the measured arm lands at $0.098\pm0.003$ against the coherent arm's
$0.072\pm0.008$, with coherent lower on every seed. The reason is not that classical feedback cannot express the
kernel, since noiselessly it nearly can. The reason is that the emulation
must pay measurement noise inside the loop, which the coherent arm never
pays. The memory advantage is the measurement-economics thesis applied to
the loop itself.

\emph{Memory-capacity sweep (extension of the main-text NARMA10 boundary result): a
falsified pre-registration and the budgeted law that replaced it.} To
convert the depth-limit account from two point failures (NARMA10;
$P_4$--$P_5$ saturation) into a curve, we pre-registered a linear
memory-capacity sweep: MC$(\ell)$ against lag $\ell=1..15$ at five
round-trip amplitudes $G\in\{0.60,0.70,0.785,0.85,0.90\}$, with the
half-capacity cutoff predicted to track $(1-G)^{-1}$
(frozen before any number was
computed). The noiseless prediction \emph{failed}: the cutoff saturates
at $9,11,14,15,15$ lags and the capacity regression reaches only
$R^2=0.77$ . The mechanism was
identified before any new run: loss attenuates the echo amplitude by
$\sim G^{\ell}$ but adds no stochasticity to an exactly read covariance
feature, so noiseless correlation persists to float precision. Noiseless
memory capacity probes information \emph{presence}; the depth limit
$M\approx(1-G)^{-1}$ is a finite-budget statement. The disclosed
addendum, a single re-run of the identical sweep under the campaign's
analytic shot model at $B=10^4$ (vacuum readout, \texttt{ridge\_rung}
verbatim), delivers the budgeted law cleanly: total capacity
MC$_{\rm tot}=0.16,0.26,0.41,0.59,0.85$ rises monotonically with $G$ and
regresses on $(1-G)^{-1}$ with slope $0.090$ and $R^2=0.992$
. Retrievable memory,
priced at the campaign budget, is proportional to the predicted depth;
information mere presence is not. The finding is a third instance of the
paper's central distinction, expressible versus affordable, applied this
time to memory itself.

\begin{table}[!htb]
\caption{Encoding-limited regime: centered order probes $P_m$, test NMSE
(three-seed means, full register). Both machines granted identical
post-processing (per-bin monomials to polynomial order 3 plus adjacent
triple
products); budget columns are shot noise at $B$ samples per feature
component, vacuum-limited for both.}
\label{tab:disp}
\begin{ruledtabular}
\begin{tabular}{lcccccc}
 & \multicolumn{3}{c}{bilinear machine} & \multicolumn{3}{c}{displacement (means)}\\
$m$ & noiseless & $10^5$ & $10^6$ & noiseless & $10^5$ & $10^6$\\
\colrule
2 & 0.0010 & 0.027 & 0.0049 & \textbf{0.0009} & \textbf{0.008} & \textbf{0.0017}\\
3 & 0.025 & 0.47 & 0.115 & \textbf{0.029} & \textbf{0.076} & \textbf{0.031}\\
4 & \textbf{0.70} & \textbf{1.57} & \textbf{1.31} & 1.21 & 1.56 & 1.56\\
5 & 1.20 & 1.58 & 1.64 & 1.22 & 1.52 & 1.51\\
\end{tabular}
\end{ruledtabular}
\end{table}

\subsection{Exploratory (post-hoc) respecifications}
\label{sec:si-exploratory}

Two mechanisms in this paper are post-hoc respecifications of falsified
pre-registered predictions, fitted to the same data that falsified their
predecessors. They carry no main-text mechanism
status; they are collected here, clearly demarcated, pending
out-of-sample confirmation.

\begin{figure}[!t]
\centering
\includegraphics[width=\columnwidth]{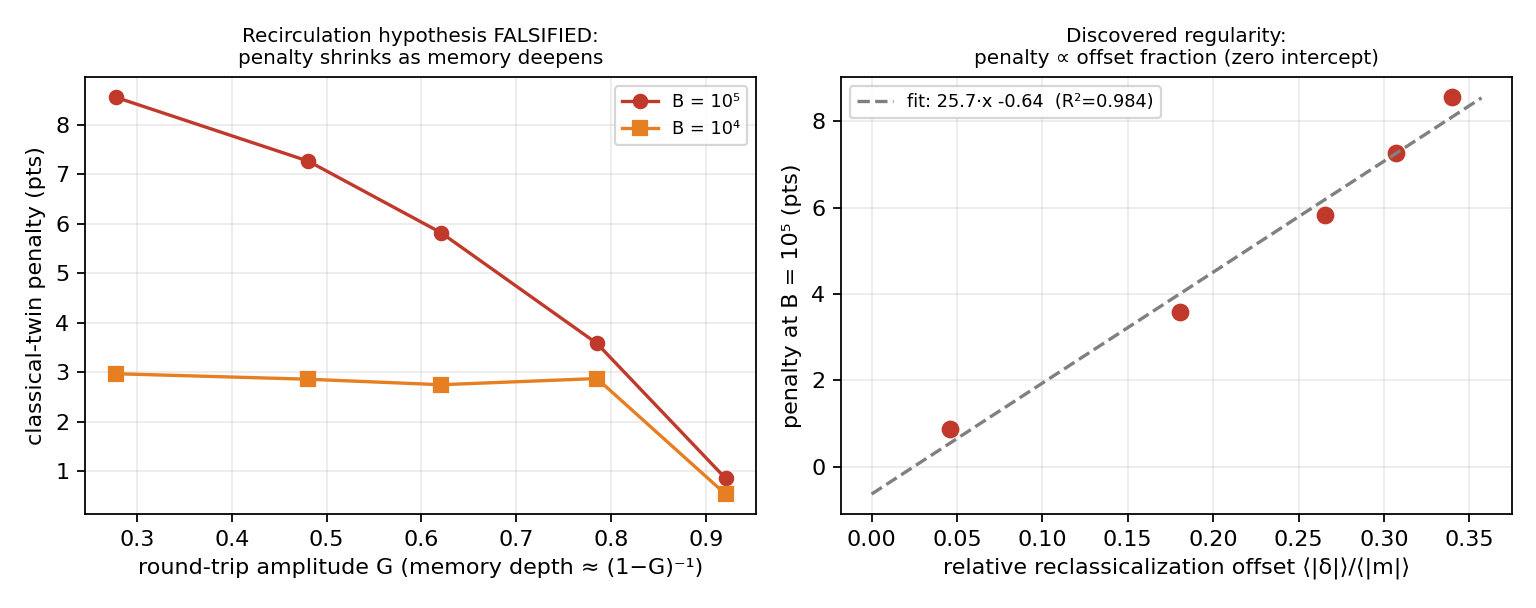}
\caption{\textbf{Falsification of the recirculation hypothesis.} Left: the
classical-light penalty falls, not rises, with memory depth. Right: it is
instead proportional to the reclassicalization offset fraction, with near-zero
intercept.}
\label{fig:gablation}
\end{figure}

\emph{(E1) The reclassicalization-offset description of the twin's
matched-noise penalty.} The pre-registered recirculation mechanism
failed (the penalty falls, not rises, with memory depth;
Fig.~\ref{fig:gablation}, left). The surviving description---penalty
proportional to the offset fraction
$\langle|\delta_{\rm cl}|\rangle/\langle|m|\rangle$ of main-text
Proposition~2(ii), with near-zero intercept
(Fig.~\ref{fig:gablation}, right)---was formulated after, and fit to,
the same $G$ sweep.

\emph{(E2) The budgeted memory-capacity law.} The pre-registered
noiseless cutoff prediction failed (details in the finding above). The
budgeted respecification, now supported at ten amplitudes
$G=0.60$--$0.92$: slope $0.085$ (bootstrap $95\%$ CI $[0.078,0.101]$),
intercept $-0.008$, $R^2=0.988$. Supported, exploratory, and
awaiting an out-of-sample amplitude set.

\subsection{Predicted measurements}

The remaining pre-registered figures are computable from the twin and
measurable on the device: \emph{(i)} the cross-time kernel map from pulse-pair probing at
separations $(m\tau,n\tau)$, imaging the delayed products of
main-text Eq.~(5) with first moments only; \emph{(ii)} the entanglement map
(log-negativity of the bin register) overlaid on the kernel map, testing whether
nonlinear memory tracks the entanglement bands; \emph{(iii)} the crossing experiment extended into the non-Gaussian tier
($\varepsilon_G$ large), the remaining regime where in-dynamics nonlinearity may
pay beyond the order-multiplier law demonstrated in the encoding-limited setting; \emph{(iv)} the node-count
optimum: at fixed $B$, per-bin estimator variance scales as $1/\Delta$, so
capacity versus $N$ has an interior optimum that readout squeezing shifts upward.

\section{Hyperparameter optimization of the quantum machine}
\label{si:tuning}

Every machine number in the main text is reported at a per-task,
per-budget optimized operating point. This section specifies the search
completely enough to reproduce it without the code bundle, states the
tuning budgets of both machines, and records the predictions that were
staked on it---including those that failed.

\subsection{Search space and constraints}

Three dials are searched: the per-bin squeeze parameter $r$, the
feedback transmission $\eta_{\rm fb}$, and the encoding gain $\beta$
(searched in $\log_2$). The mask family is held fixed at full width and
enters only as quenched disorder; all other parameters are design
frozen (Methods). Two constraints bound the box. Injectivity requires
$\beta s_{\max}<\pi$, so that the encoding map is invertible on the
drive range. Stability is \emph{not} imposed analytically: a candidate
point is admissible only if the covariance recursion converges under
period doubling, tested directly (tolerance $10^{-3}$ in vacuum units,
up to $24$ joint periods), and non-convergent points are recorded as
infeasible rather than scored. This is a deliberate change of practice:
an amplitude-based guard is a conservative certificate, not a boundary,
because under a generic mask the growth rate is that of a product of
non-commuting rotated squeezes.

\subsection{Objective and partitions}

The objective is the task metric evaluated on a validation block
(the final segment of the training partition), at the selection seed
only, averaged over three mask draws and three noise realizations, with
the regularizer fixed to the budget's shot level. Test partitions and
held-out task seeds are untouchable during the search. For maximization
metrics the objective is the corresponding loss ($C_{\max}-C$, or
$1-\text{accuracy}$). The surrogate consumes $\log(\text{loss})$
together with the per-point realization variance, so mixed-fidelity
evaluations remain consistent.

\subsection{Search procedure}

A Gaussian-process surrogate (Mat\'ern-$5/2$, automatic relevance
determination, fitted marginal likelihood) models the log-loss; a
second Gaussian process classifies feasibility from the settle-test
outcomes, and the acquisition is expected improvement weighted by the
squared feasibility probability. The schedule has two phases. A global
phase (scrambled Sobol initialization followed by constrained expected
improvement) enumerates basins; its trace is deliberately non-monotone,
since exploration of high-variance regions is the point. A refinement
phase then runs a trust region around the incumbent---expanding after
consecutive improvements, contracting after failures---escalating the
objective's fidelity (more mask draws and realizations) as the region
tightens, and finishing with a pure exploitation endgame. Fifty
evaluations are used per task and budget.

The champion is not the search's best observation. Candidates are
shortlisted by posterior mean under a distinctness filter, and each is
re-scored on a larger mask ensemble using validation data only; the
best re-scored candidate is the champion. This rule exists because it
is needed: at several points the best search-scale observation proved
unstable on unseen masks, in one case by more than three orders of
magnitude, and would otherwise have been reported.

\subsection{Tuning budgets of both machines}

The machine receives $50$ adaptive evaluations per task and budget; the
digital baseline receives a $36$-point grid, re-selected per task and
budget with the same validation protocol and averaged over five tuning
draws. The asymmetry favors the machine and is disclosed as such: no
baseline number in this work should be read as the best attainable by
its family.

\begin{figure}[!htb]
\centering
\includegraphics[width=\columnwidth]{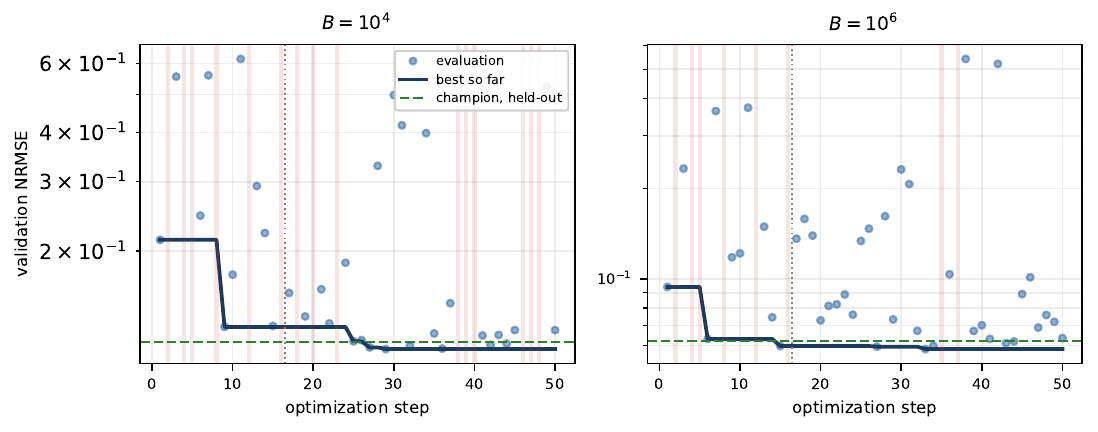}
\caption{Illustrative single runs of the hyperparameter search on
NARMA2, one per measurement budget, shown to document how the search
proceeds---not as a performance claim. Circles are individual
evaluations of the validation objective; the solid line is the best
observation so far; red bands mark evaluations that failed the
stability test and were recorded as infeasible ($15$ and $8$ of $50$
here), which is how the search maps the stability boundary
empirically; the dotted vertical line separates the global and
refinement phases; the dashed line is the champion's held-out level
after mask-ensemble re-scoring. Both runs are cold started: a run
initialized at an already-converged point produces a flat trace that
would misrepresent the search's behavior.}
\label{fig:bo}
\end{figure}

\subsection{What polynomial order does not buy}
Membership in a low sector does not imply reach at low polynomial order;
two
families make the block characterization sharp (proofs by the
interpolation bound of clause~(III); SI Sec.~S3). \emph{Parity:} the
charge-$0$, weight-$2$ character $e^{i\beta(s_a-s_b)}$ needs two
blocks of charge $\pm1$, so it lies outside $\mathcal A_1$, at
$\operatorname{dist}_\infty\ge\sin(\beta s_{\max})
\sin^2(\beta s_{\max}/8)\approx1.3\times10^{-2}$ at the campaign
point.
\emph{Per-slot:} a pattern with a slot exponent $2D{+}1$ escapes
$\mathcal A_D$ for every $D$, because $\le D$ feature factors cap
each slot at $2D$. Degree buys \emph{block count}, not sector
membership per se: the order-one projector reaches admissible
charge-one patterns of \emph{every} weight (Theorem~S1, SI, exact block
matching), while the budgeted route of the affordable-universality corollary of the main text
pays polynomial order equal to weight.

\subsection{Classical-light equivalence: clauses (i)--(ii) in detail}
In the channel basis of the one-step map, $\bm{\mathcal N}_{\rm cl}$ has
Laurent support $d\in\{-1,0,+1\}$ ($m$-channel charge $+1$,
$\mathsf J$-channel $0$, the $\bar m$ channel the forced conjugate at
$-1$): the twin is generated by the same gauge-graded transfer operator
plus a source inside the graded algebra, and every readable twin
monomial obeys the three kernel laws. The readable channel is
$m_{\rm cl}=m+\delta_{\rm cl}$ with $\delta_{\rm cl}$ the Dyson series
with one emission vertex per walk replaced by the charge-$(+1)$ source;
hence $\delta_{\rm cl}\in\mathcal A_1$ and all twin features lie in
$\mathcal A_D$, alone forbidding any noiseless twin capability outside
the machine's reach.

\subsection{Locality of the three kernel laws}
The per-slot bound and ordering law are \emph{local} (testable on
windows); the selection rule is \emph{global}, its operational test the
gauge projector (settle at $\theta_{\rm mask}+\chi$, harvest, Fourier
transform in $\chi$), not a windowed transform.

\subsection{Registered predictions and their outcomes}

The search was pre-registered; predictions and outcomes are recorded
here in full.
\emph{Confirmed:} the optimum migrates toward weaker coupling as the
measurement budget grows, in the direction the shot-cost model
predicts; mask-family performance exhibits a width knee, below which
conditioning degrades sharply.
\emph{Falsified and retracted:} (i) a stability boundary placed
analytically from an amplitude bound---direct measurement showed points
beyond that bound to be stable, and the inference was withdrawn, leaving
the amplitude condition as a conservative certificate; (ii) a
prediction that memory capacity would be independent of delay-line
length---measurement showed capacity growing with length under the
as-coded loss convention, and the prediction was retracted. Both entries
are kept because the corrected practice they produced---measure
stability and memory, never infer them from worst-case bounds---is part
of the method.

\section{Baseline constructions and the scoped classical-light control}
\label{si:baselines}

\subsection{Digital reservoir baseline}

The baseline is an echo-state network with sparse ($10\%$) Gaussian
recurrent weights rescaled to a chosen spectral radius, $\tanh$
activation, leak rate and input scaling as searched parameters, driven
by the same drive stream the machine encodes. Its readout matches the
machine's: the same feature tier, the same ridge with standardized
features and centered targets, the same washout, training and test
partitions, the same task seeds. Measurement noise is imposed at
matched per-feature signal-to-noise, referred to the machine's shot
level at its operating point; the digital machine is otherwise
noiseless. Selection is performed once per task and budget on the
validation block, averaged over five tuning draws, and frozen before
any held-out seed is touched. Performance is then reported as an
ensemble over $1000$ fresh reservoir realizations, quoted as mean with
quenched standard deviation.

For the RF corpus the baseline is architecture matched rather than
merely size matched: a bank of instances mirroring the machine's mask
bank, each emitting the same number of real read channels per bin, with
noise injected on those channels as the machine's shot noise enters its
harvests, followed by the identical feature expansion and readout, so
that both machines carry the same readout dimension.

\subsection{Classical-light twin: a scoped control}

The classical-light twin remains the control for the resource question:
the same loop, mask, transmission and readout, with the squeezed source
replaced by the closest classically representable state---the minor
axis lifted to vacuum, the major axis matched. At infinite budget it
computes the same functions, and the measurements confirm this at the
optimized operating points of both benchmark families and the RF corpus.
At finite budget its margin is an operating-point statement and not a
machine-class property: at the optimized NARMA2 points it changes sign
with budget ($-15.7\%$ at $B=10^4$, $+5.8\%$ at $10^6$), and
exploratory scans show a further sign change along the feedback axis,
correlated with a crossover in feature amplitude between the two
machines. We therefore report it as a control on the classical-light
equivalence proposition of the main text and rest no resource claim
on it. A dedicated registered campaign would be required before the
feedback-axis crossover could be used as a finding.

\section{Measurement economics and architecture walkthrough}
\label{sec:si-walkthrough}

This section retains, in full, the pricing of the two placements of
nonlinearity and the element-by-element walkthrough of the machine of
main-text Fig.~2, with the mechanism carried by each element and the
hardware basis of every quoted parameter.

\subsection{Measurement economics: pricing the two placements of nonlinearity}

Let us price the two placements of nonlinearity against each other. The
calculation is elementary, and the conclusion is stark.

Suppose we wish to estimate a feature to precision $\epsilon$ from
homodyne samples of a quadrature $X$ with single-shot variance
$\sigma_X^2=\mathrm{Var}(X)$; quadrature and mode conventions are as in
the main text. A first moment costs $\sigma_X^2/\epsilon^2$ shots.
That is a benign price, because $\sigma_X^2$ is bounded, and because for
the squeezed quadrature it can be pushed below vacuum. An $m$-th moment
costs $\mathrm{Var}(X^m)/\epsilon^2$ shots instead, and here
\begin{equation}
\mathrm{Var}(X^m) \;\sim\; (2m-1)!!\,\sigma_X^{2m}
\label{eq:si-factorial}
\end{equation}
for Gaussian-like statistics. The coefficient is the $2m$-th moment of a
zero-mean Gaussian, with $(2m-1)!!$ counting its pairings. At fixed
precision, then, order-$m$ features are exponentially expensive in shots.
The ``arbitrary nonlinear measurements'' that universality arguments
invoke so casually are unaffordable in any finite experiment.

The in-dynamics alternative computes its products before detection.
In the Heisenberg picture the same class of numbers is produced either
way, as noted in the main text. But dynamically the polynomial is evaluated in
amplitude, per shot, at optical bandwidth, and its polynomial order is
raised by
recursion rather than by measurement complexity.
Table~\ref{tab:duality} sets the two placements side by side. The readout
in our machine faces order-one statistics forever, at a variance that
the phase-sensitive amplifier renders immune to detection loss.

\begin{table}[!htb]
\caption{The two placements of nonlinearity.}
\label{tab:duality}
\begin{ruledtabular}
\small\setlength{\tabcolsep}{3pt}
\begin{tabular}{@{}lll@{}}
 & Measured~\cite{Nokkala2021} & Dynamical (here)\\
\colrule
Polynomial formed & across shots & within each shot\\
Estimator variance & factorial in order & bounded, squeezed\\
Order ceiling & fixed per shot & grows per pass\\
Feature count & $\sim N^2/2$ moments & $\sim$ few$\times N$ means\\
\end{tabular}
\end{ruledtabular}
\end{table}

Our placement carries one cost worth naming: feature
\emph{count}. First moments supply $O(N)$ numbers where covariance
readout supplies $O(N^2)$. We recover the missing multiplicity in two
ways. The first is temporal multiplexing, through the bin register. The
second is a bounded concession: we permit \emph{quadratic}
post-processing of the measured quadratures, which is the one benign rung
of the moment ladder. That concession also opens a direct classical
channel into the output covariances, complementing the coherent channel
through the $\chitwo$ transducer.

\subsection{Walking the signal path}

We now walk the signal path, giving the mechanism carried by each
element. Main-text Fig.~2 shows the machine, and Table~II of the
main-text Methods gives the fast-clock parameter set,
distinguishing hardware targets from the conservative campaign operating
point used for every quoted result.

\subsubsection{Bilinear encoding at the squeezer}

The 775-nm pump is carved into a 20-GHz pulse train, which serves as the
master clock, and phase-modulated so that bin $k$ of symbol $n$ meets the
on-chip squeezer ring with squeezing angle
\begin{equation}
\theta_k = \theta_{\mathrm{mask}}(k \bmod N') + \beta\, s_n ,
\label{eq:si-encoding}
\end{equation}
where $\theta_{\mathrm{mask}}$ is a fixed pseudorandom mask of period
$N'=N\pm1$. The clock offset braids the $N$ virtual nodes into one long
cycle~\cite{Appeltant2011}. Writing $(\bm m,\bm\Sigma)$ for the mean
vector and covariance matrix of the register's quadratures, the per-bin
symplectic update of the Gaussian state is
\begin{equation}
\bm\Sigma \to S_k\,\bm\Sigma\,S_k^{\top} + \mathcal N_k,\qquad
S_k\big|_{\rm sq} = R(\tfrac{\theta_k}{2})\,Z(r)\,R(\tfrac{\theta_k}{2})^{\top},
\label{eq:si-update}
\end{equation}
with $R(\phi)$ the $2\times2$ rotation by angle $\phi$ and
$Z(r)=\mathrm{diag}(e^{-r},e^{r})$ the squeeze along the rotated axes.
Here $S_k$ is the full per-bin symplectic update of the register;
$S_k|_{\rm sq}$ denotes its oriented-squeeze block, acting on the bin at
the squeezer, with the remaining blocks (interferometer, coupler, losses)
specified in Sec.~S2; and $\mathcal N_k$ is the noise injected by that
bin's losses.

The essential point is where the data sit. They appear \emph{inside} the
symplectic matrix $S_k$, and never as a displacement: with
squeezed-vacuum input, $\bm m = 0$ identically. Why is that a richer
channel than additive driving? A displacement \emph{adds} the input to
the state, so histories combine linearly, and the machine is a filter.
Steering the map instead \emph{multiplies} input-dependent, non-commuting
operations together, so the state after many bins is governed by the
ordered product $S_M\cdots S_1$. Because those factors do not commute,
the order in which data arrived matters, and products of inputs appear in
the state of their own accord. This is the same reason a recurrent
network with input-dependent weights outclasses a linear filter of the
same size, and it gives a function class that strictly exceeds linear
filtering before any non-Gaussian physics is invoked. Encoding in an
angle has a practical virtue too: it holds photon flux and stability
margins constant as the data vary.

\subsubsection{Cluster-chain non-Markovian memory (the coherent arm)}

The fundamental band circulates. A buried, ultra-low-loss Si$_3$N$_4$
spiral~\cite{Blumenthal2018} stores $\tau=5$~ns, giving $N=100$ bins at
$\Delta=50$~ps for about $0.1$~dB. A one-bin unbalanced interferometer
chains neighboring bins, an in-loop tuner pilot-locks the phase, and a
tunable Mach--Zehnder re-injects the delayed field into the squeezer
ring.

Re-injecting into the squeezer, rather than merely alongside it, is what
makes the memory coherent. Because the returning bin is squeezed
\emph{jointly} with fresh vacuum, the OPO correlates each bin with its
predecessor one round trip earlier. The loop is thereby the
time-multiplexed cluster-state generator of
Refs.~\cite{Yokoyama2013,Asavanant2019}, operating as a reservoir memory.

Why is the one-bin coupler necessary at all? The transducer reads
$\avg{a^2}$ of the single bin passing through it, so cross-time products
become readable only if that bin already superposes several epochs. The
loop alone correlates lags at multiples of $N$, a round-trip comb, which
would leave adjacent-symbol memory unreachable at polynomial order one.
The one-bin
interferometer adds unit lag, and the composition of the two over
successive circulations fills the comb in, at lags $mN\pm j$, producing
the banded kernels of main-text Eq.~(5). It is also what upgrades
$N$-spaced pairs into the connected cluster chain, and it does so at no
cost to the charge algebra, since a passive coupler never touches the
pump and is charge-neutral.

One dial controls the whole arrangement: the round-trip parametric
amplitude $G$, the product of pump power, MZ ratio, and loop
transmission. The echo-state property holds for $G<1$, memory depth is
$M\approx(1-G)^{-1}$ circulations, and the optimal operating point sits
just below feedback-OPO threshold. Alongside this coherent arm we also
provide a slower, electronic \emph{measured-feedback} arm, which
modulates the pump with a programmable delayed kernel of the homodyne
record, $\varepsilon_2(t)\propto\int h(t')Y(t-t')dt'$. It reaches memory
timescales photons cannot. Comparing the two arms at matched budget is
how we will isolate the operational value of entangled memory.

\subsubsection{Recursive $\chitwo$ and coherent moment down-conversion}

Each circulation passes a poled, group-velocity-matched,
\emph{single-pass} TFLN waveguide, gated by the co-propagating pump pulse
(temporal aperture $\approx$ the 2-ps walk-off over
6~mm~\cite{Wang2018,Lu2019,Nehra2022}). Being non-resonant, the segment
has no linewidth, and so accommodates the 20-GHz clock without smearing.
Its perturbative action per pass, with conversion amplitude $\mu\ll1$, is
the mean-field kick
\begin{equation}
\Delta\avg{b} = -i\mu\,\avg{a^2}
= -\tfrac{i\mu}{4}\big[(\Sigma_{XX}-\Sigma_{PP}) + 2i\,\Sigma_{XP}\big].
\label{eq:si-downconversion}
\end{equation}
Read the right-hand side carefully: it is a \emph{linear read-off of
three entries of the quadrature covariance $\bm\Sigma$}. The hardware has
done, optically, what a moment-readout scheme would have to do
statistically. (The kick itself is derived from the traveling-wave
equations in Sec.~S1; the covariance identity on the right-hand side
is derived in the main-text Methods.)

What makes those three entries interesting is that they are not local in
time. The loop and interferometer have already made the field at the
segment a coherent superposition of epochs,
$a(t)=\sum_m c_m\,\tilde a(t-m\tau)$, so the local entries in
Eq.~\eqref{eq:si-downconversion} are weighted sums of the banded cross-time
covariances $\Sigma(k,k{-}1)$, $\Sigma(k,k{-}N)$, and their compositions:
\begin{equation}
\avg{a^2(t)} = \sum_{m,n} c_m c_n\, \avg{\tilde a(t-m\tau)\,\tilde a(t-n\tau)} .
\label{eq:si-kernel}
\end{equation}
The cross terms are the machine's delayed nonlinear multiplications, that
is, its memory kernel. Past the bound of
Refs.~\cite{Duan2000,Simon2000} they are also entanglement correlators.
The division of labor is clean: linear delay \emph{assembles} the
multi-epoch superposition, instantaneous $\chitwo$ \emph{multiplies} it,
and recursion raises both the polynomial order and the temporal span of
the
products on every round trip. This is the quantum limit of the Ikeda
system~\cite{Ikeda1979}.

\subsubsection{Degree-one readout at full budget efficiency}

The 775-nm output passes an on-chip poled phase-sensitive amplifier
before leaving the chip. This is noiseless gain on the read quadrature,
and its purpose is to render detector inefficiency irrelevant. Its axis
is locked to the local oscillator, not to the rotating signal ellipse,
which is why the 20-GHz encoding poses no tracking problem (main-text Methods).
Balanced homodynes then read the field at polynomial order one, with a
775-nm local
oscillator picked off after the doubler, frequency-matched by
construction, and passively interleaved between the $X$ and $P$ bases by
a static $90^\circ$-plus-one-bin-delay Mach--Zehnder.

It is worth being concrete about what is actually read out. Each shot
returns one number: the windowed homodyne photocurrent of one quadrature
of the \emph{harmonic} field, $X_b^\vartheta$, for that bin. With
$\avg{b}\propto -i\mu\,m$, the means of the two interleaved bases are
\begin{equation*}
\avg{X_b^{0}}\propto \mu\,\Sigma_{XP},
\qquad
\avg{X_b^{\pi/2}}\propto -\tfrac{\mu}{2}\,(\Sigma_{XX}-\Sigma_{PP}),
\end{equation*}
so even bins deliver the imaginary part of $m$ and odd bins the real
part, and together they reconstruct the complex feature. One shot is not
the feature: the feature entering the trained readout is the $B$-shot
sample mean, whose standard error
$\smash{\Sigma_{X_bX_b}^{\vartheta\,1/2}/\sqrt B}$ is precisely the
variance the phase-sensitive amplifier squeezes. The detector thus
estimates a \emph{first} moment of $b$ whose value is a \emph{second}
moment of $a$: the order reduction is optical, and the statistics stay at
polynomial order one. The fundamental band itself is never detected; the
WDM routes
only the harmonic to the receivers, and parity guarantees the harmonic
mean field is the only nonzero first moment in the system.

We do not attempt simultaneous conjugate readout. The Arthurs--Kelly
cost~\cite{ArthursKelly1965} must be paid somehow, and interleaving pays
it in a halved per-quadrature rate rather than in 3~dB on every sample,
which beam splitting would cost. All timing derives from one RF
oscillator slaved to the loop's pilot tone, the synchronously pumped OPO
discipline. Envelope synchronization is therefore over-satisfied by five
orders of magnitude, simply by the phase lock that entanglement survival
demands anyway, and residual jitter appears only as small, analytically
priceable multiplicative gain noise.

\section{Session-level diagnostics: how the machine computes}
\label{sec:si-session}

This section is pedagogical and diagnostic. It contains no new claims: it
makes the encoding, harvest, and readout stages of Sec.~S1 visible on one
concrete NARMA2 session, exhibits the elementary non-commutation mechanism
that the reachability analysis of Sec.~\ref{sec:si-proofs} formalizes, and
reports how the test error of that session converges as harvest bins are
withheld from the readout---a task-level, single-session illustration of the
budgeted-universality trade-off of Sec.~\ref{sec:si-budget}. All panels use
the noiseless rung and the \emph{linear}-feature reconstruction protocol
(one symbol per mask period, $N'=61$
slots with the drive held across the symbol; washout/train/test
$100/1500/400$; features are the $61$ complex harvests, $122$ real
components; ridge with validation-selected regularization). Its noiseless
test error, NMSE $=0.0055$, reproduces the archived noiseless entry and differs from the ladder value
$0.0039$ of Sec.~\ref{sec:si-narma}, which was produced by the W6/A3
campaign configuration (quadratic feature layer, $305$ components, drive
mapped to $[-1,1]$); the difference is the feature tier, not the machine.
{\sloppy Generator scripts and their recorded outputs are bundled with
the code archive.\par}

\subsection{Non-commutation of oblique squeezes}
\label{sec:si-noncommute}

\begin{figure}[!t]
\includegraphics[width=0.78\textwidth]{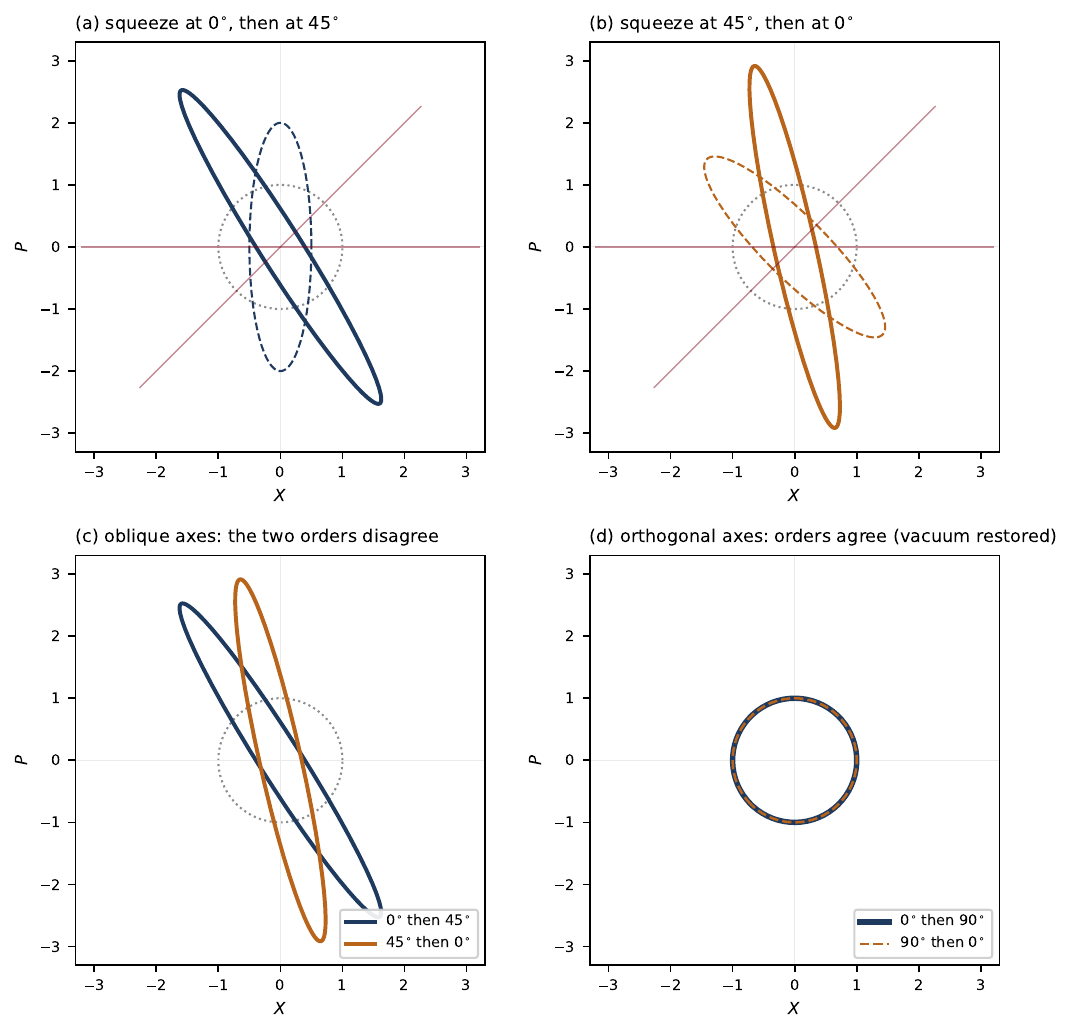}
\caption{Squeezes along different axes do not commute. Noise-ellipse
congruence $\Sigma\to S\Sigma S^{T}$ from vacuum (dotted circle), squeeze
strength $e^{r}=2$; thin lines mark the squeeze axes.
(a)~Squeeze along $0^{\circ}$ (dashed intermediate), then along
$45^{\circ}$ (solid). (b)~The same two operations in the opposite order.
(c)~Overlay of the two final ellipses: identical eigenvalues, different
tilt and shear---the difference is the commutator, made visible.
(d)~Control with orthogonal axes $0^{\circ}/90^{\circ}$: the two orders
coincide exactly and restore the vacuum, so order sensitivity requires
oblique angle differences.}
\label{fig:si-noncommute}
\end{figure}

Data enter the machine only through the squeeze angle, so the elementary
question is what composing squeezes at \emph{different} angles does that
composing squeezes at one angle cannot. Figure~\ref{fig:si-noncommute}
answers it at the level of a single mode. Write the squeeze along the axis
at angle $\phi$ with strength $e^{r}=2$ as the congruence
$\Sigma\to S_\phi\Sigma S_\phi^{T}$,
$S_\phi=R(\phi)\,\mathrm{diag}(e^{-r},e^{r})\,R(\phi)^{T}$. Starting from
vacuum ($\Sigma=\openone$), applying $S_{0}$ then $S_{45^{\circ}}$ gives
\begin{equation}
\Sigma_{0\to45}=\begin{pmatrix} 2.641 & -3.984\\ -3.984 & 6.391\end{pmatrix},
\qquad
\Sigma_{45\to0}=\begin{pmatrix} 0.531 & -1.875\\ -1.875 & 8.500\end{pmatrix},
\end{equation}
for the two orders (exact values for $r=\ln 2$): identical eigenvalues, but
different orientation and shear
[Fig.~\ref{fig:si-noncommute}(a)--(c)]. The composite state therefore
depends on the \emph{order} of the operations, and---because the angles
carry the data---on products of earlier and later inputs, not on their sum
alone. This order sensitivity is the machine's multiplication primitive: a
constant-angle schedule composes commuting maps and stores only a leaky
cumulative total of the drive, whereas an angle-diverse schedule composes
non-commuting maps whose commutators generate the rotations and shears
that make the harvested features genuinely multiplicative in the input
history (the group-theoretic form of this statement is the structure lemma
of Sec.~\ref{sec:si-prf-lemma}). Panel~(d) shows the control that sharpens
the requirement: squeezes along \emph{orthogonal} axes commute exactly (and
at equal strength cancel back to vacuum), so non-commutation demands
\emph{oblique} angle differences. A generic mask draw supplies oblique
consecutive angles automatically; this is one facet of the mask-genericity
condition invoked in Sec.~\ref{sec:si-narma}.

\subsection{Anatomy of one NARMA2 session}
\label{sec:si-anatomy}

Figure~\ref{fig:si-session} (placed in Sec.~\ref{sec:si-model}) shows
the three stages of one session on a common time axis. Panel~(a): the per-bin encoding phase
$\theta_k=\theta_{\rm mask}(k\,\mathrm{mod}\,N')+\beta s_k$ over six test symbols.
The fast intra-symbol structure is the frozen mask orbit
$\theta_{\rm mask}(1),\dots,\theta_{\rm mask}(61)$ (drawn once, i.i.d.\ uniform on
$[-\pi/2,\pi/2]$, then fixed for the whole session), repeated identically
every symbol; the data appear only as the rigid vertical offset
$\beta u_t$ between repetitions (step trace). Note the protocol
distinction: $s_k$ is the per-bin encoding sequence of the hardware, and
the per-symbol protocol sets $s_k=u_{\lfloor k/N'\rfloor}$---the task
drive held across the symbol. Panel~(b): the harvested readable channel
$m_k=(\Sigma_{XX}-\Sigma_{PP})/2+i\,\Sigma_{XP}$ of the head bin. The
waveform inherits the mask period, while its modulation across symbols is
the data- and history-dependent response of the register. Panel~(c): the
ridge readout $\hat y_t$ against the NARMA2 target over the test segment
(test NMSE $=0.0055$ for this session).

\subsection{Convergence with readout size}

\emph{Why the curve saturates.} The saturation is a span effect, not a noise effect: the participation ratio of the standardized training-feature covariance is $\approx\!2$ for every $K\ge2$ (maximum $122$ components), so successive bins re-express a small set of effective directions rather than enlarging the span. The height of the plateau at this session length is not a floor: lengthening the training block from $1{,}500$ to $6{,}000$ steps lowers the $K{=}61$ value from $0.0055$ to $0.0030$ (fixed seed, noiseless protocol), identifying a finite-sample estimation component that recedes with data while the saturation in $K$ is unchanged in mechanism.
\label{sec:si-readout-convergence}

Figure~\ref{fig:si-convergence} (placed in Sec.~\ref{sec:si-model})
withholds harvest bins from the readout of
the same session: for each $B_h\le 61$, only $B_h$ bins per symbol ($2B_h$ real
components) are retained and the identical ridge protocol is refit. The
canonical selection takes $B_h$ evenly spaced bins; a dispersion band from
ten random $B_h$-subsets per point quantifies selection sensitivity, which is
substantial only at very small $B_h$ (at $B_h=1$ the subset spread is
$0.27\pm0.07$ against the canonical $0.33$) and negligible beyond
$B_h\approx10$. The test error falls steeply---NMSE $0.33$ at $B_h=1$,
$0.0086$ at $B_h=20$---and flattens onto the full-readout floor $0.0055$
well before $B_h=61$. Two readings. First, this is the linear-readout
expressivity argument in empirical form: a single harvest supplies only
two fixed basis functions of the input history, and the $\sim\!60\times$
error reduction from $B_h=1$ to $B_h=61$ is the value of the remaining
basis. Second, it is a task-level, single-session illustration of the
budgeted-universality trade-off of Sec.~\ref{sec:si-budget}: accuracy is
purchased with readable components at a steep initial exchange rate that
saturates once the task-relevant subspace is spanned.